\documentclass[12pt, a4paper, twoside, openright]{book}
\usepackage{vuwthesis} 
\usepackage{palatino} 
\usepackage{url} 
\usepackage{graphicx}
\usepackage{epigraph}
\usepackage{bm}
\usepackage{amsmath}
\usepackage{amssymb}
\usepackage{mathrsfs}
\usepackage{titlesec}
\usepackage{tikz}
\usetikzlibrary{decorations.markings}
\usetikzlibrary{decorations.pathmorphing}
\usepackage{xcolor}
\usepackage[titles]{tocloft}
\usepackage{makeidx}
\usepackage[bookmarks=false]{hyperref}

\titleformat{\chapter}[frame]{\normalfont}{\filright CHAPTER \thechapter}{8pt}{\Huge\rm\filcenter}

\titleformat{\section} {\titlerule
\vspace{-2.8ex}%
\normalfont\sc} {\thesection.}{.5em}{}

\titleformat{\subsection} {\titlerule
\vspace{-2.8ex}%
\normalfont\sl} {\thesubsection.}{.5em}{}

\titlespacing*{\chapter}{0mm}{-40pt}{40pt}
\titlespacing*{\section}{0mm}{40pt}{10pt}
\titlespacing*{\subsection}{0mm}{40pt}{10pt}

\makeindex

\begin{document}

\def\d{{\mathrm{d}}}
\newcommand{\scri}{\mathscr{I}}
\newcommand{\sun}{\ensuremath{\odot}}
\def\J{{\mathscr{J}}}
\def\sech{{\mathrm{sech}}}
\def\T{{\mathcal{T}}}
\def\tr{{\mathrm{tr}}}
\def\diag{{\mathrm{diag}}}
\def\L{{\mathcal{L}}}

\def\Horava{Ho\v{r}ava}
\def\Aether{\AE{}ther}
\def\AEther{\AE{}ther}
\def\aether{\ae{}ther}
\def\UH{{\text{\sc uh}}} 
\def\KH{{\text{\sc kh}}} 
\def\ks{{ k_s }}
\def\vb{\mathbf{v}}
\def\n{\mathbf{n}}
\def\x{\mathbf{x}}
\def\s{\mathbf{s}}
\definecolor{gridcolor}{rgb}{0.3 0.4 0.4} 
\definecolor{gridlabelcolor}{rgb}{0.03 0.07 0.09} 

\def\rnd#1{
    \pgfmathprintnumberto[precision=2]{#1}{\temp}\temp
}

\frontmatter

\title{Strange Horizons: \\
\Large Understanding Causal Barriers Beyond General Relativity}
\author{Bethan Cropp \\
\vspace{1cm}
Supervisor: Stefano Liberati}

\subject{Astroparticle Physics}

\abstract{This thesis explores two avenues into understanding the physics of black holes and horizons beyond general relativity, via analogue models and Lorentz violating theories. Analogue spacetimes have wildly different dynamics to general relativity; this means time-independent black hole solutions have fewer symmetries, allowing the possibility of non-Killing horizons in stationary solutions. Surface gravity is one of the most important quantities characterizing black holes, with many physically distinct definitions. In the case of non-Killing horizons these different definitions of surface gravity are truly different quantities. This also has application to modified theories of gravity, where there is no reason to expect all horizons to be Killing horizons. In Lorentz violating theories, the situation becomes even stranger, as Killing horizons are at best low energy barriers, but for superluminal dispersion relations a true causal barrier, the universal horizon, may be present. Universal horizons are extremely interesting as they seem to be linked to the thermodynamic consistency of Lorentz-violating theories. Hence, we investigate the nature of these universal horizons via a ray tracing study, and delve into what happens near both the universal and Killing horizons. From this study we determine the surface gravity of universal horizons by the peeling properties of rays near the horizon. As the surface gravity is strongly linked to the properties of Hawking radiation, we investigate whether, and at what temperature these horizons radiate. Finally, we combine our investigations of universal horizons and analogue spacetimes, and ask why we have not seen a universal horizon in studies of analogue gravity. We examine some possibilities to include an \aether\ distinct from the velocity flow characterizing analogue spacetimes, laying the groundwork for an analogue universal horizon.}

\addcontentsline{toc}{chapter}{Abstract}

\phd


\maketitle

\newpage
\thispagestyle{empty}
\mbox{}

\chapter{Acknowledgments}

First and foremost, I wish to thank my supervisor, Stefano Liberati, for his patient guidance, advice, lively interest and helpful suggestions. 

I am much indebted to my other wonderful collaborators: Matt Visser, Arif Mohd, and Rodrigo Turcati. You who have been a great pleasure to work with, and a wealth of knowledge.

For other useful advice and fruitful discussions I wish to thank Thomas Sotiriou, David Mattingly, Ted Jacobson, Ian Vega, and Jishnu Bhattacharrya.  

The rest of the gravity group of SISSA, past and present: Eolo Di Casola, Dario Bettoni, Alessio Belenchia, Ramit Dey, Marco Letizia, and Dionigi Benincasa I wish to thank for the cheerful and friendly atmosphere. 

Finally, to all the friends and family who have supported me through this adventure. I thank you staying in touch from the other side of the world, or attempting to make me feel at home through cultural and language barriers, and putting up with me chattering about research. 

\newpage
\thispagestyle{empty}
\mbox{}

\renewcommand{\cftchapfont}{%
  \fontsize{13}{13}\bf\selectfont}  
  
\renewcommand{\cftsecfont}{%
  \fontsize{12}{13}\rm\selectfont}  
  
\renewcommand{\cftsubsecfont}{%
  \fontsize{11}{13}\sl\selectfont}  

\tableofcontents

\mainmatter

\chapter*{Preface}

\epigraph{If the semi-diameter of a sphere of the same density as the Sun were to exceed that of the Sun in the proportion of 500 to 1, a body falling from an infinite height towards it would have acquired at its surface greater velocity than that of light, and consequently...all light emitted from such a body would be made to return towards it by its own proper gravity}{John Michell (1724 - 1793)}

\noindent Black holes, while generally considered as objects predicted by general relativity (GR), also exist in many other theories of gravity. The key aspect of a black hole is that it is enclosed by a surface, the horizon, beyond which there may be no communication with the outside world. Due to this lack of visible structure, black holes are some of the simplest objects in our universe and, in many theories, are describable by just a handful of parameters.

This thesis is about black holes beyond the well-explored world of general relativity. In particular, it will consider two particular aspects; black holes in analogue gravity, and those in Lorentz violating theories of gravity. 

Analogue gravity emerged out of the realization that, at low energies, perturbations moving in condensed matter systems feel an effective curved metric, and thus can model the kinematics of gravity. Black hole geometries can be set up in such systems, and important effects such as Hawking radiation can be probed both theoretically and experimentally within this framework.

Importantly, the \emph{dynamics} of such systems differs wildly from dynamics of gravitational theories, meaning the analogue black holes typically do not have to obey the symmetries of stationary black holes in GR. This in turn can lead to novel features which are worth exploring both in their own right from the perspective of an experimental realization of these analogue black holes, and as a warm up to possible extended theories

The first part of this thesis is concerned with these more general black hole solutions, in particular studying the surface gravities of these systems. While in general relativity symmetries imposed on stationary solutions by the dynamics enforce strict equivalence of all possible definitions of surface gravity, in analogue systems the possibility that event horizons are no longer Killing allows for a plethora of possibly inequivalent definitions of the surface gravity. While these quantities are explicitly spelt out for analogue systems, this has wilder applicability for studying black holes in alternative theories of gravity, which may also lack the on-horizon symmetries of general relativity. 

In the case of alternative theories of gravity, it is illuminating to look at Lorentz-violating theories, which may admit superluminal dispersion relations. This necessarily alters the notion of a black hole \emph{is}, and hence what a horizon is. In particular we will examine black holes in Einstein-\Aether\  gravity and \Horava\ -Lifshitz Gravity. Surprisingly, in these theories a notion of black hole remains, separated from the outside world by a ``universal horizon". 

The new degrees of freedom present in the theories make it unclear how to apply some of the standard purely geometrical notions for these horizons to extract physical information about these black holes. One way to bypass this is to use ray tracing, giving a clear physical picture of these black holes. From this picture we can obtain a peeling notion of surface gravity, which is expected to relate closely to the temperature of Hawking radiation. In turn we shall show that this quantity appears to correspond to a suitable generalization to universal horizons of one of the standard definitions of surface gravity. 

Hawking radiation in this context is of particular interest because of potential problems with black hole thermodynamics in these theories, allowing processes that violate the Generalized Second Law of thermodynamics. At the time of writing, there are two calculations of Hawking radiation with opposite conclusions. These issues are also explored herein, and some new insight is provided albeit inconclusive so far.

Finally, as noted above, analogue models of gravitational systems have proved very useful in gaining an understanding of classical and semi-classical aspects of spacetimes. Therefore an analogue model of Einstein--\Aether\ black holes would be invaluable. Such a construction poses new difficulties with respect to standard general relativity, precisely because coupling to the extra degrees of freedom is essential to the physical relevance of the universal horizon. Here, we work on developing such an analogue model, using the little-explored system of relativistic Bose-Einstein condensates. In doing so we have accidentally solve a long-standing problem of how to create an analogue model for a system that includes vorticity. We further provide a candidate system for simulating universal horizons.

\section*{Arrangement of this thesis}

This thesis is arranged into six chapters. The first is a general introduction to the subjects of black hole thermodynamics, analogue gravity, and Lorentz violating theories which is necessary background to the research presented later. 

The other chapters are primarily composed of original research. The second chapter considers black hole solutions in analogue gravity, and the surface gravities of these black holes. This chapter is based on \cite{Cropp:2013zxi}, work done in collaboration with Stefano Liberati and Matt Visser. The third chapter is based on \cite{Cropp:2013sea}, and is concerned exploring universal horizons via ray-tracing, research with was carried out with Arif Mohd, Stefano Liberati and Matt Visser. 

Chapters four and five are based on currently unpublished research. Chapter four discusses numerous forays into deciding whether, and at what temperature universal horizons radiate. Though ultimately inconclusive, these investigations shed light on which approaches might work for shedding light on Hawking radiation in Lorentz-violating black holes. This is based on ongoing research with Arif Mohd and Stefano Liberati.  

Chapter five presents ongoing work with Stefano Liberati and Rodrigo Turcati, and discusses analogue black holes using relativistic Bose-Einstein condensates. Firstly, presenting black holes in these systems, which have not been investigated in detail, before going on to discuss how to create and analogue \aether\ field.

The final chapter draws some conclusions and discusses further possible avenues of research.

\chapter{Introduction}
\epigraph{Gravity is a habit that is hard to shake off}{Terry Pratchett, \it{Small Gods}}

\section{Black holes --- the hydrogen atom of quantum gravity}
\label{bhs}

It is clear the general relativity, though a wildly successful theory, and passing all experimental tests to date (see, for instance \cite{Will}), cannot be valid at all energies. Fundamentally, the problem is that is it unclear how to combine gravity with quantum mechanics. From the QFT point of view, general relativity is a perturbatively non-renormalizable theory, and at best can be regarded as an effective field theory valid at low energies. This is not enough to answer questions about, for instance, singularity resolution. There also remain deep philosophical questions about how to reconcile a theory of dynamical spacetime to standard quantum mechanics which takes place in a spacetime. 

Black holes are an ideal arena in which to explore the failures and possible extensions of GR. They are typically regions of strong gravity which are as simple as possible whilst being non-trivial, and they exist (in the sense that there is something dense, dark and small, that is well approximated by a black hole) for us to potentially study \cite{bhevidence}. 

Let us take a historical tour of the hints that black holes have given us on the nature of gravity and spacetime. 

It was first understood by Planck, well before the formulation of general relativity, that using the constants $c$, $\hbar$, and $G_N$ one can create natural units of length, time, and mass. These units are too small or big for our current technology to even approach directly measuring these scales. 

Such units became more interesting after Schwarzschild's discovery of the black hole solution to general relativity \cite{Schwarzschild:1916uq}. The Planck mass is precisely the mass at which, the Schwarzschild radius is the Compton wavelength. This implies that at this scale, both gravity and quantum mechanics are crucially important. Naively, locating a Planck-mass black hole  within the accuracy of its radius means that the momentum has such an uncertainty that one could create another black hole of equal size. 

The Schwarzschild solution is the simplest black hole: a static, spherically symmetric, vacuum solution to Einstein's equations. By Birkhoff's theorem \cite{birkhoff, jebsen} it is the unique metric for the spacetime outside any spherically symmetric asymptotically flat source for GR (this has also been proven for f(R) theories \cite{Nzioki:2013lca} and Lovelock gravity \cite{Zegers:2005vx}).
 In Schwarzschild coordinates (also sometimes referred to as curvature coordinates) it takes the form
\begin{equation}
\d s^2 = -f(r)\d t^2 + \frac{1}{f(r)}\d r^2 +r^2\d \Omega^2, \qquad f(r) = 1-\frac{2M}{r}.
\end{equation}
At the surface $r= 2M$ this solution appears to break down (as well as at $r=0$). This is a signal, not of a true singularity, but of an event horizon. One can also work in horizon-penetrating coordinates such as Eddington--Finkelstein coordinates
\begin{equation}
v= t + r_* ; \qquad u= t - r_*;\qquad r_* = \int \frac{\d r}{1-\frac{2M}{r}},
\end{equation}
so the metric takes the form
\begin{equation}
\label{eq:ef}
\d s^2 = -f(r)\d v^2 + 2\d v\d r +r^2\d \Omega^2.
\end{equation}
The existence of such coordinates shows that $r=2M$ is only a coordinate singularity (as opposed to the singularity at $r=0$ which is a true curvature singularity, as can be computed by, for instance, checking $R_{abcd}R^{abcd}$). However the reason a coordinate singularity appears in such natural coordinate systems is that $r=2M$ is a special surface, the event horizon, the surface from which light cannot escape. One way to see this is to note that the surface $r=2M$ is defined by constant $u$, and is thus null. Further, as the Schwarzschild is static, it is equipped with the timelike Killing vector $\chi = \frac{\partial}{\partial t}$. At the Killing horizon $\chi^2 = 0$ so is null, and inside $\chi$ becomes spacelike.

Note that Eddington-Finkelstein coordinates are only one of many possible horizon penetrating coordinates. Another common choice are Painleve -- Gullstrand coordinates, which shall be useful later. Defining
\begin{equation}
T = t + \int \frac{\sqrt{2M/r}}{1-\frac{2M}{r}} \d r,
\end{equation}
the line element transforms to
\begin{equation}
\label{PGcoords}
\d s^2 = -f(r)\d T^2 + 2\sqrt{1-f(r)}\d T\d r + \d r^2 +r^2\d \Omega^2.
\end{equation}



The Schwarzschild solution is highly idealized, and many doubted objects such as black holes would exist in the real world (as real matter twists, shears, and has pressure). However, in the 60s and 70s a series of singularity theorems were developed proving that general and physically reasonable matter configurations can form singularities \cite{HawkingEllis}. 
Further, in the 1960s Kerr discovered a rotating black hole solution \cite{kerrsol} that is both more complicated and more astrophysically relevant. It is a axisymmetric, stationary, vacuum solution for a black hole. Significant work over a number of years has gone into proving that this is the unique solution. However, there is no equivalent of the Birkhoff theorem: outside a rotating star, the metric is not uniquely Kerr (but should approximate Kerr at large distances). 

In Boyer-Linquist coordinates the solution is
\begin{eqnarray}
\d s^2 &=&-\left(1-\frac{2M}{r\left(1+ \frac{a^2\cos^2\theta}{r^2}\right)} \right)\d t^2 - \frac{4M \sin^2 \theta}{r\left(1+\frac{a^2\cos^2\theta}{r^2}\right)} \d t\d \phi \nonumber \\
&+& \frac{1+ \frac{a^2\cos^2\theta}{r^2}}{1-\frac{2M}{r}+\frac{a^2\cos^2\theta}{r^2}}\d r^2 +r^2\left(1+ \frac{a^2\cos^2\theta}{r^2}\right)\d \theta^2 \nonumber \\
&+&r^2 \sin^2 \theta \left(1+ \frac{a^2}{r^2}+\frac{2M \sin^2 \theta}{r^3\left(1+ \frac{a^2\cos^2\theta}{r^2}\right)}\right)\d \phi^2.
\end{eqnarray}
Here $a = \frac{J}{M}$, where $J$ can be identified as the angular momentum of the black hole. If $a = 0$ we recover the Schwarzschild solution, while $M=0$ is 
\begin{eqnarray}
\d s^2 &=& -\d t ^2 + \frac{1+ a^2\cos^2\theta/r^2}{1+a^2/r^2}\d r^2 +r^2\left(1+a^2\cos^2\theta/r^2\right)\d \theta^2 \nonumber \\
&+&r^2\sin^2\theta\left(1+\frac{a^2}{r^2}\right)\ d \phi^2
\end{eqnarray}
which can be seen to be (in very odd coordinates) Minkowski space. 

There are event horizons when $r_\pm =M \pm \sqrt{M^2-a^2}$. The two Killing vectors are, $\xi = \frac{\partial}{\partial t}$ and $\psi = \frac{\partial}{\partial \phi}$, but more useful is the combination $\chi= \xi + \Omega_H \psi$ where $\Omega_H =\frac{a}{r_+^2+ a^2}$, is the rotation of the horizon. This is the (unique) Killing vector that is null on the event horizon.

Furthermore, there is a new surface, where the timelike Killing vector, $\xi$, becomes null, known as the ergosurface, at $r_E =M + \sqrt{M^2-a^2\cos^2\theta}$. This is the point at which it is no longer possible to ``stand still".

For $a=M$ the black hole is maximally rotating, meaning the inner and outer horizons coincide. For $a >M$ the horizons vanish and the black hole becomes a naked singularity.

\subsection{Penrose process and superradiance}
\label{penrose}

In the late 1960s Penrose \cite{Penrose:1969pc} devised a mechanism for extracting energy from these rotating black holes. Start with particle $A$, with energy, as measured at infinity, of $E_A \equiv -\chi\cdot p_A$, where $\chi$ is the timelike Killing vector. This quantity is conserved along geodesics and is the energy measured at infinity. Let $A$ drop into the ergosphere of a black hole, where it splits into particles $B$ and $C$, which have $E_B \equiv -\chi\cdot p_B$ and $E_C \equiv -\chi\cdot p_C$ respectively. As we are in the ergosphere, $\chi$ is now spacelike and thus, along certain geodesics, $E_B$ can be negative. We let $B$ fall behind the horizon, carrying this negative energy, while $C$ escapes. As the energies are conserved, $E_C > C_A$ and we have extracted energy from the black hole. 

This process also necessarily causes the black hole to lose angular momentum, in turn shrinking the volume of the ergoregion. If repeatedly carried out, eventually all the rotational energy will be extracted, the ergoregion will vanish, and one will be left with a Schwarzschild black hole. As with all processes for extracting energy one can ask how efficient can this process be made. It turns out the the maximally efficient way to extract energy in the Penrose mechanism is through a process by which the area of the horizon (which depends on both the mass and the angular momentum) is unchanged. 

A similar process exists for fields incident on a spinning black hole, known as superradiance  can occur (for an extensive review see \cite{Brito:2015oca}). The simplest way to see this, (following \cite{Wald}) is to take a scalar field of form $\psi = \Re [\psi_0(r, \theta)e^{i(-\omega t+ m\phi)}]$ the energy current is given by
\begin{equation}
J_a = -T_{ab}\xi^b\, ; \quad T_{ab}= (\nabla_a\psi)(\nabla_b\psi)- \frac{1}{2}g_{ab}(\nabla_c\psi \nabla^c\psi +m^2\psi^2) 
\end{equation} 
The flux out of a region containing the horizon can be calculated by the flux over the horizon by the innner product with the current and the normal, $n$, to the horizon
\begin{eqnarray}
\left\langle J_an^a\right\rangle &=& \left\langle J_a\chi^a\right\rangle =\left\langle T_{ab}\chi^a\xi^b\right\rangle \nonumber \\
&=& \left\langle (\chi^a\nabla_a\psi)(\xi^b\nabla_b\psi) \right\rangle =\frac{1}{2}\omega (\omega -m\Omega_H)\left|\psi_0\right|^2
\end{eqnarray}
Thus if $\omega < m\Omega_H$ the energy flux into the black hole becomes negative. 
Note how crucially the difference between the time translation Killing vector, $\xi$, and the linear combination of radial and time Killing vectors, $\chi$, which defines the horizon, comes in. 

This seems like a simulated emission, as in a laser, and the question arises of whether spontaneous emission can also occur. These two techniques for extracting energy from a black hole paved the way to black hole mechanics, Hawking radiation, and finally black hole thermodynamics. 

\subsection{Surface gravity}
\label{introsurfgrav}

Before we go on to state the four laws of black hole thermodynamics, we need to introduce one useful quantity associated to the horizon, the surface gravity, which is essentially a measure of the strength of the pull of gravity at the horizon. This section will mostly follow \cite{Wald}. 

At the horizon $\chi$ is both null and perpendicular to the horizon, meaning $\nabla^a(\chi^b\chi_b)$ is also perpendicular to the horizon, so there is a function of proportionality such that
\begin{equation}
\nabla^a(\chi^b\chi_b)= -2\kappa\chi^a.
\label{kappanormalintro}
\end{equation} 
This $\kappa$ is the surface gravity, and can be shown to be a constant over the horizon. 

The properties of the symmetry of the Kerr solution means that on the horizon, the Killing vector is geodesic and foliates the horizon, meaning it obeys the geodesic equation, so one may re-express $\kappa$ via 
\begin{equation}
\chi^a\nabla_a \chi^b = \kappa \chi^b
\label{kappageointro}
\end{equation}
One can physically understand this by considering the force on a (co-rotating, ideal) rope held at infinity. Through a series of slightly tedious but straightforward manipulations using the Killing, and geodesic equations and the Frobenius theorem \cite{Wald}, one can arrive at
\begin{equation}
\kappa^2=\lim_H \frac{\left(\chi^b\nabla_b \chi^c \right)\left(\chi^a \nabla_a \chi_c \right)}{\chi^d\chi_d}
\end{equation}
where the limit is taken on-horizon. This can be see as a combination of the redshift, $\chi^a\chi_a$, and the acceleration, 
\begin{equation}
A^a = \frac{\chi^a\nabla_a \chi^b}{\chi^c\chi_c} , 
\end{equation}
which is the force needed to hold a object static against gravity. 
One can rewrite $\kappa$ as
\begin{equation}
\kappa = \lim_H \sqrt{-\chi^2}\left\|A\right\|.
\end{equation}
Now the reason for the name surface gravity becomes clear: the force to hold a mass stationary at the horizon is infinite, but so is the redshift. Holding a rope at infinity attached to a test mass at the horizon gives a sensible, finite calculation of the force of gravity at the horizon surface.

Note that many of these calculations hinge on exact geometrical properties, and once any of these properties fail to hold, notions of surface gravity become a lot murkier. Even in the case of spherically symmetric evolving black holes, a quite significant amount of work has been devoted to considering extensions to the usual notion of surface gravity that would be suitable for dynamical situations in standard general relativity, such as a forming or evaporating black hole (see, for instance~\cite{Nielsen:2005af, Nielsen:2007ac, Pielahn:2011ra} and~\cite{Hayward:1993wb, Fodor:1996rf, Hayward:1997jp,  Booth:2003ji, Booth:2006bn}). The difficulty of the task is not so surprising; in dynamical scenarios the event horizon may not even exist and many notions of a locally defined horizon can be constructed \cite{Schnetter:2006yt, dynamichorizons}. 

Consider the simplest possible scenario of a \emph{slowly} evolving black hole;
there are essentially two basic conceptions of surface gravity, related to the inaffinity of null geodesics \emph{on the horizon} \ref{kappanormalintro}, and the peeling off properties of null geodesics \emph{near the horizon} \ref{kappageointro}, respectively. 
For stationary Killing horizons these two notions coincide, but even in the simplest case of a spherically symmetric dynamical evolution these are two quite distinct quantities. We will work thorough a brief calculation, adapted from~\cite{Barcelo:2010xk} (see also~\cite{Barcelo:2010pj}), as an example.
Without loss of generality, write the metric  for an evolving spherically symmetric black hole in the form
\begin{equation}
\d s^2 = - e^{-2\Phi(r,t)} [1-2m(r,t)/r)] \d t^2 + {\d r^2\over1-2m(r,t)/r} + r^2 \{ \d\theta^2+\sin^2\theta\; \d\phi^2\} ,
\end{equation}
and define the ``evolving horizon'', $r_H(t)$,  by the location where $2m(r,t)/r = 1$. 
(Working from the Kodama vector, a ``geometrically natural'' justification for interest in this particular form of the line element is presented in \cite{Abreu:2010ru}.)

\subsubsection{Peeling off properties of null geodesics}\label{peeling}

A radial null geodesic satisfies
\begin{equation}
\left({\d r\over\d t}\right) =  \pm e^{-\Phi(r,t)} [1-2m(r,t)/r)].
\end{equation}
If the geodesic is near $r_H(t)$, that is  $r \approx r_H(t)$, then we can Taylor expand
 \begin{equation}
{\d r\over\d t} =  \pm {e^{-\Phi(r_H(t),t)} [1-2m'(r_H(t),t)]\over r_H(t)} \; [r(t) - r_H(t)] +{\cal O}\left( [r(t) - r_H(t)]^2 \right),
\end{equation}
where the dash indicates a radial derivative. That is, \emph{defining}
\begin{equation}
\kappa_\mathrm{peeling}(t) =  {e^{-\Phi(r_H(t),t)} [1-2m'(r_H(t),t)]\over 2 r_H(t)},
\end{equation}
which, in the static case, reduces to the standard result~\cite{Visser:1992qh}
\begin{equation}
\kappa = {e^{-\Phi_H} (1-2m'_H)\over 2 r_H},
\end{equation}
we have
 \begin{equation}
{\d r\over\d t} =  \pm 2 \kappa_\mathrm{peeling}(t) \; [r(t) - r_H(t)] +{\cal O}\left( [r(t) - r_H(t)]^2 \right).
\end{equation}
Then, for two null geodesics $r_1(t)$ and $r_2(t)$ on the \emph{same} side of the evolving horizon
\begin{equation}
{\d |r_1-r_2|\over\d t} \approx 2 \kappa_\mathrm{peeling}(t) \; |r_1(t) - r_2(t)|,
\end{equation}
(automatically keeping track of all the signs), so
\begin{equation}
|r_1(t)-r_2(t)| \approx |r_1(t_0)-r_2(t_0)| \; \exp \left[ 2\int \kappa_\mathrm{peeling}(t) \d t \right].
\end{equation}
This makes manifest the fact that $\kappa_\mathrm{peeling}$ as we have defined it is related to the exponential peeling off properties of null geodesics \emph{near the horizon}.

\subsubsection{Inaffinity properties of null geodesics}

Consider the outward-pointing radial null vector field
\begin{equation}
\ell^a = \left( 1,  e^{-\Phi(r,t)} (1-2m(r,t)/r), 0, 0\right).
\end{equation}
In a static spacetime, this null vector field is very simply related to the Killing vector, 
\begin{equation}
\ell^a = \chi^a + \epsilon^a{}_b \chi^b,
\end{equation}
where $\epsilon_{ab}$ is a 2-form acting on the $r$--$t$ plane, normalized by $\epsilon^{ab} \, \epsilon_{ab} = - 2$. The radial null vector field $\ell^a$ is automatically geodesic. 
Hence the inaffinity  $\kappa_\mathrm{inaffinity}(r,t)$ can be defined by 
\begin{equation}\label{radialinaffinity}
\ell^a \nabla_a \ell^b = 2 \kappa_\mathrm{inaffinity}(r,t) \; \ell^b,
\end{equation}
which \emph{always} exists, everywhere throughout the spacetime. 
This construction naturally extends the notion of on-horizon geodesic inaffinity, defined in a static spacetime as
\begin{equation}
\chi^a \nabla_a \chi^b = \kappa_\mathrm{inaffinity} \; \chi^b.
\end{equation}
That is,  equation (\ref{radialinaffinity}) naturally defines a notion of surface gravity even for a time-dependent geometry.
A brief calculation shows that at the evolving horizon~\cite{Nielsen:2005af, Abreu:2010ru},
\begin{eqnarray}
\kappa_\mathrm{inaffinity}(r_H(t),t) &=& {e^{-\Phi(r_H(t),t)} [1 - 2 m'(r_H(t),t)] \over 2r}  - {1\over2}\dot\Phi(r_H(t),t),  \nonumber\\
&=& \kappa_\mathrm{peeling}(t)  - {1\over2}\dot\Phi(r_H(t),t) .
\end{eqnarray}
While we do not \emph{a priori} know exactly where the event horizon (absolute horizon) is, we can certainly assert that when asymptotically approaching a  quasi-static situation the event horizon will be close to the evolving horizon. We then have
\begin{equation}
r_E(t) \approx r_H(t),
\end{equation}
in which case we can expand in a Taylor series
\begin{equation}
\kappa_\mathrm{inaffinity}(r_E(t),t)\approx  \kappa_\mathrm{inaffinity}(r_H(t),t) +  \kappa_\mathrm{inaffinity}'(r_H(t),t) [r_E(t)-r_H(t)].
\end{equation}
That is
\begin{equation}
\kappa_\mathrm{inaffinity}(r_E(t), t)\approx   \kappa_\mathrm{peeling}(t)  - {1\over2}\dot\Phi(r_H(t),t)   +  \kappa_\mathrm{inaffinity}'(r_H(t),t) [r_E(t)-r_H(t)].
\end{equation}
In particular, for sufficiently slowly evolving horizons the two concepts are for all practical purposes indistinguishable.

\subsection{Four laws of black hole thermodynamics}
\label{fourlaws}

In addition to the Penrose process and superradiance, some interesting hints towards possible new physics came from a series of ideas in the 1970s, starting with Hawking noting that in collisions between black holes the (total) area of the black hole horizons never decreases \cite{Hawking:1971tu}. Shortly thereafter, Wheeler proposed a gedanken experiment: mix hot and cold tea together, then drop it behind the horizon of a black hole. All traces of your ''crime'' of increasing the entropy of the universe are now hidden (this also works for hiding evidence of other crimes very effectively).  In response, Bekenstein \cite{Bekenstein:1972tm}, \cite{Bekenstein:1973ur}, \cite{Bekenstein:1974ax}, inspired by processes of energy extraction from black holes, and black hole collisions in which the area must increase, proposed associating an entropy to the horizon area. That it is the \emph{area}, as opposed to the \emph{volume} is of particular interest. 

If an object has an entropy that implies some sort of microstructure, and therefore it should also have a temperature. It was eventually proved by Hawking in \cite{Hawking:1974sw} that black holes do indeed have a temperature. Such Hawking radiation relies on the fact that in curved spacetime different one obcserver may measure nothing with their particle detector, while another obsever can measure particles for the same state (further discussion of Hawking radiation is reserved for chapter \ref{Hawking}). This last piece in the puzzle means the analogy with the laws of thermodynamics seems to be complete.   

\begin{itemize}

	\item Zeroth law: The surface gravity is constant over the horizon. \\

This is non-trivial and can be proven in two ways. Either use the field equations of general relativity and the dominant energy condition, and the assumption of stationarity \cite{HawkingEllis}. Alternatively, assuming that the horizon is Killing, and that the spacetime is either static or $t-\phi$ symmetric \cite{Racz:1995nh}.
	
	\item First law: $\d M = \frac{\kappa}{8\pi G_N}\d A + \Omega \d J$. \\

	This is in direct correspondence to the usual first law $\d U=T\d S + P\d V$, with Hawking temperature $T=\frac{\kappa}{2\pi}$ and black hole entropy $S=\frac{A}{4}$. Note that the efficiency of the Penrose process of section \ref{penrose} is maximum when the horizon area does not change, i.e. for a reversible process.

	\item Generalized Second Law: In any process $\delta S_{BH} +\delta S_{\mathrm{outside}} \geq 0$, where $S_{BH}$ is calculable from observable properties of the black hole alone. 
	
	This is basically a statement that entropy of black holes can be calculated by features observable to an outside observer --- in the case of general relativity, proportional to the horizon area. The net entropy of the outside universe and black hole entropies cannot decrease.
	This is the generalization of the fact that area cannot decrease in processes where the strong energy condition s obeyed. In processes such as Hawking radiation the area will decrease, but the entropy of  the universe is increased. Note that, with reasonable assumption for the scaling of the entropy of the black hole (the best we can do in absence of a full quantum theory), without the generalized second law one can theoretically build some types of perpetual motion machines. While an understanding of the number of quantum gravity microstates could fix this issue, it is highly distasteful.
	Several proofs of the generalized second law in general relativity have been proposed, for an informative review see \cite{Wall:2009wm}. 
		
	\item Third Law: The surface gravity of a black hole cannot be reduced to zero in a finite number of steps. 
	
	This is the equivalent of the ``process version'' of the third law of thermodynamics: that is is impossible to reach absolute zero in a finite number of steps. An alternative version, which states that the entropy of a system cannot be reduced to zero has no clear correspondence to black hole thermodynamics. 
	
Although the third law has been proven under certain assumptions \cite{isreal}, it is not clear whether this covers all scenarios of physical interest	 see discussions in \cite{Hubeny:1998ga, Belgiorno:2002pm, deFelice:2001wj, Jacobson:2009kt, Richartz:2011vf, Chirco:2010rq, Gao:2012ca, Saa:2011wq, BouhmadiLopez:2010vc, Bouhmadi-Lopez:2011lra, Colleoni:2015afa, Barausse:2011vx, Barausse:2010ka}).
\end{itemize}

For further details on various aspects of black hole thermodynamics see the reviews \cite{Jacobson:2003vx, Padmanabhan:2009vy, Wald:1999vt},  and various references therein.
 
Black hole thermodynamics are tantalizing, but one may wonder if they are properties of these particular solutions of general relativity, or are manifestations of a deeper thermodynamic character of the theory. 

\section{The Einstein Equation as a Thermodynamic Relation}

In an attempt to strengthen the relation between general relativity and thermodynamics Jacobson \cite{Jacobson:1995ab} showed that the Einstein equations can be derived from thermodynamic identities on the Rindler horizon. Here we shall succinctly review the basic steps that are key to this derivation. 

\subsection{The Rindler horizon}

Take standard Minkowski space 
\begin{equation}
\d s^2 = -\d t^2 + \d z^2 +\d x^i \d x_i .
\end{equation} 
With the transform
\begin{equation}
t=\xi \sinh(\kappa \tau)\, ; \qquad z = \xi \cosh(\kappa \tau)
\end{equation}

\begin{figure}[!htb]

\begin{minipage}[c]{0.7\textwidth}
\begin{tikzpicture}[%
    scale=5,%
		xshift=5cm,
    maingrid/.style={draw=gridcolor,very thick},%
    subgrid/.style={draw=gridcolor,thin},%
    tlabels/.style={pos=0.88,above,sloped,yshift=-.3ex,gridlabelcolor},%
    label/.style={%
        postaction={%
            decorate,%
            transform shape,%
            decoration={%
                markings,%
                mark=at position .65 with \node #1;%
            }%
        }%
    },%
]%
    \pgfmathdeclarefunction{arcosh}{1}{\pgfmathparse{ln(#1+sqrt(#1+1)*sqrt(#1-1))}}
    \pgfmathsetmacro{\Xmax}{1.2}
    \pgfmathsetmacro{\Tmax}{1.2}
    \pgfmathsetmacro{\g}{1}
    \newcommand\mylabelstyle\tiny

    \foreach \t in {-3,-2.75,...,3}{%
        \path[maingrid] (0,0) -- (\Xmax,{\Xmax*tanh(\g*\t)});
    }   

   
    \foreach \xx in {0.2,0.4,...,1}{%
        \path[maingrid]
            plot[domain=-{arcosh(\Xmax/\xx)/\g}:{arcosh(\Xmax/\xx)/\g}]
            ({\xx*cosh(\g*\x)},{\xx*sinh(\g*\x)});  
    }

    \foreach \t in {-1,0,...,1}{%
        \path (0,0) -- (\Xmax,{\Xmax*tanh(\g*\t)})
            node[tlabels] {$\tau=\t$};
    }
    \foreach \xx in {0.6,1.01}{%
        \path[gridlabelcolor,label={[above]{$\xi=\rnd{\xx}$}}]
            plot[domain=-{arcosh(\Xmax/\xx)/\g}:{arcosh(\Xmax/\xx)/\g}]
            ({\xx*cosh(\g*\x)},{\xx*sinh(\g*\x)});  
    }

    \draw[thick,-stealth] (0,0) -- (\Xmax,0) node[below] {$z$};
    \draw[thick,-stealth] (0,-\Tmax) -- (0,\Tmax) node[left] {$t$};
    \draw[dashed] (0,0) -- (\Xmax,\Tmax) 
        node[pos=0.37,above,sloped,yshift=-.3ex] {\mylabelstyle$x=0$}
        node[tlabels,black] {\mylabelstyle$\xi=\infty$};
    \draw[dashed] (0,0) -- (\Xmax,-\Tmax)
        node[tlabels,black] {\mylabelstyle$\xi=-\infty$};     
	
\end{tikzpicture}

\end{minipage}\hfill
\begin{minipage}[c]{0.3\textwidth}

\hspace{-2cm}

    \caption{The Rindler Wedge: higher acceleration corresponds to a path closer to the $z=|t|$ surface. }
  \end{minipage}

\label{fig:rindler}
\end{figure}

One arrives at 
\begin{equation}
\d s^2 = -\kappa^2\xi^2\d \tau^2 + \d \xi^2 +\d x^i \d x_i.
\label{rindlerform}
\end{equation}

The $\xi = \mathrm{constant}$ surfaces can be associated with uniformly accelerating observers (see figure \ref{fig:rindler}) with proper time $\tau$. Note these coordinates do not cover the whole of Minkowski space, but only one section: the Rindler wedge, where $z > |t|$ with the surface $z=|t|$ forming a forms a causal barrier, the Rindler horizon; light emitted behind such a surface can never reach the accelerating observer. 

Similarly to how a black hole horizon radiates, a Rindler horizon might be expected to radiate also. this fact was independently discovered by Fulling \cite{Fulling}, Davies \cite{Davies} and Unruh \cite{Unruh:1976db}. Indeed, an observer constantly accelerating with $\kappa$ would experience a thermal bath of 
\begin{equation}
T = \frac{\hbar \kappa}{2\pi}.
\end{equation}

\subsection{Deriving the Einstein Equations}

Pick a point p, with an associated 2-surface in the directions orthogonal to $t, z$, called $\mathscr{P}$, such that its past null normal has vanishing shear and expansion (at first order). This is essentially the condition of equilibrium. Now the past horizon for $\mathscr{P}$ is the local Rindler horizon of $\mathscr{P}$.  
Consider a heat flow across this horizon,
\begin{equation}
\delta Q = \int_H T_{ab}\chi^a\d \Sigma^b.
\end{equation}
for
\begin{equation}
\d \Sigma^a = k^a \d \lambda \d A.
\end{equation}
where $k$ is tangent to the horizon generators.
So one may write the heat flux as 
\begin{equation}
\delta Q = -\kappa \int_H \lambda T_{ab} k^a k^b \d \lambda \d A.
\end{equation}

An important assumption, that the entropy is entanglement entropy and therefore 
\begin{equation}
\delta S = \alpha \delta A
\label{eq:entropy}
\end{equation}
where the area variation is given by 
\begin{equation}
\delta A = \int_H \theta \d \lambda \d A
\end{equation}
where theta is the expansion of the horizon generators. We can re-express this using the (null) Raychaudhuri equation,
\begin{equation}
\frac{\d \theta}{\d \lambda} = -\frac{1}{2}\theta^2 -\sigma^2 -R_{ab}k^a k^b.
\end{equation}

If the Rindler horizon is instantaneously stationary at point $\mathscr{P}$ the $\theta^2$ and $\sigma^2$ terms are of order $\lambda^2$ and can be neglected, so
\begin{equation}
\delta A = -\int_H R_{ab}k^ak^b \lambda \d \lambda \d A
\end{equation}

Now we use the fundamental thermodynamic relation and the assumption \eqref{eq:entropy}
\begin{equation}
\delta Q =T \d S =\alpha T\delta A
\end{equation}
to arrive at
\begin{equation}
T_{ab}k^ak^b = \frac{\hbar \kappa}{2\pi}\alpha R_{ab}k^ak^b.
\end{equation}
Given that $k^a$ is an arbitrary null vector, this means
\begin{equation}
T_{ab}k^ak^b = \frac{\hbar \kappa}{2\pi}\alpha R_{ab}k^ak^b +fg_ab.
\end{equation}
What form can $f$ take? We need $T_{ab}$ to be conserved, so $f= -\frac{R}{2} +\Lambda$ for some constant $\Lambda$. This lead us to the end result: the Einstein equations 
\begin{equation}
R_{ab} - \frac{1}{2}R g_{ab} + \Lambda g_{ab}= T_{ab}. 
\end{equation}
For further elucidation of these ideas see \cite{Padmanabhan:2009vy}. 

\subsubsection{Near-equilibrium thermodynamics}

What if we work slightly out of equilibrium, including dissipative terms? What meaning would these terms have gravitationally? This was considered in \cite{Chirco:2009dc}. Work from a generalized Clausius relation
\begin{equation}
\d_e S + \d_i S = \frac{\delta Q}{T} + \delta N.  
\end{equation}
Here $\d_e S$ and $\d_i S$ are the entropy exchange with the surrounding and internal entropy production respectively, and $N$ is the ``uncompensated heat", the heat lost the the internal degrees of freedom. 
From the irreversible part one can find 
\begin{equation}
T \delta N = \frac{\alpha T}{\kappa}\int_H  \epsilon ||\sigma||_p^2 \d v
\end{equation}
which is a Hartle-Hawking tidal heating. This term describes the energy loss associated to the emission of gravitational waves, which dissipates a shear on the horizon. As such, we see that the gravitational fluxes are analogous to heat fluxes in the spacetime thermodynamics approach. 

\section{Emergent Gravity}

The analogy above seems a powerful hint that some microstructure, other than the simple quanta of gravitation, should account for the fabric of spacetime. What sort of microstructure? We can quantize electromagnetism to arrive at QED. But if we quantize hydrodynamics nothing quite so simple happens. 

Which sort of theory is gravity? Naive quantization fails. Such a standard picture could still be rescued, for instance by the ongoing programme into asymptotic safety (see \cite{asymsafety}). However, many researchers are looking into the idea of the continuum geometry and equation of motions ``emerging'' from a fundamentally different picture at microscopic scales. Possibly the earliest incarnation of this idea is due to Sakharov \cite{Sakharov:1967pk}. 

However, note there are general restrictions on such theories \cite{Weinberg:1980kq, Marolf:2014yga} under certain assumptions. 

A word of caution is necessary when talking about ``emergent gravity''. This term has been used extensively in the past few decades, and means very different things to different researchers. Some interesting perspectives can be found in \cite{Barcelo:2001tb, Barcelo:2014yna, Carlip:2012wa}

In the broadest sense, emergence will imply that the symmetries of general relativity could merely be a low-energy effect, and violated at high energies. Some implementations of this idea will be discussed in section \ref{HLAE} 

The next section will be devoted to one encouraging feature that has been developed in the last few decades i.e. the discovery that it is possible to emerge \emph{some} features of gravity from a diverse range of physical systems. This analogue gravity programme has been successful in opening both experimental and theoretical aspects in gravitational physics.

\section{Analogue Gravity}
One arena where some - but very importantly not all - of the features of gravitational solutions and Lorentz symmetry can emergence is analogue gravity. The basic idea of analogue gravity is that for some range of energy/length scales, perturbations feel a metric that is \emph{not} the lab Minkowski metric, but instead a curved metric. This was first realized in water waves \cite{Unruh:1980cg}, but has since been extended to many systems.

\subsection{The Basic Picture}
\label{analogue-basic}
Consider a sound ray, in a fluid that is moving at velocity $\vec{v}$ with respect to a laboratory observer. The velocity of the sound ray, by Galilean Relativity, is
\begin{equation}
\frac{\d x}{\d t}= c \vec{n} + \vec{v}.
\end{equation}
Which we can rewrite as
\begin{equation}
\left(\frac{\d x}{\d t}+ \vec{v}\right)^2= c^2, 
\end{equation}
using the fact that $n$ is a unit normal vector. This in turn is equivalent to 
\begin{equation}
-\left(c^2+v^2\right)\d t^2-2\vec{v}d\vec{x}\d  t + \d \vec{x}^2 = 0.
\end{equation}
This equation is precisely that for a null ray moving in the metric 
\begin{equation}
g_{ab} = \Omega^2 \left( \begin{array}{cc}
c^2-v^2 & v_i  \\
v_j & \delta_{ij} \end{array} \right) . 
\end{equation}
where $\Omega$ is undetermined at this level of analysis, as the lightcone structure is conformally invariant. 

This very general and simple derivations shows that lightcones ``feel'' a curved Lorentzian metric. This can be extended to perturbations of many systems. And is enough for discussing the causal structure of the emergent spacetime. However, when dealing with more general other features we will need physical acoustics, examining the equation of motion for perturbation of a given background for a particular physical system.

\subsection{Bose Einstein Condensates}
\label{BEC}

One of the most physically interesting analogue gravity systems is that of Bose Einstein Condensates (BECs), first studied in \cite{Garay:1999sk}. These are interesting as one of the main motivations behind the analogue gravity programme was to potentially observe semiclassical curved space effects (such as Hawking radiation and cosmological particle production), in the laboratory. For such a direct observation to be feasible we would like system at low temperature, with a high level of quantum coherence and a low speed of sound (see the discussion in \cite{Barcelo:2001ca}). BECs are good candidates for all these requirements. The derivation in this section will mostly follow the presentation in \cite{Barcelo:2000tg}. 

In a dilute gas, the evolution of a BEC can be described by a many-body Hamiltonian
\begin{eqnarray}
H &=& \int \d x\, \hat{\Psi}^\dag(x, t)\left(-\frac{\hbar^2}{2m}\nabla^2+V_{\mathrm{ext}}\right)\hat{\Psi}(x, t) \nonumber \\
 &+& \frac{1}{2}\int \d x \d x' \hat{\Psi}^\dag(x, t)\hat{\Psi}^\dag(x', t)V(x-x')\hat{\Psi}(x, t)\hat{\Psi}(x', t)  
\end{eqnarray}
We then separate out $\Psi$ into background condensate, $\Psi_s= \left\langle \Psi \right\rangle$  (the Boboluibov decomposition)
\begin{equation}
\hat{\Psi} = \Psi_s + \psi
\end{equation}
Using this and assuming that $V = \lambda \delta (x-x')$, we can arrive at the  Gross-Pitaevskii equation
\begin{equation}
i\hbar \partial_t \Psi = \left(\frac{-\hbar^2}{2m}\nabla^2 + V + \frac{4\pi a \hbar^2}{m}|\Phi|^2\right)\Psi
\end{equation}
which is also sometimes known as the non-linear Schr{\"o}dinger equation. 

Now  $\Phi_s$ can be written in the Madelung representation
\begin{equation}
\Psi_s= \sqrt{\rho (x)}\exp{\left(\frac{i\theta (x)}{\hbar}\right)},
\end{equation}
where the density is
\begin{equation}
\rho = \left|\Psi\right|^2
\end{equation}

and we can define a speed of sound
\begin{equation}
c(x)=\frac{\hbar}{m}\sqrt{4\pi a \rho}
\end{equation}
and the background velocity 
\begin{equation}
v(x) = \frac{\hbar}{m}\nabla \theta. 
\end{equation}
Plugging the Madelung representation into the Gross-Pitaveskii equation, taking the real and imaginary parts, and using the definitions of $c$ and $v$, one arrives at two equations: a continuity equation
\begin{equation}
\partial_t \rho +  \nabla (\rho v) = 0,
\label{eq:continuity}
\end{equation}
and a type of Hamilton-Jacobi equation
\begin{equation}
\partial_t \theta +\frac{1}{2m}\left(\nabla \theta\right)^2 + V_{\mathrm{ext}}+ \frac{1}{2}\lambda \rho^2 -\frac{\hbar^2}{2m}\left(\frac{\nabla^2 \sqrt{\rho}}{\sqrt{\rho}}\right) 
\end{equation}
These are very similar to the equations for a irrotational inviscid fluid (discussed in \cite{Visser:2013bga, Barcelo:2005fc}), the difference being in the presence of the quantum potential
\begin{equation}
T_\rho \equiv \frac{\hbar^2}{2m}\left(\frac{\nabla^2 \sqrt{\rho}}{\sqrt{\rho}}\right) 
\label{non-relqp}
\end{equation}
The presence of the $\hbar^2$ means that the quantum potential is suppressed,and for the present purposes can be safely neglected. For further discussion see \cite{Barcelo:2000tg}. Neglecting this term, we can now follow the steps of \cite{Barcelo:2005fc}, and rewrite the Euler equation as 
\begin{equation}
\partial_t v= v \times (\nabla \times v)-\frac{1}{\rho}\nabla p + \nabla\left(\frac{1}{2}v^2\right)
\end{equation}
Note that $v$ can be expressed as the derivative of a scalar by definition, it is irrotational (vorticity free). Hence the first term on the RHS of the previous equation is zero. 

Now linearize around this solution
\begin{equation}
\rho = \rho_0 +\epsilon \rho_1 , \qquad \theta =\theta_0 +\epsilon \theta_1
\end{equation}
From the continuity equation
\begin{eqnarray}
\partial_t \rho_0 &+&  \nabla (\rho_0 \nabla \theta_0) = 0 \nonumber \\
\partial_t \rho_1 &+&  \nabla (\rho_0 \nabla \theta_1+\rho_1 \nabla \theta_0) = 0.
\end{eqnarray}
and from the Hamilton-Jacobi equation (after we have thrown away the quantum potential)
\begin{eqnarray}
\partial_t \theta_0 &+& \frac{1}{2m}\left(\nabla \theta_0\right)^2 + V_{\mathrm{ext}}+ \frac{1}{2}\lambda \rho_0^2 \nonumber \\
\partial_t \theta_1 &+& \frac{1}{m}\left(\nabla \theta_0\right)\left(\nabla \theta_1\right) + V_{\mathrm{ext}}+ \frac{1}{2}\lambda \rho_0 \rho_1 
\end{eqnarray}
Now we can solve the linearized Hamilton-Jacobi equation for $\rho_1$, and substitute into the continuity equation.
\begin{eqnarray}
&-&\partial_t\left(\frac{2}{\lambda}\left(\partial_t \theta_1+\frac{1}{m}\nabla\theta_0\cdot \nabla\theta_1 \right)\right) \nonumber \\
&+&\nabla\left(\rho_0 \nabla \theta_1 - \nabla \theta_0\left[\frac{2}{\lambda}\left(\partial_t +\frac{1}{m}\nabla\theta_0\nabla\theta_1\right)\right]t\right) = 0.
\label{eq:perturbationeqn}
\end{eqnarray}

Now consider the matrix 
\begin{equation}
f^{\mu\nu} = \left( \begin{array}{cc}
f^{tt} & f^{ti}  \\
f^{tj} & f^{ij} \end{array} \right) ,
\end{equation}
where
\begin{eqnarray}
f^{tt} &=&-\frac{2}{\lambda};\nonumber \\
f^{ti} &=& -\frac{2}{\lambda} \nabla^i \theta; \nonumber \\
f^{ij} &=&  \frac{\rho_0\delta^{ij}}{m} -\frac{2}{\lambda m}\nabla^i\theta \nabla^j\theta.
\end{eqnarray}
By inspection, one can see that equation \ref{eq:perturbationeqn} is equivalent to
\begin{equation}
\partial_\mu \left(f^{\mu\nu}\partial_\nu \theta_1 \right)=0.
\end{equation}
We can further identify $f^{\mu\nu}$ as a tensor density, $f^{\mu\nu} = \frac{1}{\sqrt{-g}}g^{\mu\nu}$ so this can be rewritten as the standard form of the Klein-Gordon equation in curved spacetime.  
\begin{equation}
\frac{1}{\sqrt{-g}}\partial_\mu \left(\sqrt{-g}g^{\mu\nu}\partial_\nu \theta_1 \right)=0.
\end{equation}
Therefore we can read off the metric as
\begin{equation}
g^{ab} =  \frac{m}{\rho_0 c_s}\left( \begin{array}{cc}
-1 & v_0^j \\
v_0^i & c_s^2\delta^{ij}-v_0^iv_0^j \end{array} \right), 
\end{equation}
\begin{equation}
g_{ab} = \frac{\rho_0}{c_s m}\left(\begin{array}{cc}
c^2-v_0^2 & v_{0j}  \\
v_{0i} & \delta^{ij}\ \end{array} \right) .
\end{equation}
Note there are two physically inequivalent ways of deriving a analogue mspacetime from a BEC: Here we are linearizing the  Gross-Pitaveskii equation essentially dealing with perturbations of the classical background field, whereas one could look explicitly at the quantum excitations, as in \cite{Barcelo:2000tg}. For further details of mimicking spacetimes with BECs see \cite{Garay:2000jj, Garay:1999sk, Barcelo:2000tg, Garay:2002kz}.  

\subsection{Dispersion relation}

So far we have neglected the quantum potential. What is the effect of including it? Take the eikonal approximation on the perturbation
\begin{equation}
\rho_1 =A_\rho \exp\left(-i\left(\omega t + k x\right)\right); \quad \theta_1 =A_\theta \exp\left(-i\left(\omega t + k x\right)\right).
\label{eikonal}
\end{equation}
One can maintain the quantum potential throughout the calculation to arrive at a more complicated version of eqn \ref{eq:perturbationeqn} (see \cite{Barcelo:2000tg}). Plugging eqn \ref{eikonal} into this, one arrives at the dispersion relation for the BEC
\begin{equation}
\omega= v^ik_i + \sqrt{c_s^2k^2+ \left(\frac{\hbar}{2m}\right)k^4}.
\end{equation}
Neglecting the quantum potential, one would have instead have arrived at $\omega= v^ik_i +c_sk$. Obviously such systems only have an approximate Lorentz symmetry; the dispersion relation incorporates the information that at sufficiently high energies the physics governing such perturbations is quite different. 

\subsubsection{An Emergent Analogue Gravity}
 
Usually in analogue gravity it is only the spacetime, not the dynamics that emerge in the low-energy limit. However in the case of Bose--Einstein condensates it is possible to have a type of gravitational dynamics \cite{Girelli:2008gc} (but see also \cite{Belenchia:2014hga, Girelli:2008qp, Finazzi:2011zw}). These attempts emerge Newtonian and Nordstr{\"o}m gravity respectively (Given the degrees of freedom available one cannot expect to derive a tensor equation for gravity). This is a tantalizing hint that one could delve into the dynamics of general relativity and other theories of gravity in a similar way (see however, \cite{Marolf:2014yga} for some non-trivial requirements on the fundamental theory in order to give rise to a truly background independent gravitational theory).

\section{Einstein-{\AE}ther and \Horava\ --Lifshitz Gravity}\label{HLAE}

Einstein--{\AEther} and \Horava--Lifshitz gravity are two well-studied theories of gravity that are diffeomorphism invariant but violate local Lorentz invariance. Einstein--{\AEther} theory violates Lorentz invariance by introducing a preferred frame $u^a$, while \Horava--Lifshitz gravity introduces a preferred foliation defined by a scalar field $\tau$ called the khronon. 


\subsection{Einstein--{\AE}ther gravity}

Originally proposed in modern form in \cite{Jacobson:2000xp} (but see also \cite{Gasperini:1987nq}, \cite{Gasperini:1998eb}), Einstein--{\AE}ther ({\AE}) gravity was developed as a general framework for probing Lorentz violating effects, while maintaining the useful features of general relativity of diffeomorphism invariance and not having higher than second-order terms in the Lagrangian. This is achieved by introducing a unit, timelike vector field, the \aether\ .

The most general action possible under these constraints is 
\begin{equation}
S=\frac{1}{16\pi G}\int \d^4 x\sqrt{-g} (R+\mathcal{L}_{ae})\,;  
\label{aeaction}
\end{equation}
\begin{equation}
\mathcal{L}_{ae}=-Z^{ab}{}_{cd}\,(\nabla_au^c)(\nabla_b u^d)+\lambda(u^2+1).
\end{equation}
Here $\lambda$ is a Lagrange multiplier, enforcing the unit timelike constraint on $u^a$, and $Z^{ab}{}_{cd}$ couples the \aether\ to the metric through four distinct coupling constants:
\begin{equation}
Z^{ab}{}_{cd}=c_1g^{ab}g_{cd}+c_2\delta^a{}_c\delta^b{}_d+c_3\delta^a{}_d\delta^b{}_c-c_4 u^au^bg_{cd}.
\end{equation}
The unit constraint ensures that the \aether\ can never vanish. In many contexts it is convienent to use additive combinations of these coeficients, for which a easy shorthand is $c_{12}=c_1 +c_2$ and so forth.

The action \ref{aeaction} is invariant under disformal transformations 
\begin{equation}\label{disformal}
 \bar{g}_{ab}= g_{ab}+(s^2-1)u_au_b ; \qquad  \bar{u}^a=\frac{1}{\sqrt{s^2}}u^a.
\end{equation}
This transformation changes the effective metric to a mode moving with speed $s$ rather than $c (=1)$. 


A nice way of rewriting this action, as pointed out in \cite{Jacobson:2013xta}, is to decompose the \aether\ vector as one would for the Raychaudhuri equation
\begin{equation}
 \nabla_au_b=-\frac{1}{3}\theta h_{ab}+\sigma_{ab}+\omega_{ab}+u_aa_b.
\end{equation}
We can then rewrite the action of the \aether\ in terms of these variables
\begin{equation}
 S_{ae}=\frac{M_{\rm Pl}^{2}}{2}\int \sqrt{g}\, \d^4x \, {}^{(3)}R+\frac{1}{3}c_\theta \theta^2+c_\sigma \sigma^2+c_\omega \omega^2+c_a a^2 
\label{aesheartwist}
\end{equation}
where ${}^{(3)}R$ is the spatial Ricci scalar constructed with ${}^{(3)}g$ and the coupling constants are combinations of the $c_i\,'s$ previously in \ref{aeaction}.

\subsection{\Horava\ --Lifshitz Gravity}

One way to improve the renormalizability of gravity is to include higher order terms (this was first attempted in \cite{Stelle:1976gc}). However, due to the presence of higher order time derivatives, these theories become non-unitary. The way around this would be to add higher order spatial derivatives, but not time derivatives. 

The natural way to implement promoting time and space to different footings is via the Arnowitt--Deser--Misner (ADM) construction. Note that once we include such terms as the extrinsic curvature in the action we have changed from using the ADM construction as a natural system to express a given metric, to something much more fundamental, a physical foliation of space and time. The ADM decomposition involves splitting a metric into a lapse function, $N$, shift vector, $N^i$, and a spatial metric ${}^{(3)}g_{ij}$
\begin{equation}
 \d s^2=-N^2\d t^2+{}^{(3)}g_{ij}\left(\d x^i+N^i\d t\right)\left(\d x^j+N^j\d t\right).
\end{equation}
A natural quantity to consider is the extrinsic curvature of the spatial metric embedded in the foliation, 
\begin{equation}
 K_{ij}=\frac{1}{2N}\left(\frac{\d {}^{(3)}g_{ij}}{\d t}-2\nabla_{(i} N_{j)} \right)
\end{equation}
where $\nabla_i$ is the covariant derivative associated with the spatial metric (note also that some definitions of extrinsic curvature differ on overall sign). 
Further note that, in this formalism, one can write the action for general relativity as
\begin{equation}
 S=\frac{M_P^2}{2}\int \d^3x \d t N\sqrt{{}^{(3)}g}\left(K^{ij}K_{ij}-K^2+{}^{(3)}R \right)
\end{equation}
Note that the the term involving the extrinsic curvature are the only ones with time derivatives and can be regarded as the kinetic term, while ${}^{(3)}R$ contains only spatial derivatives and can be regarded as the potential energy. It is this potential term that \Horava\ --Lifshitz gravity changes. 

If we include a preferred foliation, the theory can no longer be invariant under the full set of diffeomorphisms allowable in general relativity. The best one can do it to maintain invariance under a restricted set of diffeomorphisms
\begin{equation}
 t \to \bar{t}(t)\,; \qquad x^i \to \bar{x}(t, x^i)
\end{equation}

Now if one wants to construct a theory that is only second order in time derivatives, one must consider what invariants to include. $K_{ij}$ has one time derivative so both $K^{ij}K_{ij}$ and $K^2$ are second order in time. Any invariants constructed with $K_{ij}$ and $N^i$ would not preserve the restricted set of diffeomorphisms desired. Therefore, the most general action possible is
\begin{equation}
  S=\frac{M_P^2}{2}\int d^3x dt N\sqrt{{}^3g}\left(K^{ij}K_{ij}-\lambda K^2+F({}^{(3)}g_{ij}, N) \right). 
\end{equation}

In principle, $F(g_{ij}, N)$ could have infinitely many terms. However, there is a minimum number of spatial derivatives for achieving a power-counting renormalizable theory. In three spatial dimensions, if can be shown that including terms with up to order six operators renders the theory power-counting renormalizable \cite{Horava:2009uw, Visser:2009fg}. 

\subsubsection{Projectable and detailed balance versions}

Due to the unmanageably large number of potential terms in the full theory, often two possible restrictions are considered, the projectable version and the detailed balance versions of \Horava\ --Lifshitz gravity. Neither of these restrictions are strongly physically motivated, and a more for calculations convenience.

\begin{itemize}

 \item{\it{Detailed Balance}\\} 
This is the original version proposed by \Horava\ \cite{Horava:2009uw}.  The detailed balance version states that $F(g_{ij}, N)$ must come from a superpotential, inspired by critical phenomena. 


 \item{\it{Projectable version}\\}
This version makes the assumption that the lapse is only a function of time, $N=N(t)$, thus removing any derivatives of $N$ from $F(g_{ij}, N)$. See \cite{Weinfurtner:2010hz}

\end{itemize}

\noindent For many purposes, one may want to consider the full theory, which has the action
\begin{equation}
S_{HL}= \frac{M_{\rm Pl}^{2}}{2}\int \d t\, \d^3x \, N\sqrt{{}^{(3)}g}\left(L_2+\frac{1}{M_\star^2}\;L_4+\frac{1}{M_\star^4}\;L_6\right)\,,
\label{hlac}
\end{equation}
where $h$ is the determinant of the induced metric $h_{ij}$ on the spacelike hypersurfaces, while
\begin{equation}
L_2=K_{ij}\,K^{ij} - \lambda K^2 + \xi\, {}^{(3)}\!R + \eta a_ia^i\,,
\end{equation}
with $a_i=\partial_i \ln N$. The quantities $L_4$ and $L_6$ denote a collection of $4^{\mathrm{th}}$ and $6^{\mathrm{th}}$ order operators respectively, and $M_\star$ is the scale that suppresses these operators (which does not coincide {\em a priori} with $M_{\rm Pl}$). This full version of the theory, also sometimes known as the ``extended" or ``healthy" version \cite{Blas:2009yd, Blas:2009qj, Blas:2009ck}, has the possibility to be phenomenologically viable \cite{Sotiriou:2009gy} (but see also \cite{Papazoglou:2009fj}, which contrains $M_\star$ as the theory is strongly coupled in the IR), and has been extensively studied. 

\subsection{Relation between {\AE} and HL gravity}

Given that both Einstein--{\AE}ther and \Horava --Liftshitz gravity are Lorentz violating theories people considered that there may be some relation between the two. Indeed this is the case. Taking the \AE\ action \ref{aeaction} and imposing that
\begin{equation}
u_a = \frac{\partial_a \tau}{\sqrt{g^{bc}\partial_b \tau\partial_c \tau}},
\end{equation}
and taking the variation of the action with respect to $\tau$, one obtains the $L_2$ part of the action in \ref{hlac}.

A nice way of viewing this, as pointed out in \cite{Jacobson:2013xta} is using equation \ref{aesheartwist}. If $u^a$ is hypersurface orthogonal, 
$\omega_{ab} \equiv \nabla_{[a}u_{b]}-u_{[a}a_{b]}$ becomes antisymmetrizations of partial derivatives, and is therefore zero. Thus we see that (low-energy) \Horava\ --Lifshitz gravity is twist-free Einstein--{\AE}ther gravity. This automatically implies we have one fewer coupling constants in low energy HL gravity. 

Note that this equivalence is only when the hypersurface orthogonality is taken \emph{before} variation to find the equations of motion, as discussed in \cite{Jacobson:2013xta, Barausse:2012qh}. In particular, though any solution of \AE\ theory for which the aether is hyperssurface orthogonal is also a solution of HL gravity, but the converse is not true (see section 2.1 of \cite{Barausse:2013nwa}). 

\subsection{Dispersion relations}
\label{dispersion}

In a Lorentz-violating scenario, particles will generically be coupled to the preferred frame, and such a coupling will imply modified dispersion relations which can be naturally assigned in the preferred, \aether\ frame. If, as in the case of Einstein--\Aether\ and \Horava\ --Lifshitz the {\ae}ther is dynamical, modified dispersion relations do not cause any inconsistency in the Bianchi identities as discussed in Ref.~\cite{Kostelecky:2003fs}.

Let us stress that Lorentz-violation in the gravitational sector is expected to percolate into the matter sector via radiative corrections, at least in an effective field theory framework.  
This implies that even starting with a  Lorentz invariant matter sector our theory will end up  providing modified dispersion relations for all particles. In the matter sector, dispersion relations are well constrained \cite{Mattingly}, however there is a mechanism of protection due to the weakness of the gravitational coupling. Indeed, if there is a large separation between the Lorentz breaking scale in gravity and the Planck scale $M_* \ll M_P$ then the theory is viable \cite{Pospelov:2010mp, Liberati:2013xla}. Further note that we often only consider UV modifications of the matter dispersion relations, while of course also in this case one might also expect radiative corrections to produce IR modifications (for e.g., inducing particle-dependent coefficients of the order $\ks^2$ terms), see for e.g. Ref.~\cite{Collins:2004bp}. Such effects are even more highly constrained, and several protection mechanisms to suppress them have been devised in the literature \cite{Liberati:2013xla}. Modified dispersion relations in the context of Einstein--{\AE}ther theory are also discussed in Ref.~\cite{Jishnu_thesis}.\par

\section{Lorentz Violation and Thermodynamics}
\label{perpetualmotion}

Let us come full circle and discuss black hole thermodynamics in the presence of Lorentz violations. One of the motivations for initially considering Lorentz violations was, based on hints from black hole thermodynamics, the idea that mircostructure of quantum gravity may be based on very different physics than the macrostructure, and thus have different symmetries. However, when one considers black hole thermodynamics in the Lorentz violating case, one runs into potential problems.

This was initially observed in \cite{Dubovsky:2006vk}, which studied the ghost condensate model (for background on this model see \cite{ArkaniHamed:2003uy}). For our purposes the key point is that it is natural in this theory for different particles to propagate at different speeds, that is, the $k^2$ term in the dispersion relation can have different coeffiecents. Thus different species have different Killing horizons for the same black hole, being at a larger radius for subluminal particles and a smaller radius for superluminal particles. Already at this point it is unclear what horizon area one might associate to an entropy. Further, different horizons radiate at different temperatures. To see just how worrying this feature is \cite{Dubovsky:2006vk} used the following gedanken experiment to generate a perpetuum mobile of the $2^{nd}$ kind (that is, a machine whose sole result is the transfer of energy from a cold object to a hot object):

Consider two particles, $\phi_1$ and $\phi_2$, with speeds $v_1$ and $v_2$ respectively with $v_1 < v_2$, and therefore the horizons for these fields have temperatures $T_1 > T_2$. Now surround this black hole with two shells, $A$ and $B$ such that shell $A$ interacts only with $\phi_1$ and shell $B$ only with $\phi_2$. Let the temperatures of these shells satisfy 
\begin{equation}
T_2 > T_B > T_A > T_1.
\end{equation}
As $T_1 < T_A$ there will be a flow of heat from shell $A$ into the black hole, and as $T_B < T_2$ there will be a low of heat from the black hole to shell $B$. We can therefore choose the temperatures such that the net heat flow out of the black hole is zero, leaving the state of the black hole the same for an outside observer. Thus a perpetual motion machine has been created. 

In response to this \cite{Eling:2007qd} (but see also \cite{Jacobson:2008yc}), considered a number of possible ways this conclusion might be evaded, but found only effects that would be negligible for large black holes. They further constructed another gedanken experiment to violate the GSL, involving a classical extraction of energy through a process similar to the Penrose process.   

For this process we must make some reasonable assumptions on the form of the entropy $S(M)$, namely: i). For 
\begin{equation}
M_1 > M_2 \quad \Longrightarrow \quad  S(M)_1 >S(M)_2
\end{equation}
and ii). For a large enough M, thew entropy carried away by radiation can be a sufficiently small fraction of $S(M)$. Both of these conditions are satisfied if $S(M) \propto M^\alpha$ for $\alpha >0$. 

Take a particles $A$ and $B$ with limiting speeds $v_A < v_B$, so that the $A$ horizon is outside the $B$ horizon. Let a system made of both $A$ and $B$ particles fall through the $A$ horizon, into the $A$ ergoregion, and then split, such that the $A$ falls across the $B$ horizon  \emph{carrying negative Killing energy}, and $B$ exits from the $A$ horizon. By energy conservation carrying more energy than when the system entered the A horizon. Thus the mass, and hence entropy, of the black hole has decreased. It is possible to have $A$ and $B$ in a pure state, carrying no entropy at all. Hence by this process the total entropy of the universe has decreased.

\section{\AE\ and HL Black Holes and Universal Horizons}
\label{ae-bhs}

In the special case of spherically symmetric, static, asymptotically flat solutions are the same in \Horava\ --Lifshitz and Einstein--\Aether\ gravity \cite{Barausse:2013nwa}. Such solutions have been extensively considered in recent years (see for example Refs.~\cite{Barausse:2011pu, Blas:2011ni, Berglund:2012bu, Berglund:2012fk, Eling:2006ec, Mohd:2013zca}). 

In addition to the spherically symmetric solutions mentioned here, rotating black holes in three dimensional \Horava\ --Lifshitz \cite{Sotiriou:2014gna}, and slowly rotating black holes in both theories have been studied \cite{Barausse:2012qh, Barausse:2012ny, Barausse:2011pu}, as well as black holes which are asymptotically de Sitter/Anti de Sitter \cite{Bhattacharyya:2014kta}. 

Among the most striking results concerning these solutions was the realization --- in the (static and spherically-symmetric) black-hole solutions of both Einstein--{\AE}ther and \Horava--Lifshitz gravity ---  that they seem generically to be endowed a new structure that was soon christened the universal horizon~\cite{Barausse:2011pu, Blas:2011ni}. 

These universal horizons can be described as compact surfaces of constant khronon field and radius. As nothing singular happens to the metric, and the khronon, which diverges, can be reparameterized to be regular, this is not a singularity. Given that the khronon field defines an absolute time, any object crossing this surface from the interior would necessarily also move back in absolute time (the {\ae}ther time), something forbidden by the definition of causality  in the theory. Another way of saying this is that even a particle capable of instantaneous propagation, (light cones opened up to an apex angle of a full 180 degrees, something in principle possible in Lorentz-violating theories), would just move around on this compact surface and hence be unable to escape to infinity. 
This explains the name of universal horizon; even the superluminal particles would not be able to escape from the region it bounds.

One way to grasp the universal horizon is to consider the disformal transformation of eqn \ref{disformal} and consider what happens to the Killing horizon when we move to a mode of speed $s$. Note we must also rescale
\begin{equation}
\bar{\chi}^a=\frac{1}{\sqrt{s^2}}\chi^a
\end{equation}
so $\bar{u}$ and $\bar{\chi}$ are equal at infinity. Now
\begin{equation}
\bar{g}_{ab}\bar{\chi}^a\bar{\chi}^b=\frac{1}{s^2}\chi^2+\frac{s^2-1}{s^2}(\chi \cdot u)^2. 
\end{equation}
We can see that depending on the value of $s$ the condition for the Killing horizon will move inwards (for $s>1$) or outwards (for $s<1$). Now consider taking the limit $s \to \infty$. 
\begin{equation}
\lim_{s \to \infty}\tilde{\chi}^a\tilde{\chi}^b\tilde{g}_{ab}=\lim_{s \to \infty}\frac{s^2-1}{s^2}(\chi \cdot u)^2=(\chi \cdot u)^2
\end{equation}
So, in the limit of infinite velocity, the horizon is where $u\cdot \chi=0$, which is precisely the condition for a universal horizon. 

\begin{figure}[!htb]
\centering
\includegraphics[scale=1.0]{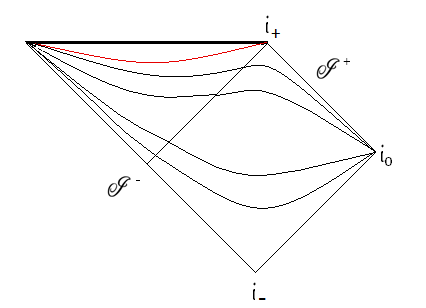}
\caption{Conformal diagram of black hole with Universal horizon, showing lines of constant khronon field, with the Universal horizon shown in red.}
\label{fig:conformal}
\end{figure}

The causal structure of these black hole spacetimes is as shown in Fig.~\ref{fig:conformal}. The indicated hypersurfaces are constant khronon hypersurfaces. 
The special hypersurface behind the Killing horizon --- where the {\ae}ther and the Killing vector fields become orthogonal --- is the Universal horizon. Note that the Killing vector generates the time-translation isometry outside the Killing horizon, while inside it is spacelike. 

Universal horizons can be shown to form from (idealized) gravitational collapse \cite{Saravani:2013kva}, and their stability perturbations have been studied \cite{Blas:2011ni}. 

For some specific combinations of the coefficients there are explicit, exact solutions for black holes. In particular, two exact solutions for static, spherically symmetric black holes have been found. 
As we will use these solutions extensively throughout this paper, we will briefly summarize some of their relevant details. For more information and background we refer the reader to Ref.~\cite{Berglund:2012bu}. Both solutions, in Eddington--Finkelstein coordinates, can be written as 
\begin{equation}
\d s^2 =-e(r)\;\d v^2 +2\,\d v\,\d r +r^2 \; \d \Omega^2.
\end{equation}
It is precisely the case in which the metric can be fully characterized by just one function that these exact solutions have been found \cite{Jishnu_thesis}.  
Here the form of the \aether\ is
\begin{equation}
u^a=\left\lbrace \alpha(r), \beta(r), 0, 0 \right\rbrace; \qquad u_a=\left\lbrace \beta(r)-e(r)\alpha(r), \alpha(r), 0, 0\right\rbrace.
\end{equation}
Note from the normalization condition, $u^2=-1$, there is a relation between $\alpha(r)$ and $\beta(r)$:
\begin{equation}
\beta(r)=\frac{e(r)\alpha(r)^2-1}{2\alpha(r)}.
\end{equation}
We can also define a spacelike vector $s^a$ (either inwards or outwards pointing), such that
\begin{equation}
s^au_a=0; \qquad s^2=1.
\end{equation}
Explicitly, for the inward pointing,
\begin{equation}
s^a=  \left\{\alpha(r),e(r)\alpha(r)-\beta(r),0,0\right\} =   \left\lbrace \alpha(r), \frac{e(r)\alpha(r)^2+1}{2\alpha(r)}, 0, 0\right\rbrace,
\label{eq:s}
\end{equation}
which clearly ensures $s^2=1$.
The two known exact black-hole solutions to Einstein--{\AEther} theory correspond to the special combinations of coefficients $c_{123} = 0$ and $c_{14}=0$.
\begin{itemize}
 \item Solution 1: $c_{123} = 0$.

For this solution we have
\begin{equation}
e(r)=1-\frac{r_0}{r}-\frac{r_u(r_0+r_u)}{r^2}; 
\end{equation}
where
\begin{equation}
r_u = \left[\sqrt{\frac{2 - c_{14}}{2(1 - c_{13})}} - 1\right]\frac{r_0}{2}.
\end{equation}
Here is $r_0$ is essentially the mass parameter, which can be directly related to the ADM mass \cite{Berglund:2012bu}. Furthermore
\begin{equation}
\alpha(r)=\left(1+\frac{r_u}{r}\right)^{-1}; \qquad \beta(r)=-\frac{r_0+2r_u}{2r}.
\end{equation}
It is also useful to decompose the Killing vector along $u$ and $s$ using the relations
\begin{equation}
\chi \cdot u=-1+\frac{r_0}{2r}\,; \qquad \chi\cdot s =\frac{r_0+2r_u}{2r}.
\end{equation}
For this particular exact solution, the Killing horizon and universal horizons are located at 
\begin{equation}
r_{\KH}=r_0+r_u, \quad \mathrm{and} \quad r_{\UH}=\frac{r_0}{2}
\end{equation}
respectively. 

\item Solution 2: $c_{14} = 0$.

For this solution we have
\begin{equation}
e(r) = 1 - \frac{r_0}{r} - \frac{c_{13}r_{\ae}^4}{r^4}; \qquad r_{\ae} = \frac{r_0}{4}\left[\frac{27}{1-c_{13}}\right]^{1/4};
\end{equation}
\begin{equation}
\alpha(r) = \frac{1}{e(r)}\left(-\frac{r_{\ae}^2}{r^2} +\sqrt{e(r) + \frac{r_{\ae}^4}{r^4}}\right); \qquad \beta(r) = -\frac{r_{\ae}^2}{r^2}~.
\end{equation}

Furthermore, the Killing vector is decomposed as 
\begin{equation}
\chi \cdot u =-\sqrt{1-\frac{r_0}{r}+\frac{(1-c_{13})r_{\ae}^4}{r^4}}\, ; \qquad \chi \cdot s =\frac{r_{\ae}^2}{r^2}~.
\end{equation}
The Killing horizon is located by solving the quartic polynomial $e(r)=0$, so
\begin{equation}
r_{\KH}=\frac{r_0}{4}+\frac{A}{8}+\frac{1}{8}\sqrt{8r_0^2-{\frac {6c_{13}r_0^2}{{\zeta}^{2}}}+{\frac {6r_0^2\zeta^2}{1-c_{13}}}+\frac{16 r_0^3}{A}},
\end{equation}
where
\begin{equation}
A = r_0 \sqrt{ 4 + {6c_{13}\over\zeta^2} + {6\zeta^2\over (c_{13}-1)}}; 
\end{equation}
\begin{equation}
\zeta=  \sqrt[6]{c_{13}\left((c_{13}-1)^2+\sqrt{\left(1-c_{13} \right)^{3}} \right)}
\end{equation}
The Universal horizon is located at 
\begin{equation}
r_{\UH}=\frac{3 r_0}{4}.
\end{equation}

\end{itemize}

These universal horizons are interesting from the thermodynamics point of view, and have been shown to obey a first law \cite{Berglund:2012bu}, and there are hints they may radiate \cite{Berglund:2012fk}. Could these surfaces offer a solution to the problems with the generalized second law in Lorentz violating theories?

\section{Final Remarks}

We have now essentially come full circle: black hole thermodynamics and a thermodynamic derivation of the Einstein equations has led us to consider the possibility that the fundamental symmetries of general relativity, in particular Lorentz invariance, may only be low-energy effects. We are strengthened in this supposition by an understanding of analogue models of spacetimes, where we can express motion of perturbations in terms of movement of a curved geometry. 

One exciting feature of such analogue gravities is the solutions can have more (and less) freedom, and the dynamics are completely different. This is an aspect we will explore further in the next chapter. We saw that BECs were ideal systems for analogue gravity. We will later return to a different sort of BEC in an attempt to model new features in analogue gravity. 

If the inspiration that Lorentz symmetry may only be a low-energy symmetry, it makes sense to study Lorentz violating theories of gravity, and in particular, black holes, our ``hydrogen atom" in such theories. Surprisingly, we find there are problems with the thermodynamics of black holes in such therories, the same hint that brought us here in the first place! Chapters three and four will look into universal horizons, the thermodynamics of which could solve this issue.

\chapter{Surface Gravities for Non-Killing Horizons}
\label{surfgrav}
\epigraph{Space-time is like some simple and familiar system, which is both intuitively understandable and precisely analogous, and if I were Richard Feynman I'd be able to come up with it}{Randall Munroe,  XKCD, \it{Teaching Physics}}

\section{Introduction}

Surface gravity, as discussed in section \ref{introsurfgrav}, is an important quantity in classical general relativity, which plays a vital role in black hole thermodynamics and semi-classical aspects of gravity, being closely related to the temperature of Hawking radiation. 
However,  in a large number of situations, the surface gravity cannot be calculated unambiguously, as standard definitions rely on the existence of a stationary spacetime with a Killing horizon.
 
Despite the recent work on developing notions of surface gravity for an evolving black hole (\cite{Nielsen:2005af, Nielsen:2007ac, Pielahn:2011ra, Hayward:1993wb, Fodor:1996rf, Hayward:1997jp,  Booth:2003ji, Booth:2006bn}), much less effort has been devoted to stationary scenarios where the horizon is no longer a Killing horizon. 
The explanation for this is simple: 
For the standard case of general relativity, due to the rigidity theorem (see, for instance~\cite{HawkingEllis, carter1, carter2} and~\cite{Heusler, Friedrich:1998wq, Robinson:2004, Kodama:2011}), in stationary spacetimes all event horizons are automatically Killing horizons (\emph{i.e.}, the spacetime must possess a Killing field which is normal to the event horizon). 
However, this result hinges on the Einstein field equations, and in modified gravity, or in the arena of analogue spacetimes, there is no \emph{a priori} reason to expect this result will continue to hold.
We will address a number of scenarios where the standard calculations for surface gravities either will not hold, or will give rise to distinct quantities. 

This technical heart of the chapter is essentially divided into three sections. In the first section, we will briefly present the standard general relativity case, 
and run through several quite standard ways to calculate the surface gravity in stationary spacetimes, as presented (for instance) by Wald in reference~\cite{Wald}, drawing explicit attention to the assumptions built into the calculations; assumptions that we shall then relax in subsequent discussion. 
As a first step in this relaxation process we consider the conformal Killing horizons of Jacobson and Kang~\cite{Jacobson-Kang}. 

The second section is devoted to the analogue spacetime case, focussing specifically on acoustic horizons. 
In this context, all horizons are null surfaces, (in fact, they are even geodesic null surfaces), but in the case of non-zero rotation, (non-zero vorticity, or more precisely non-zero helicity), can nevertheless be non-Killing. 
We demonstrate that the different definitions of the surface gravity will in this context lead to physically and mathematically distinct quantities, and discuss which is the most relevant one in the case of analogue horizon thermodynamics.

The third section will be devoted to discussing the universal horizons of section \ref{ae-bhs}. Such horizons are spacelike instead of null surfaces, and are not Killing horizons. Thus they seem to require new techniques to calculate. We will present a first attempt at understanding the surface gravity of such horizons.

Finally we end with a brief discussion putting our calculations in context. 
In particular, while for definiteness in this chapter we will discuss non-Killing horizons in analogue spacetime and in Einstein-\Aether\ and \Horava\ --Lifshitz contexts, the issues raised are much more general --- similar considerations will apply in various modified gravity models where modification of the Einstein equations generically eliminates the rigidity theorems so non-Killing horizons are likely to be generic. 
This has importance for mmany areas of physica and, for instance, non-Killing horizons have very recently become of interest both in AdS/CFT~\cite{Fischetti:2012vt} and holographic~\cite{Figueras:2012rb} situations.

\begin{table}[!htdp]
\caption{Some of the multiple notions of surface gravity}. 
\vspace{-15pt}
\begin{center}
\def\arraystretch{1.4}
\setlength{\tabcolsep}{1.5em}
\begin{tabular}{|c| |c|}
\hline
\hline
Name & Key features \\
\hline
\hline
peeling & peeling off properties for null geodesics near horizon\\  
\hline
inaffinity & inaffinity properties for null geodesics on horizon\\
\hline
\hline
normal & null normal to a null surface\\
\hline
generator & anti-symmetrized derivatives of horizon generators\\
\hline
tension & tension in an ideal massless rope\\
\hline
expansion & geodesic expansion transverse to the horizon\\
\hline
\hline
Euclidean & elimination of angle deficit at horizon\\
\hline
\hline

\end{tabular}
\\[5pt]
Some of these definitions require specific simplifying assumptions. \\
Others are (or can be made to be) more general. \\
All definitions are equivalent for Killing horizons. 
\end{center}
\label{default}
\end{table}%





\section{Standard general relativity --- stationary case}\label{standardcase}
Let us now consider stationary horizons in standard general relativity, so that (in view of the classical rigidity theorems \cite{HawkingEllis, carter1, carter2, Heusler, Friedrich:1998wq, Robinson:2004, Kodama:2011}) all horizons are automatically Killing. 
\begin{itemize}

\item 
The peeling definition of surface gravity $\kappa_\mathrm{peeling}$ is somewhat messy to write down in the general stationary case, though it is already clear from the spherically symmetric discussion in the previous chapter that it will almost certainly equal $\kappa_\mathrm{inaffinity}$.

\item
In contrast, for stationary horizons the inaffinity definition of surface gravity is typically restricted to an explicitly on-horizon version, and given by a simple explicit formula. In terms of the Killing vector $\chi$ (see for example Wald~\cite{Wald}):
\begin{equation}
\chi^a \nabla_a \chi^b = \kappa_\mathrm{inaffinity} \; \chi^b,
\end{equation}
where this formula now makes sense only on the horizon.

\item
A third notion of surface gravity is that of the null normal derivative evaluated on the horizon (see for example Wald~\cite{Wald}):
\begin{equation}
\nabla^a(\chi^b\chi_b)=-2\kappa_\mathrm{normal}\;  \chi^a.
\label{normal}
\end{equation}
Equivalently, 
\begin{equation}
\chi^b \nabla_a \chi_b = - \kappa_\mathrm{normal}\; \chi_a.
\end{equation}
Using Killing's equation we see $ \kappa_\mathrm{normal}= \kappa_\mathrm{inaffinity} $, but this equality will generically fail once we move to consider non-Killing horizons. 
(We shall exhibit explicit failure of this equality for acoustic horizons later on in the chapter.)

\item
As a fourth notion of surface gravity Wald~\cite{Wald} furthermore argues that it is useful to define the equivalent of
\begin{equation}
\kappa_\mathrm{generator}^2 = -{1\over2} (\nabla^{\left[ a\right. } \chi^{\left. b\right] }) (\nabla_{\left[ a\right. } \chi_{\left. b\right] }),
\end{equation}
(this name is chosen because the integral curves of the vector field $\chi^a$ generate the horizon.)
This definition makes sense everywhere throughout the spacetime. 
A brief calculation~\cite{Wald} demonstrates that on the (Killing) horizon
\begin{equation}
\left.\kappa_\mathrm{generator}\right|_H = \kappa_\mathrm{inaffinity} .
\end{equation}
Again,  this inequality will generically fail once we move to consider non-Killing horizons.
(Also in this case we shall exhibit explicit failure of this equality for acoustic horizons later on in the chapter.)

\item
A fifth notion of surface gravity can be formulated in terms of the tension in an ideal massless rope holding a unit mass steady just above the Killing horizon:
\begin{equation}
\kappa_\mathrm{tension} = \lim_H \sqrt{-\chi^2}\, \Vert A\Vert .
\end{equation}
Here $\Vert A\Vert $ denotes the magnitude of the 4-acceleration. 
Wald demonstrates that for Killing horizons $\kappa_\mathrm{tension} =  \kappa_\mathrm{generator} =\kappa_\mathrm{inaffinity}$, but this equality will again generically fail once we move to consider non-Killing horizons.
(Again, we shall demonstrate explicit failure of this equality for acoustic horizons later in the chapter.)

\item
A sixth notion of surface gravity recently developed by Jacobson and Parentani \cite{Jacobson:2008cx} is based on relating the surface gravity to the expansion of the 2-d surface drawn by (timelike) geodesic congruences orthogonal to the horizon. Define
\begin{equation}
\label{E:theta-0}
\theta_{2d} = h^a{}_b \, \nabla_a u^b,
\end{equation}
for 
\begin{equation}
h^a{}_b = u^au_b -s^as_b, 
\label{eq:h}
\end{equation}
is the surface projector onto the 2-d surface generated by the congruence with tangent $u$, and $s$ is (spacelike) vector orthogonal to $u$, which can always be expressed as
\begin{equation}
s^a =\frac{(\chi \cdot u)u^a -\chi^a}{\chi\cdot s}.
\end{equation}
We pick an appropriate congruence Lie dragged by the Killing flow such that
\begin{equation}
\chi^a \nabla_a u^b=u^a\nabla_a \chi^b,
\end{equation}
and, using \eqref{eq:h}, we can write this 2-d expansion as 
\begin{equation}
\theta_{2d}  =\frac{\frac{1}{2} u^a\nabla_a\chi^2}{\chi^2-(\chi\cdot u)^2}.
\end{equation}
Then on-horizon, where $\chi^2=0$, we have
\begin{equation}
\left.\theta_{2d}\right|_H  = -\frac{\frac{1}{2} u^a\nabla_a\chi^2}{(\chi\cdot u)^2}.
\end{equation}
It is then most useful to normalize by defining
\begin{equation}
\label{E:theta-n}
\kappa_\mathrm{expansion} = \left.\left\{ (\chi\cdot u) \; \theta_{2d}\right\}\right|_{H},
\end{equation}
which in the case of standard general relativity automatically implies, as can be seen by taking \ref{normal} and decomposing $\chi$ into $u$ and a  vector orthogonal to $u$
\begin{equation}
\kappa_\mathrm{expansion}= \kappa_\mathrm{normal}. 
\end{equation}
This notion of surface gravity has been explicitly constructed so that $\kappa_\mathrm{expansion} =  \kappa_\mathrm{normal}$, and hence, in this case, is also equal to $\kappa_\mathrm{inaffinity}$.
This derivation relies on the construction of a geodesic congruence that is invariant under the flow of a Killing vector, and so cannot, 
without suitable alterations, be extended to non-Killing horizons that might be present in modified gravity or analogue spacetimes.

\item
Finally, a seventh notion of surface gravity can be based on Euclidean continuation (Wick rotation), and demanding the elimination of the deficit angle at what used to be the horizon in Lorentzian signature (see chapter 6 of \cite{frolov2011introduction}). 
This construction of $\kappa_\mathrm{Euclidean}$ is extremely delicate, implicitly requiring constancy of the surface gravity over the horizon (and so implicitly appealing to the rigidity theorems) to even make sense --- but when it works this Euclideanization procedure has the virtue that it automatically forces all quantum fields into an equilibrium thermal bath at the Hawking temperature $k T_H = \hbar \kappa_\mathrm{Euclidean}/2\pi$. 
This procedure works best for static spacetimes, and is already somewhat delicate for stationary non-static spacetimes. 
We will not explore this particular approach any further.

\end{itemize}
While all of these notions of surface gravity are degenerate in the case of Killing horizons, the situation for non-Killing horizons is much more complex.  
\begin{itemize}
\item 
In standard general relativity it is a well-known result that the surface gravity is constant over the event horizon. 
This result can be proven without recourse to the field equations \emph{if} the horizon is assumed to be Killing~\cite{Racz:1995nh}, but for modified gravity (with field equations that differ from the Einstein equations) one may encounter non-Killing horizons. 
Alternatively, in standard general relativity,  constancy of the surface gravity can be proved using stationarity, the Einstein field equations, and the dominant energy condition for matter~\cite{Hawking:1971vc}. 
(However, note that the dominant energy condition is known to be violated by vacuum polarization effects~\cite{Barcelo:2002bv}.) 
In short, this result strongly hinges on the classical equations of motion, and as such, we have no reason to believe this will hold for modified gravity or in analogue spacetime scenarios. 
\item
As a first step beyond standard general relativity, note that even in the case of conformal Killing horizons four of the definitions given in section \ref{standardcase} (inaffinity, normal, generator, tension) do \emph{not} generically coincide. 
This case was considered by Jacobson and Kang~\cite{Jacobson-Kang}, motivated by theories, such as Brans-Dicke, where two conformally related metrics are of physical interest. 
The key point is that Jacobson and Kang distinguish several slightly different notions of surface gravity, all of which happen to coincide for Killing horizons (see also~\cite{Nielsen:2012xu, Devecioglu:2011yi}).

The key result (from our current perspective) can be summarized as follows: 
For a conformal Killing vector by definition one has
\begin{equation}\label{conformalkilling}
2\nabla_{(a}\chi_{b)}= \L_\chi g_{ab}= 2F\, g_{ab}.
\end{equation}
Then the relationship between the various surface gravities defined above is
\begin{equation}
\kappa_\mathrm{normal}=\kappa_\mathrm{inaffinity}-2F=\kappa_\mathrm{generator}-F,
\end{equation}
where we have altered their notation to correspond to ours. 
Only one of the definitions can be a true conformal invariant, which they find to be $\kappa_\mathrm{normal}$, while the others will at best be conformally invariant only for those conformal transformations that are constant on the horizon. 
Furthermore $\kappa_\mathrm{tension}$ will be invariant for this special class of transformations, but loses its interpretation for more general conformal transformations.
\end{itemize}
These results, in and of themselves, already provide a clear warning against unrestrictedly interchanging the definitions of surface gravity when working in non-general relativity contexts. 

We shall now discuss two explicit examples of stationary but non-Killing horizons --- one based on the analogue spacetime programme, and the other on universal horizons.

\section{Analogue spacetimes}\label{adefs}

While we considered the specific system of Bose Einstein condensates and derived an analogue spacetime from the perturbation in the system, as argued in section \ref{analogue-basic}, for the case of non-relativistic acoustics in the limit of geometrical acoustics, which is enough for our purposes, we can write the metric as
\begin{equation}
g_{ab} = \Omega^2 \left[\begin{array}{c|c} -(c_s^2-v^2) & -v_j \\   \hline -v_i & \delta_{ij} \end{array} \right],
\end{equation}
where (for now) the quantities $v_i$ and $c_s$ are position (but not time) dependent. 
The corresponding inverse metric is:
\begin{equation}
g^{ab} = \Omega^{-2} \left[\begin{array}{c|c} -1/c_s^2 & -v^j/c_s^2 \\   \hline -v^i/c_s^2 & \delta^{ij} - v^i v^j / c_s^2\end{array} \right].
\end{equation}
Equivalently, the line element is given by
\begin{equation}
\d s^2 = \Omega^2 \left(-c^2_s\d t^2+ (\d x^i -v^i \d t)(\d x^j -v^j \d t)\delta_{ij} \right).
\end{equation}
For later convenience also set
\begin{equation}
\tilde g_{ab} = \left[\begin{array}{c|c} -(c_s^2-v^2) & -v_j \\   \hline -v_i & \delta_{ij} \end{array} \right]; \qquad 
\tilde g^{ab} = \left[\begin{array}{c|c} -1/c_s^2 & -v^j/c_s^2 \\   \hline -v^i/c_s^2 & \delta^{ij} - v^i v^j / c_s^2\end{array} \right].
\end{equation}
Note that indices on $v$ are raised and lowered using $\delta^{ij}$ and $\delta_{ij}$.

\subsection{Horizons}
Because of the definition of event horizon in terms of phonons (which are null geodesics of the analogue spacetime) that cannot escape
the acoustic black hole, the event horizon is automatically a null surface, and the generators of
the event horizon are automatically null geodesics.

Stationary horizons are surfaces, located for definiteness at some $f(\x)=0$, that are defined by the 3-dimensional spatial condition 
\begin{equation}
\vec\nabla f \cdot \vb = c_s \; \Vert \vec\nabla f\Vert .
\end{equation}
That is, on a horizon the \emph{normal component} of the fluid velocity equals the speed of sound, thereby either trapping or anti-trapping the acoustic excitations (resulting in black holes or white holes). 

On the horizon we have 
\begin{equation}
(\vec\nabla f \cdot \vb)^2 = c_s^2 \; \Vert \vec\nabla f\Vert ^2,
\end{equation}
which we can rewrite in 3-dimensional form as 
\begin{equation}
g^{ij} \, \partial_i f \, \partial_j f = 0,
\end{equation}
that is, 
\begin{equation}
[\delta^{ij}-v^iv^j/c_s^2] \, \partial_i f \, \partial_j f = 0.
\label{eq:}
\end{equation}
Since the conformation, and location, of the horizon is time independent this statement can be bootstrapped to 3+1 dimensions to see that \emph{on the horizon}
\begin{equation}
g^{ab} \; \nabla_a f \; \nabla_b f = 0. 
\end{equation}
That is, the 4-vector $\nabla f$ is null on the horizon. In fact, on the horizon, where in terms of the (inward-pointing) 3-normal $\n$ we can decompose $\vb_H = c_H \; \n +\vb_\parallel$ (where the subscript $H$ indicates on-horizon), we can furthermore write
\begin{equation}
 \left(g^{ab} \; \nabla_b f\right)_H  =  {\Vert \vec\nabla f\Vert \over \Omega^2_H \; c_H} \; \left( 1; \vb_\parallel\right)_H.
\end{equation}
That is, not only is the 4-vector $\nabla f$ null on the horizon, it is also a 4-tangent to the horizon (note this means we can always apply the Frobenius theorem) --- so, as in general relativity, the horizon is ruled by a set of null curves. 
Furthermore, extending the 3-normal $\n$ to a region surrounding the horizon (for instance by taking $\n = \vec\nabla f/ \Vert \vec\nabla f\Vert $) we can quite generally write
$\vb = v_\perp \; \n + \vb_\parallel$.  
Then away from the horizon
\begin{equation}
g^{ab} \; \nabla_a f \; \nabla_b f  =    { (c_s^2 - v_\perp^2)\;  \Vert \vec\nabla f\Vert ^2  \over \Omega^2 \; c_s^2}.
\end{equation}
That is, the 4-vector $\nabla f$ is spacelike outside the horizon, null on the horizon, and timelike inside the horizon.

\subsection{ZAMOs} 
A rotating analogue black hole (to be more precise: an analogue black hole where the fluid velocity is not 3-orthogonal to the horizon),  need not be equipped with the same Killing vectors as the Kerr black hole (and in fact it can be shown it is impossible to reproduce the exterior of a Kerr black hole with analogue models \cite{Visser:2004zs, ValienteKroon:2003ux, ValienteKroon:2004gj, Garat:2000pn}). 
In particular, the usual theorems whereby stationarity implies axial symmetry need no longer apply.
To attempt to generalize the constructions in Wald~\cite{Wald}, we want a natural vector that is timelike outside, spacelike inside, and null on the horizon. 
For this we will consider a vector describing an observer similar to a ZAMO (zero angular momentum observer, see for instance~\cite{raine-thomas}). 
To capture a suitable notion of ``comoving with the horizon'' let us define
\begin{equation}
Z^a = (1; \; \vb_\parallel); \qquad Z_a = -\Omega^2  ( c_s^2-v_\perp^2; \; v_\perp \n).
\end{equation}
Then we have 

\begin{equation}
g_{ab} \; Z^a \; Z^b=  - \Omega^2 (c_s^2- v_\perp^2),
\end{equation}
which is null on the horizon.
Furthermore $Z^a \partial_a f \equiv 0$, so these vector fields $Z^a$ foliate the constant-$f$ surfaces, $f(\x)=C$, and in particular foliate the horizon at $f(\x)=0$. 

In the current context the vector $Z^a$ is the closest we can get to a horizon-foliating Killing vector; it is at least horizon-foliating, even if it is not necessarily Killing.

For later convenience, we also define 
\begin{equation}
\tilde Z^a = (1; \; \vb_\parallel)= z^a; \qquad \tilde Z_a = - ( c_s^2-v_\perp^2; \; v_\perp \n)=\frac{Z_a}{\Omega^2}.
\end{equation}

\subsection{The on-horizon Lie derivative}

Note the Lie derivative
\begin{equation}
(\mathcal{L}_Z g)_{ab} = Z_{a;b} + Z_{b;a} = Z^c{}_{,a} g_{cb} + Z^c{}_{,b} g_{ca}  + Z^c{} \partial_c  g_{ab},
\end{equation}
evaluates to
\begin{equation}\label{liederiv}
(\mathcal{L}_Z g)_{ab} = \Omega^2 (\mathcal{L}_{\tilde Z} \tilde g)_{ab} + 2 (\vb_\parallel\cdot \vec\nabla \ln\Omega) g_{ab}.
\end{equation}
Explicitly
\begin{eqnarray}
(\mathcal{L}_Z g)_{ab} &=&
\Omega^2 \left[ \begin{array}{c|c}  - \vb_\parallel\cdot\vec\nabla (c^2-v^2)  & - v_\parallel{}^k{}_{,i} v_k - v_\parallel{}^k \partial_k v^i\\
\hline
- v_\parallel{}^k{}_{,j} v_k - v_\parallel{}^k \partial_k v^j &  v_{\parallel\, i,j} + v_{\parallel\, j,i} \end{array} \right] \nonumber \\
 &+& 2 (\vb_\parallel\cdot \vec\nabla \ln\Omega) g_{ab}.
\end{eqnarray}
It is the fact that this quantity is non-vanishing that makes the horizon non-Killing. 
The $(\vb_\parallel\cdot \vec\nabla \ln\Omega)$ term is just a conformal Killing contribution, hence more or less ``trivial'' (apply the Jacobson--Kang~\cite{Jacobson-Kang} argument). Now, on-horizon,
\begin{eqnarray}
(\mathcal{L}_Z g)^H_{ab} &=&
\Omega^2 \left. \left[ \begin{array}{c|c}  \vb_\parallel\cdot\vec\nabla (v_\parallel^2)    & - v_\parallel{}^k{}_{,i} v_k - v_\parallel{}^k \partial_k v_i\\
\hline
- v_\parallel{}^k{}_{,j} v_k - v_\parallel{}^k \partial_k v_j &  v_{\parallel\, i,j} + v_{\parallel\, j,i} \end{array} \right]\right|_{H} \nonumber \\
 &+& \left. 2 (\vb_\parallel\cdot \vec\nabla \ln\Omega) g_{ab} \right|_{H}.
\end{eqnarray}
We can write this in terms of the 3-d spatial Lie derivative (with respect to $\vb_\parallel$) as
\begin{equation}
(\mathcal{L}_Z g)^H_{ab} =
\Omega^2 \left. \left[ \begin{array}{c|c}  \L_{v_\parallel} (v_\parallel^2)    & - \L_{v_\parallel} v_i\\
\hline
- \L_{v_\parallel}v_j & + \L_{v_{\parallel}} \delta_{ij} \end{array} \right]\right|_H
 + \left. 2 (\vb_\parallel\cdot \vec\nabla \ln\Omega) g_{ab} \right|_H .
\end{equation}
This makes it obvious that it is the in-horizon symmetries (or lack thereof) which governs whether or not the horizon is Killing. Such symmetries will naturally be enhanced by solutions with particular symmetries (such as spherical symmetry).

From this perspective, the key reason for the degeneracy of surface horizon definitions in general relativity is that the field equations impose symmetries \emph{on horizon}.
Comparing equation (\ref{liederiv}) to equation (\ref{conformalkilling}) we can clearly see how our how our results in the next section correspond to and extend those of Jacobson and Kang~\cite{Jacobson-Kang}.

\subsection{Surface gravities}

We shall now evaluate the various definitions of surface gravity by explicit calculation.

\subsubsection{Geodesic peeling}

In the spherically symmetric case, we previously considered the peeling properties of \emph{radial} null geodesics in section \ref{peeling}. 
In contrast, here we want \emph{corotating} null geodesics, that is, outgoing null geodesics that are as close as possible to ZAMOs. Furthermore, as these geodesics emerge from the region near the horizon, their 3-velocity will have a normal component, the ``speed'' with which it is escaping ``vertically''. 
That is, take
\begin{equation}
k^a = (1, -\dot h \n + \vb_\parallel);
\end{equation}
here $h$ denotes a normal height above the horizon, and dot indicates a time derivative. 
The null condition,
\begin{equation}
g_{ab} k^a k^b=0,
\end{equation} 
yields
\begin{equation}
-(c_s^2-v^2) - 2(-\dot h v_\perp + v_\parallel^2) + \dot h^2 + v_\parallel^2 = 
- c_s^2 + (\dot h + v_\perp)^2 = 0.
\end{equation} 
Thence we have the very simple and physically plausible result
\begin{equation}
\dot h =  \pm c_s - v_\perp.
\end{equation}
For those null curves that are just escaping, near the horizon we have
\begin{equation}
\left.\dot h = c_s - v_\perp \approx - {\partial(c_s-v_\perp)\over\partial n}\right|_H \; h.
\end{equation}
(Remember $\n$ is inward pointing.) 
Let us define:
\begin{equation}
\left.\kappa_\mathrm{peeling} = -{\partial(c_s-v_\perp)\over\partial n}\right|_H =  c_H \; {\partial M_\perp\over\partial n},
\end{equation}
where $M_\perp = v_\perp/c_s$ is the transverse Mach number. 
Note this quantity $\kappa_\mathrm{peeling}$ is manifestly conformally invariant. 

Further note that $\kappa_\mathrm{peeling}$ is \emph{not} necessarily constant over the horizon; the steepness of the Mach number is \emph{not} constrained automatically to be the same everywhere along the horizon.
Then
\begin{equation}
h \approx h_*\; \exp({\kappa_\mathrm{peeling}[t-t_*]}).
\end{equation}
This is clearly related to the peeling off ($e$-folding) properties of escaping null curves near the horizon. 

\subsubsection{Null gradient normal to horizon} 
(It is best to consider this particular notion slightly ``out of order'', as $\kappa_\mathrm{normal}$ will prove useful when discussing $\kappa_\mathrm{inaffinity}$.) 
The gradient normal definition of surface gravity always works for acoustic horizons as we have defined them above, as on the horizon $Z^b Z_b =0$, and so its gradient is normal to the horizon. 
\emph{If we have already decided that the horizon is a null surface}, then its null normal must lie in the horizon, and so be proportional to $Z$. 
Then there must be a scalar $\kappa_\mathrm{normal}$ such that: 
\begin{equation}\label{normalkappa}
\nabla_a (Z^b Z_b) = - 2 \kappa_\mathrm{normal}\; Z_a.
\end{equation}
Equivalently
\begin{equation}\label{normalkappa-bis}
Z^b \nabla_a Z_b = - \kappa_\mathrm{normal} \; Z_a.
\end{equation}
But by explicit computation we now see
\begin{eqnarray}
\left.\nabla_a (Z^b Z_b)\right|_H &=& \nabla_a \left[ -\Omega^2 (c_s^2- v_\perp^2)\right] \nonumber \\
&=& - 2\Omega^2 \left( 0 ; \; c_s \nabla_i (c_s-v_\perp)\right) \nonumber \\ 
&=& - 2\Omega^2 c_s {\partial(c_s-v_\perp)\over\partial n}\left( 0; \; \n\right) \nonumber \\
&=& 2 {\partial(c_s-v_\perp)\over\partial n} \; Z_{a|H},
\end{eqnarray}
where
\begin{equation}
Z_{a|H} =  - \Omega^2 c_H (0;\; \n).
\end{equation}
Therefore with  this definition:
\begin{equation}
\kappa_\mathrm{normal} = -{\partial(c_s-v_\perp)\over\partial n} =  c_H \; {\partial M_\perp\over\partial n} = \kappa_\mathrm{peeling}.
\end{equation}
So we explicitly see that the peeling and normal gradient notions of surface gravity are still degenerate for acoustic horizons.

\subsubsection{Inaffinity}

Now consider the inaffinity definition of surface gravity. We would like to be able to write
\begin{equation}
Z^b \nabla_b Z_a =  \kappa_\mathrm{inaffinity} \; Z_a.
\end{equation}
Our first problem is that, although $Z^a$ is null, we have no  \emph{a priori} reason to expect $Z^b \nabla_b Z_a $ to be null, despite being automatically orthogonal to $Z^a$. 
We need to show that our horizon is what we will term as ``geodesic'', that is, foliated by null geodesics. 
Note that (on horizon) we \emph{always} have:
\begin{eqnarray}\label{geodesichorizon}
Z^b \nabla_b Z^a &=& Z^b (\nabla_b Z^a +\nabla^a Z_b) - {1\over2}\nabla^a (Z^b Z_b)
\nonumber \\
 &=& (\L_Z g)^a{}_b Z^b + \kappa_\mathrm{normal} Z^a.  
\end{eqnarray}
(The occurrence of the quantity $(\L_Z g)^a{}_b$ above is the explicit signal of a possible non-Killing horizon, and the reason we discussed and evaluated this quantity previously.)
On the horizon $Z^a$ is guaranteed null; both $Z^b \nabla_b Z^a$ and $(\L_Z g)^a{}_b Z^ b$ are guaranteed to be orthogonal to $Z$, but without further assumptions we cannot guarantee that they are null.
\emph{If (for now) we simply assume the horizon is geodesic}, that is, foliated by null geodesics, then
\begin{equation}
Z^b \nabla_b Z^a = \kappa_\mathrm{inaffinity} \; Z^a,
\end{equation}
and then
\begin{equation}
(\L_Z g)^a{}_b Z^ b = (\kappa_\mathrm{inaffinity}-\kappa_\mathrm{normal} ) \; Z^a = \Delta \kappa \; Z^a.
\end{equation}
Note the condition $(\L_Z g)^a{}_b Z^ b = \Delta \kappa \; Z^a$ is equivalent to demanding
\begin{equation}
(\L_Z g)_{ab} = \Delta \kappa \; g_{ab} + \zeta \; Z_a Z_b + \xi\; P^\perp_{ab}.
\end{equation}
Here 
\begin{equation}
P^\perp_{ab}=g_{ab}+{Z_aZ_b\over\Vert Z\Vert^2}.
\end{equation}

\bigskip
\noindent
This construction defines a hierarchy of possible horizons: 
\begin{itemize}
\item 
Killing ($\L_Z g=0$, the standard GR case); 
\item
conformally Killing ($\Delta\kappa\neq0$, $\zeta=\xi=0$, the Jacobson--Kang generalization); 
\item
``Kerr--Schild-like''  ($\Delta\kappa\neq0$, $\zeta\neq0$, $\xi=0$); 
\item general geodesic  ($\Delta\kappa\neq0$, $\zeta\neq 0$, $\xi\neq0$, our current case). 
\end{itemize}
The horizons termed ``Kerr--Schild-like'', where $(\L_Z g)_{ab}$ is of Kerr--Schild form on the horizon, have not to the best of my knowledge, been separately studied. 
We will now \emph{prove} that all the acoustic horizons we are considering are geodesic horizons, a fact that will also be used in the analysis of the next definition ($\kappa_\mathrm{generator}$).

We see from equation (\ref{geodesichorizon}) that
\begin{eqnarray}
\label{geodesichorizon2}
Z^b \nabla_b Z^a &=& Z^b (\nabla_b Z^a -\nabla^a Z_b) + {1\over2}\nabla^a (Z^b Z_b)
\nonumber \\
 &=&  Z^b (\nabla_b Z^a -\nabla^a Z_b)  - \kappa_\mathrm{normal} Z^a.  
\end{eqnarray}
Thus the horizon is geodesic iff (on the horizon)
\begin{equation}
Z^b (\nabla_b Z_a -\nabla_a Z_b)  \propto Z_a.
\end{equation}
Recall the definitions of $Z^a$, $g_{ab}$, $\tilde Z^a$ and $\tilde g_{ab}$ given in section (\ref{adefs}). 
We note
 \begin{equation}
  \nabla_{[a} Z_{b]} =  \Omega^2 \nabla_{[a} \tilde Z_{b]} + 2\nabla_{[a} \ln\Omega \; Z_{b]},
\end{equation}
where on the horizon
  \begin{equation}
  (\nabla_{[a} \tilde Z_{b]})_H = 
  -\left[\begin{array}{cc} 0 & c_H \,\kappa_\mathrm{normal} \, n_j \\ 
  - c_H \,\kappa_\mathrm{normal} \,n_i&  (v_\perp n_{[i})_{,j]}\end{array}\right].
  \end{equation}
But by definition we have $n_i = \partial_i f /\Vert \partial f\Vert $, so
\begin{equation}
(v_\perp n_{[i})_{,j]} = -(v_\perp/\Vert \partial f\Vert )_{[,i}  f_{,j]} = -\Vert \partial f\Vert  (v_\perp/\Vert \partial f\Vert )_{[,i} n_{j]} =c_H \tilde s_{[i} n_{j]},
\end{equation}
where we now define
\begin{equation}
\tilde s_i \equiv  - {\Vert \partial f\Vert \over c_H} \; (v_\perp/\Vert \partial f\Vert )_{,i} = - {v_{\perp,i}\over c_H} + \partial_i \ln \Vert \partial f\Vert ,
\end{equation}
with dimensions $[\tilde s] = {1/[L]}$. Therefore
 \begin{equation}
  (\nabla_{[a} \tilde Z_{b]})_H = \left[\begin{array}{cc} 0 & -c_H  \,\kappa_\mathrm{normal} \,  n_j \\  
  c_H  \,\kappa_\mathrm{normal} \,  n_i &  c_H n_{[i} \tilde s_{j]}\end{array}\right].
  \end{equation}
Now defining $\tilde S_a=(2\kappa_\mathrm{normal}, \tilde s_i)$,  again on the horizon
 \begin{equation}
  (\nabla_{[a} \tilde Z_{b]})_H = \tilde Z_{[a} \tilde S_{b]}.
 \end{equation}
 Thence, defining $S_a = \tilde S_a  - 2\nabla_{a} \ln\Omega $ we see that on the horizon
 \begin{equation}
 (\nabla_b Z_a -\nabla_a Z_b)_H = Z_a S_b - S_a Z_b.
 \end{equation}
 But then
 \begin{equation}
 (\nabla_b Z_a -\nabla_a Z_b)_H Z^b = Z_a (S_b Z^b) =  (2\kappa_\mathrm{normal} + \vb_\parallel \cdot \s) Z_a.
 \end{equation}
This observation is already enough to guarantee that the horizon is geodesic.

But now that we have shown that the horizon is geodesic, it follows immediately that we have the even stronger statement:
\begin{eqnarray}
\kappa_\mathrm{inaffinity}  &=&   (S_b Z^b) - \kappa_\mathrm{normal} = \kappa_\mathrm{normal} + \vb_\parallel \cdot \s \nonumber \\
&=& \kappa_\mathrm{normal} - 2\vb_\parallel\cdot\nabla \ln\Omega + \vb_\parallel \cdot \tilde{\s}. 
\end{eqnarray}
But now
\begin{equation}
 \vb_\parallel \cdot \tilde{\s} =  -{\vb_\parallel \cdot \nabla v_\perp\over c_H} + { \vb_\parallel \cdot \nabla \ln\Vert \partial f\Vert } 
 =  -\vb_\parallel \cdot \nabla \ln c_H +  \vb_\parallel \cdot \nabla \ln\Vert \partial f\Vert .
\end{equation}
Furthermore
\begin{eqnarray}
\vb_\parallel \cdot \nabla \ln\Vert \partial f\Vert  &=& {1\over2} \vb_\parallel \cdot \nabla \ln[\Vert \partial f\Vert ^2] 
= {v_\parallel^i \; f_{,ij} \; f_j \over \Vert \partial f\Vert ^2} 
= - {v_\parallel^i{}_{,j} \; f_{,i} \; f_j \over \Vert \partial f\Vert ^2} 
\nonumber\\
&=&  -v_\parallel{}^{i,j} n_i n_j =  -v_\parallel{}^{(i,j)} \; n_i n_j.
\end{eqnarray}
Pulling it all together
\begin{equation}
\kappa_\mathrm{inaffinity}  = \kappa_\mathrm{normal} - 2\vb_\parallel \cdot\nabla \ln\Omega - \vb_\parallel \cdot \nabla \ln c_H -v_\parallel{}^{(i,j)} \; n_i n_j.
\end{equation}
The last term is an internal horizon \emph{shear}. 
This quantity $\kappa_\mathrm{inaffinity}$ is manifestly \emph{not} a conformal invariant. 
One can also express this as
\begin{equation}
\kappa_\mathrm{inaffinity}  = \kappa_\mathrm{normal} - \vb_\parallel \cdot \nabla \ln [c_H \Omega^2] -v_\parallel{}^{(i,j)} \; n_i n_j. 
\end{equation}
This is consistent with the Jacobson--Kang analysis, as for them, automatically, the in-horizon shear is taken to be zero.

\subsubsection{Generator-based} 
We shall define
\begin{equation}
\kappa_\mathrm{generator}^2 = -{1\over2} (\nabla^{[a} Z^{b]})_H (\nabla_{[a} Z_{b]})_H.
\end{equation}
We can always \emph{define} this quantity into existence, the question is how does it relate to the previous two definitions? 

We have already shown that at the analogue horizon
 \begin{equation}
 (\nabla_b Z_a -\nabla_a Z_b)_H = Z_a S_b - S_a Z_b.
 \end{equation}
 But then
\begin{equation}
(\nabla_b Z_a -\nabla_a Z_b)_H (\nabla^b Z^a -\nabla^a Z^b)_H = - 2(S_a Z^a)^2.
\end{equation}
Therefore
\begin{eqnarray}
\kappa_\mathrm{generator}^2 &=& -{1\over2} (\nabla^{[a} Z^{b]})_H (\nabla_{[a} Z_{b]})_H \nonumber\\
&=&  -{1\over8}  (\nabla^b Z^a -\nabla^a Z^b)_H (\nabla_b Z_a -\nabla_a Z_b)_H \nonumber\\
&=& \; \; \, {1\over4} (S_a Z^a)^2,
\end{eqnarray}
and so
\begin{equation}
\kappa_\mathrm{generator} = {1\over2} (S_a Z^a) = {\kappa_\mathrm{normal}+\kappa_\mathrm{inaffinity}\over2}.
\end{equation}
Pulling it all together we see
\begin{equation}
\kappa_\mathrm{generator}  = \kappa_\mathrm{normal} - \vb_\parallel \cdot\nabla \ln\Omega - {1\over2}\vb_\parallel \cdot \nabla \ln c_H -{1\over2}v_\parallel{}^{(i,j)} \; n_i n_j.
\end{equation}
Alternatively,
\begin{equation}
\kappa_\mathrm{generator}  = \kappa_\mathrm{normal}- {1\over2}\vb_\parallel \cdot \nabla \ln [c_H\Omega^2] -{1\over2}v_\parallel{}^{(i,j)} \; n_i n_j.
\end{equation}
This quantity is manifestly \emph{not} conformally invariant.

\subsubsection{Tension in a rope}

There is a nice argument leading to a tidy physical interpretation of the surface gravity in terms of tension in an ideal massless rope held at infinity. 
In the current context we would want to evaluate
\begin{equation}
\kappa_\mathrm{tension} = \lim_H \sqrt{-Z^2}\, \Vert A\Vert ,
\end{equation}
with $A$ the magnitude of the 4-acceleration of the integral curves of $Z^a$. Define
\begin{equation}
V^a = {Z^a\over\sqrt{-Z^b Z_b}}, \qquad A^a = V^b \nabla_b V^a,
\end{equation}
as the velocity and acceleration of an orbit of $Z^a$. Now using
\begin{equation}
Z^a = \sqrt{-Z^b Z_b}\; V^a,
\end{equation}
we see
\begin{equation}
Z_b \nabla^b Z^c =  (-Z^b Z_b) A^c + {1\over2} {Z^b \nabla_b (-Z^2)\over (-Z^2)} Z^c. 
\end{equation}
Then working outside the horizon, where $A$ and $Z$ are 4-perpendicular, and $Z$ is timelike while $A$ is spacelike,  we have
\begin{equation}
  (-Z^b Z_b) \Vert A^c\Vert ^2 = { \Vert Z_b \nabla^b Z^c \Vert ^2 \over (-Z^2) }+  {1\over4} {[Z^b \nabla_b \ln (-Z^2)]^2}.
\end{equation}
Now, as we approach the horizon
\begin{equation}
{\Vert Z_b \nabla^b Z^c \Vert ^2 \over (-Z^2) }  \to  {0\over0}.
\end{equation}
Since this is indeterminate it is useful to consider
\begin{eqnarray}
{ \nabla_a\Vert Z_b \nabla^b Z^c \Vert ^2 \over \nabla_a (-Z^2) } 
&\to& 
{ 2(Z_b \nabla^b Z^c)\nabla_a(Z_b \nabla^b Z^c)  \over -2\kappa_\mathrm{normal} Z_a} \nonumber\\
&=& 
{ (\kappa_\mathrm{inaffinity} Z_c)\nabla_a(Z_b \nabla^b Z^c)  \over -\kappa_\mathrm{normal} Z_a} \nonumber\\
&=& 
{ \kappa_\mathrm{inaffinity} (\nabla_a( Z_c Z_b \nabla^b Z^c) - (\nabla_a Z_c)  (Z_b \nabla^b Z^c)) \over -\kappa_\mathrm{normal} Z_a} \nonumber\\
&=& 
{ \kappa_\mathrm{inaffinity} (\nabla_a(0) - (\nabla_a Z_c)  (\kappa_\mathrm{inaffinity} Z^c)) \over -\kappa_\mathrm{normal} Z_a} \nonumber\\
&=& 
{ -\kappa_\mathrm{inaffinity}^2 (\nabla_a Z_c) Z^c \over -\kappa_\mathrm{normal} Z_a} \nonumber\\
&=& 
\kappa_\mathrm{inaffinity}^2 \; \left({ \kappa_\mathrm{normal} Z_a \over \kappa_\mathrm{normal} Z_a}\right) \nonumber\\
&=& 
\kappa_\mathrm{inaffinity}^2.
\end{eqnarray}
So by the l'Hospital rule:
\begin{equation}
\lim_H \left\{  { \Vert Z_b \nabla^b Z^c \Vert ^2 \over \Vert Z\Vert ^2 } \right\} = \kappa_\mathrm{inaffinity}^2.
\end{equation}
Furthermore, as we approach the horizon
\begin{equation}
Z^2 = -2\Omega^2(c_s^2-v_\perp^2) \approx  -\Omega^2 c_H \kappa_\mathrm{normal} \times (\hbox{normal 3-distance to horizon}).
\end{equation}
So 
\begin{equation}
\lim_H \left(Z^d \nabla_d \ln(-Z^2) \right)  = \vb_\parallel\cdot\nabla\ln[\Omega^2 c_H \kappa_\mathrm{normal}] .
\end{equation}
(Remember that for an acoustic horizon there is no need to believe in a zeroth law, there is no need for $\kappa_\mathrm{normal}$ to be constant over the horizon).
Pulling everything together
\begin{equation}
\kappa_\mathrm{tension}^2 = \lim_H \{(-Z^2) \Vert A^c\Vert ^2\} =  
\kappa_\mathrm{inaffinity}^2 +  {1\over4} \left(\vb_\parallel\cdot\nabla\ln[\Omega^2 c_H \kappa_\mathrm{normal}] \right)^2.
\end{equation}
That is:
\begin{equation}
\kappa_\mathrm{tension} = \sqrt{\kappa_\mathrm{inaffinity}^2 +  {\textstyle{1\over4}} \left(\vb_\parallel\cdot\nabla\ln[\Omega^2 c_H \kappa_\mathrm{normal}] \right)^2}.
\end{equation}
Now using
\begin{equation}
\kappa_\mathrm{inaffinity}  = \kappa_\mathrm{normal} - \vb_\parallel \cdot \nabla \ln [c_H \Omega^2] -v_\parallel{}^{(i,j)} \; n_i n_j ,
\end{equation}
we have
\begin{eqnarray}
\kappa_\mathrm{tension} &=& \left(\left(\kappa_\mathrm{normal} - \vb_\parallel \cdot \nabla \ln [c_H \Omega^2] -v_\parallel{}^{(i,j)} \; n_i n_j \right)^2\right. \nonumber \\
&+& \left. {\textstyle{1\over4}} \left(\vb_\parallel\cdot\nabla\ln[\Omega^2 c_H \kappa_\mathrm{normal}] \right)^2\right)^{1/2}.
\end{eqnarray}
So also this quantity is manifestly \emph{not} a conformal invariant.

\subsubsection{2-d expansion}

Finally, we consider the definition relating the surface gravity to the expansion of a suitably defined congruence of timelike geodesics normal to the horizon~\cite{Jacobson:2008cx}. See earlier discussion and equations (\ref{E:theta-0})--(\ref{E:theta-n}).
The key point here, is once again this equality relies on the existence of an appropriate geodesic congruence invariant under the flow of a Killing (or Killing-like) vector, and so cannot be applied blindly to modified gravity or analogue gravity scenarios.

For an acoustic horizon we would want to pick a congruence dragged by $Z^a$,
\begin{equation}
Z^a \nabla_a u^b=u^a\nabla_a Z^b.
\end{equation}
\emph{If} it is possible to construct such a congruence, then from equation (\ref{normalkappa}), we know that 
\begin{equation}
\theta_{2d}  =\frac{(u\cdot Z) \kappa_\mathrm{normal}}{(Z\cdot u)^2}.
\end{equation}
And hence, now for an acoustic horizon,
\begin{equation}
\kappa_\mathrm{expansion}=(u \cdot Z)\theta_{2d}= \kappa_\mathrm{normal}. 
\end{equation}

\subsubsection{Summary}

For an acoustic horizon we generically have
\begin{equation}
\kappa_\mathrm{normal} = \kappa_\mathrm{peeling} = \kappa_\mathrm{expansion}.
\end{equation}
On the other hand $\kappa_\mathrm{inaffinity}$, $\kappa_\mathrm{generator}$, and $\kappa_\mathrm{tension}$ are generically distinct from each other, and from the preceding three items.

\section{Modified gravity}
While in the previous section we have been interested in the framework of analogue gravity, the concerns we have are also of vital importance for modified gravity. Some general points to consider:

\begin{itemize}

\item The usual situation, where the final state of a black hole is either static, or stationary and axisymmetric, depends critically on the standard Einstein equations (and ``reasonable'' matter sources). 
This could easily fail in modified gravity.

\item The usual situation, where black hole horizons are Killing horizons,  depends critically on the standard Einstein equations (and ``reasonable'' matter sources), which could also easily fail in modified gravity~\cite{HawkingEllis, carter1, carter2, Heusler, Friedrich:1998wq, Robinson:2004, Kodama:2011}.

\item The usual situation, where black holes satisfy the zeroth law (constancy of $\kappa$), depends critically on the ``effective stress energy'', in the sense $G^{ab} \propto T^{ab}_\mathrm{effective}$, satisfying some form of classical energy condition. Again, this could easily fail in modified gravity~\cite{Racz:1995nh,Hawking:1971vc}. 
\end{itemize}
In short, the distinctions between the various surface gravities can also easily become important outside of the analogue spacetime framework. Here we will work through one specific example within the framework of Lorentz-violating theories to demonstrate this. 


\subsection{Universal horizons and their surface gravities}

As discussed in section \ref{ae-bhs}, universal horizons, defined by $(u\cdot \chi)=0$ are present in solutions to Einstein-\Aether\ and \Horava -Lifshitz gravity. From our point of view for the present, these universal horizons are interesting because they provide examples of non-Killing horizons, and furthermore these horizons are \emph{not} null surfaces, unlike the cases we have previously been looking at. 
Relevant questions are:
\begin{itemize}
\item Which of the definitions of surface gravity can be extended to these universal horizons?
\item Are these all identical? If not, how do they differ?
\end{itemize}
These are non-trivial questions, important for other open issues such as whether or not Hawking radiation exists for such theories, from what surface it originates, and further the wider issues surrounding the thermodynamics of such spacetimes.

\subsubsection{Generator-based}

This is the quantity calculated in reference~\cite{Berglund:2012bu}; we reproduce the most salient aspects of the derivation here. 
(We will carefully work through this definition first, as it is the one used in previous literature, and our subsequent constructions rely heavily on this set-up).

Set up a tetrad of unit vectors, the timelike vector given by the aether, $u^a$, then two spacelike vectors $m^a$ and $n^a$, which are mutually orthogonal and lie in the tangent plane of two-spheres, and a spacelike unit vector is provided by the outward-pointing $s^a$ (our radial vector). 
Further, any rank-two tensor can be expanded in terms of the quantities $u_au_b$, $u_{(a}s_{b)}$, $u_{[a}s_{b]}$, $s_as_b$, and $\hat g_{ab}$; where $\hat g_{ab}$ is projection tensor onto the spatial two-sphere surface.

As we have spherical symmetry any physical vector should have components only along $u^a$ and $s^a$. 
Also note the acceleration will only have a component along $s^a$. That is, $a^a=(a\cdot s)s^a$. 
Further note that at the universal horizon $s^a$ is, by definition, parallel to $\chi^a$.
We therefore have the useful relations:
\begin{equation}
\nabla_a u_b = -(a\cdot s)\; u_a s_b + K^{(u)}_{a b}\,; \qquad
K^{(u)}_{a b} = K_0 \; s_a s_b + \frac12{\hat{K}^{(u)}}\;\hat{g}_{a b},
\end{equation}
\begin{equation}
\nabla_a s_b = K_0 \; s_a u_b + K^{(s)}_{a b}\,; \qquad
K^{(s)}_{a b} = -(a\cdot s)\; u_a u_b + \frac12{\hat{K}^{(s)}}\;\hat{g}_{a b}.
\end{equation}
Here $K^{(u)}_{a b}$ is the extrinsic curvature of the hypersurfaces orthogonal to the aether flow $u^a$, while $ K^{(s)}_{a b}$ is the extrinsic curvature of the hypersurfaces orthogonal to $s^a$,  and $\hat{K}^{(u)}$ and $\hat{K}^{(s)}$ are the traces of the extrinsic curvatures of the 2-spheres due to their embeddings in these two hypersurfaces, respectively. Finally $K_0$ is related to the 4-acceleration of the integral curves of $s^a$ by $s^a\nabla_a s_b = K_0 \; u_b$.

Now consider an arbitrary vector of form
\begin{equation}
A_a=-fu_a+hs_a,
\end{equation}
where $f$ and $h$ are arbitrary functions respecting the symmetries of the spacetime, so in particular $A$ is Lie dragged by the Killing vector $\chi$.
By spherical symmetry
\begin{equation}\label{derivphysvector}
\nabla_{\left[a\right.}A_{\left.b\right]}= - Q_A \; u_{\left[a\right.}s_{\left.b\right]},
\end{equation}
(as this is the only possible fully anti-symmetric choice possible within spherical symmetry), with
\begin{equation}\label{generalq}
Q_A=-f(a\cdot s)-s^a\nabla_a f +hK_0+u^a\nabla_a h.
\end{equation}
We have chosen an opposite sign convention to \cite{Berglund:2012bu} to minimize subsequent sign flips.

Our Killing vector is 
\begin{equation}
\chi^a=-(u\cdot \chi)u^a +(s\cdot \chi)s^a.
\end{equation}
And from the results above, and the Killing equation, we have
\begin{equation}\label{killingderiv}
\nabla_a\chi_b=- \frac{Q_\chi}{2}(u_as_b-s_au_b),
\end{equation}
where now
\begin{eqnarray}
\label{E:tricky}
Q_\chi &=&- (u\cdot \chi)(a\cdot s) +(s\cdot \chi)K_0-s^a\nabla_a(u\cdot \chi)+u^a\nabla_a(s\cdot \chi) \nonumber\\
&=& -2\left\{ (u\cdot \chi)(a\cdot s) -(s\cdot \chi)K_0 \right\}.
\end{eqnarray}
The second equality follows from the fact that for any $A$ respecting the symmetries of the spacetime
\begin{eqnarray}
\nabla_a (A\cdot \chi) 
&=& (\nabla_a\chi^b) A_b + \chi^b \nabla_a A_b 
\nonumber\\
&=&-\chi^b\nabla_b A_a + \chi^b \nabla_a A_b 
\nonumber \\
&=& -Q_A \{ (s \cdot \chi) u_a- (u \cdot \chi) s_a\}.
\end{eqnarray}
Specializing this relation to our case we have
\begin{eqnarray}
s^a\nabla_a(u \cdot \chi)  &=& Q_{u}(u \cdot \chi) = (a\cdot s) \, (u \cdot \chi); \\
u^a\nabla_a(s \cdot \chi)  &=& Q_{s}(s \cdot \chi) = K_0\; (s \cdot \chi).
\end{eqnarray}
Combining these results we obtain the second line of (\ref{E:tricky}).

We can now identify $\kappa_\mathrm{generator}$ with $|Q_\chi|/2$, as given in equations (22) and (23) of reference \cite{Berglund:2012bu}, since, provided (\ref{killingderiv}) holds true, we have
\begin{equation}
\kappa_\mathrm{generator}=\sqrt{-\frac{1}{2}(\nabla_a\chi_b)(\nabla^a\chi^b)} = \frac{|Q_\chi|}{2}.
\end{equation}
Therefore (at any point in the spacetime)
\begin{equation}
\kappa_\mathrm{generator} = \frac{|Q_\chi|}{2} = \Big|(u\cdot \chi)(a\cdot s) -(s\cdot \chi)K_0\,\Big|.
\end{equation}
At the universal horizon, $(u\cdot \chi=0)$ by definition, and thus $\chi$ and $s$ are parallel. Therefore
\begin{equation}\label{kappauh}
\left.\kappa_\mathrm{generator}\right|_\mathrm{UH}=K_{0|\mathrm{UH}}\;  \Vert\chi\Vert_\mathrm{UH},
\end{equation}
where the absolute value and the explicit minus sign can safely be removed given that both $K_0$ and $(s\cdot\chi)$ are both positive on the universal horizon.
Indeed, this is consistent with  \cite{Berglund:2012bu} from which, by confronting our equation (\ref{killingderiv}) with equation (22) of \cite{Berglund:2012bu}, one can deduce $\kappa_\mathrm{generator}=Q_\chi/2$. 
In closing let us stress that this derivation relies \emph{very heavily} on the special symmetries of the solution and that $ ||\chi||_\mathrm{UH}\neq 0$ on the universal horizon.

\subsubsection{Peeling}

A specific class of spherically symmetric black holes was examined in \cite{Barausse:2011pu}, which in Eddington--Finkelstein coordinates take the form
\begin{equation}
\d s^2 = -e(r) \,\d \nu^2 +2f(r)\,\d\nu\,\d r +r^2\,\d\Omega.
\end{equation}
First, in analogy with section (\ref{peeling}), change this into Schwarzschild coordinates. 
Set
\begin{equation}
\d t = \d \nu -\frac{f(r)}{e(r)}\,\d r,
\end{equation}
so that
\begin{equation}
\d s^2 = -e(r)\, \d t^2 +\frac{f(r)^2}{e(r)}\,\d r^2 +r^2\d\Omega.
\end{equation}
Consider an out-going null ray
\begin{equation}
e(r) \d t^2 =  \frac{f(r)^2}{e(r)}\,\d r^2,
\end{equation}
so that
\begin{equation}
\frac{\d r}{\d t}= \frac{e(r)}{f(r)}.
\end{equation}
For $r_1$ and $r_2$ close to the universal horizon at $r=r_{UH}$ 
\begin{equation}
\left.\frac{\d (r_1-r_2)}{\d t}= \frac{\d}{\d r}\left(\frac{e(r)}{f(r)}\right)\right\vert_{\mathrm{UH}} (r_1-r_2)+ O\left([r_1-r_2]^2\right),
\end{equation}
and so for a generic universal horizon we can define
\begin{equation}
\kappa_\mathrm{peeling}= \left.\frac{1}{2}\frac{\d}{\d r}\left(\frac{e(r)}{f(r)}\right)\right\vert_{\mathrm{UH}},
\end{equation}
in general. 

Let us now apply this construction to the simplest explicit example we can find. 
Taking a look at section (4.2) in reference~\cite{Berglund:2012fk}, we see an example of an exact solution with
\begin{equation}
e \left( r \right) =1-{\frac {r_0}{r}}-{\frac {r_u \left( r_0+r_u \right) }{{r}^{2}}}; 
\qquad f(r)=1.
\end{equation}
Here
\begin{equation}
r_u=\left( \sqrt{C}-1\right) \frac{r_0}{2},
\end{equation}
with $C$ a constant depending on the coupling constants of the theory. 
Plugging this into the above, we find, that for this specific example
\begin{equation}
\kappa_\mathrm{peeling}=\frac{2C}{r_0}.
\end{equation}
In \cite{Berglund:2012bu} the surface gravity was computed to be $\frac{Q_\chi}{2}$ which we can see from inspection is equal to $\frac{2C}{r_0}$.
Thus, (at least in situations where they can both meaningfully be defined), $\kappa_\mathrm{peeling} = \kappa_\mathrm{generator}$ for universal horizons. 
We do not wish to apply this construction to the general solutions in terms of asymptotic expansions presented in that paper, as those are only valid for large $r$, and as such, ill-adapted to this calculation.
%

\subsubsection{Null normal}

Let us now see if it is possible to extend the notion $\kappa_\mathrm{normal}$ to a universal horizon, at least in a highly symmetric case. 
First, define a vector $\lambda$, everywhere orthogonal to $\chi$, by
\begin{equation}
\lambda^a=(s\cdot \chi)u^a -(u\cdot \chi)s^a.
\end{equation}
(There is a sign ambiguity in this definition depending on whether you want the inwards or outwards pointing unit spacelike vector at infinity.) 
Note also, that on the \emph{Killing} horizon, $(u\cdot \chi)=(s\cdot \chi)$, so $\lambda^a =\chi^a$.
Now, by spherical symmetry
\begin{eqnarray}
\nabla_a(\chi^2)&=& \nabla_a(\chi^b)\chi_b+\chi^b\nabla_a\chi_b \nonumber\\
&=& -\frac{Q_\chi}{2}\chi_b(u_as^b-s_au^b)-\frac{Q_\chi}{2}\chi^b(u_as_b-s_au_b) \nonumber \\
&=&Q_\chi\,(u\cdot\chi)s_a-Q_\chi\,(s\cdot\chi)u_a \nonumber\\
&=&- Q_\chi \lambda_a \nonumber\\
&=& -2\kappa_\mathrm{generator} \; \lambda_a,
\end{eqnarray}
everywhere in the spacetime. 
Off the Killing horizon, this seems to provide the most natural \emph{definition} of $\kappa_{\mathrm{normal}}$, and it is equal to $\kappa_\mathrm{generator}$.

\subsubsection{Inaffinity}

Likewise, for null horizons we have defined $\kappa_{\mathrm{inaffinity}}$ by
\begin{equation}
\chi^b\nabla_b\chi_c =\kappa_{\mathrm{inaffinity}} \chi_c.
\end{equation}
But, (as we have already seen), by spherical symmetry, 
\begin{equation}
\nabla_a\chi_b=-\frac{Q_\chi}{2} (u_a s_b-u_bs_a),
\end{equation}
so, now evaluating on the universal horizon, we have
\begin{eqnarray}
\chi^b\nabla_b\chi_c&=&- \frac{Q_\chi}{2} \chi^b(u_bs_c-u_cs_b) \nonumber\\
&=&-\frac{Q_\chi}{2}\{ (u\cdot \chi)s_c-(s\cdot \chi)u_c\} \nonumber \\
&=&\frac{Q_\chi}{2} \;\lambda_c \nonumber \\
&=& \kappa_{\mathrm{inaffinity}} \; \lambda_c,
\end{eqnarray}
where the last line is our \emph{definition} of $\kappa_\mathrm{inaffinity}$, which is now seen to be the same as $\kappa_\mathrm{normal}$ and $\kappa_\mathrm{generator}$.

\subsubsection{Tension in a rope}

Note that it is not at all obvious there should be any possible calculation for the tension in a rope at infinity, as our universal horizon is inside the Killing horizon, where nothing can stay still, so we certainly must abandon the notion of $\kappa_\mathrm{tension}$ directly relating to the tension on a rope held at infinity.

However, if we want to mathematically continue this idea, we want to calculate
\begin{equation}
\kappa_\mathrm{tension}^2 = \lim_\mathrm{UH} \left( \Vert \chi\Vert ^2 \; \Vert A\Vert ^2 \right)
\end{equation}
Because the universal horizon is not a null surface the limit is straightforward, and it is easy to see that
\begin{equation}
\kappa_\mathrm{tension}^2 = \lim_\mathrm{UH} \left\{\frac{-(\chi^b\nabla_b\chi^c)(\chi^a\nabla_a\chi_c)}{-\chi^a\chi_a} \right\}.
\end{equation}
But, we can again use equation (\ref{killingderiv}), so that
\begin{eqnarray}
\kappa_\mathrm{tension}^2&=&\left.\frac{1}{4}
\frac{-Q_\chi^2\{(\chi \cdot u)s_a-(\chi \cdot s)u_a\}\,\{(\chi \cdot u)s^a-(\chi \cdot s)u^a\}}
{-\chi^2}\right|_{UH}\nonumber\\
&=&\left.\frac{Q_\chi^2}{4}\right|_\mathrm{UH}.
\end{eqnarray}
That is
\begin{equation}
\kappa_\mathrm{tension} = \frac{|Q_\chi|}{2}. 
\end{equation}
Again we note that many of these definitions degenerate.

\subsubsection{2-d expansion}

Another possibility is to consider the quantity defined by Jacobson and Parentani~\cite{Jacobson:2008cx}. 
Instead of the form given in that paper, for spacelike regions (such as we have under consideration here) it is better to start from the basic definition 
\begin{equation}
\theta_{2d}=\frac{\frac{1}{2}u^a\nabla_a \chi^2}{\chi^2-(\chi\cdot u)^2},
\end{equation}
use the fact that $(\chi\cdot u)\to 0$ on the universal horizon, and expand the numerator to obtain
\begin{equation}
\left.\theta_{2d}\right|_\mathrm{UH} = \frac{u^a\chi^b\nabla_a\chi_b}{\chi^2}.
\end{equation}
Now we can again use our useful symmetries, and note that on the universal horizon $\chi^2 = (\chi\cdot s)^2$, to see
\begin{equation}
\left.\theta_{2d}\right|_\mathrm{UH} =  \frac{-Q_\chi u^a \chi^b (u_as_b-u_bs_a)}{2\,\chi^2}=\frac{Q_\chi\,(\chi\cdot s)}{2(\chi\cdot s)^2}
 = {Q_\chi\over 2(\chi\cdot s)}.
\end{equation}
We see that, whereas for Killing horizons, where we relate this quantity to the surface gravity through normalization with an appropriate horizon-crossing timelike vector $\chi\cdot u$, here we want to normalize with an appropriate spacelike vector $\chi\cdot s$.  
Specifically, for universal horizons we can define
\begin{equation}
\kappa_\mathrm{expansion} = \left.\left\{(\chi\cdot s) \,\theta_{2d}\right\}\right|_\mathrm{UH} =  \frac{Q_\chi}{2}.
\end{equation}
In particular, comparing with equation (\ref{kappauh}), we see that $\theta_{2d}= K_0$ at the universal horizon.  

So the equivalence of the usual plethora of sruface gravities does not necessarily require the horizon to be a Killing one. It would be interesting to see if such a degeneracy remains for universal horizons of stationary black holes, for which there is no solution in 4D.

\subsubsection{Summary}

For a spherically symmetric universal horizon, and subject to the definitions adopted above, we have
\begin{equation}
\kappa_\mathrm{generator} = \kappa_\mathrm{normal} = \kappa_\mathrm{inaffinity}=\kappa_\mathrm{tension}=  \kappa_\mathrm{expansion}.
\end{equation}
In the exact solutions, where it is possible to calculate $\kappa_\mathrm{peeling}$, we find  $\kappa_\mathrm{peeling}=\kappa_\mathrm{generator}$.

Note that it is only by using tricks of spherical symmetry that we have been able to define some extension of $\kappa_\mathrm{normal}$ and $\kappa_\mathrm{inaffinity}$. 
The most natural notions for such horizons seem to be $\kappa_\mathrm{generator}$, $\kappa_\mathrm{expansion}$ and $\kappa_\mathrm{peeling}$, as they do not \emph{a priori} require a null surface. 
In the case of our modified gravity scenario, the symmetries of the problem seem to have reduced the plethora of surface gravities we have. 
Likewise, in analogue cases, if we have enough symmetry in the set up, the number of distinct surface gravities should collapse.

Indeed the calculations presented in this section rely so heavily on the spherical symmetry, that for a stationary non-static solution possessing a universal horizon, it seems that a completely new mode of attack would need to be developed. 
It is far from obvious which if any of these degeneracies would remain in such a case, and it seems somewhat  unlikely that the notions of $\kappa_\mathrm{inaffinity}$ and  $\kappa_\mathrm{normal}$ could be developed at all. 
Overall, the best statement seems to be this: There are many possible  definitions of surface gravity, identical in cases of high symmetry. 

\section{Discussion}

In this chapter we have considered a number of different definitions of surface gravity, all of which reduce to the standard case in stationary general relativity. 
We have shown in the case of stationary analogue black holes how these different surface gravities can be calculated, and how they are related. 

These concerns are also important for modified gravity, and we have illustrated this with one example involving the so-called ``universal horizon''. 
In general,  the differences between these definitions, and appropriate choices of which to use, will become more relevant the less symmetry there is in the case under consideration. 

In the case of the universal horizon, it seems that the surface gravity calculated disagrees with the temperature derived in \cite{Berglund:2012fk}. We shall see this is due to the fact that all these definitions are purely geometrical, whereas the very existence of these horizons depends crucially on the \aether\ structure. We will return to such concerns in the next chapter.  

The symmetries in question might be obvious ones (spherical symmetry, axial symmetry), but might also be less obvious --- such as the enhanced conformal symmetry at general relativity horizons that is at least partly connected with the specific field equations and inter-twined with the rigidity theorem and zeroth law. 

Once one moves away from standard general relativity the situation becomes \emph{much} more complicated than one might have naively expected.

\chapter{Ray tracing Einstein--{\AE}ther black holes}

\epigraph{The law that entropy always increases holds, I think, the supreme position among the laws of Nature. 
If someone points out to you that your pet theory of the universe is in disagreement with Maxwell's equations -- then so much the worse for Maxwell's equations. If it is found to be contradicted by observation -- well, these experimentalists do bungle things sometimes. 
But if your theory is found to be against the second law of thermodynamics I can give you no hope; there is nothing for it but to collapse in deepest humiliation.}{Arthur Eddington (1882 –- 1944)}

\section{Introduction}
\parskip 3 pt

\noindent As discussed in section \ref{perpetualmotion}, the GSL runs into problems in the presence of Lorentz violation. One may have the hope that the universal horizon, if it is possible to associate thermodynamic variables to it, can solve this problem. Indeed, the Universal horizon is found to satisfy the first law of black-hole mechanics~\cite{Berglund:2012bu}. Not only that, the calculation of Ref.~\cite{Berglund:2012fk} to calculate the temperature, using the tunneling formalism of Ref.~\cite{Parikh:1999mf}, seems to predict the emission of a thermal flux from the Universal horizon (later, a second calculation was performed which disputes this conclusion \cite{Michel:2015rsa}). We will discuss both of these calculations for Hawking radiation in more detail in the next chapter. But is this temperature associated with the Universal horizon relevant for observers outside the black hole? Or is the emitted radiation somehow reprocessed at the Killing horizon, in an energy and species dependent way? Is this the key to recovering a healthy thermodynamic behavior of black holes in Lorentz-violating theories? In order to answer these questions, as a preliminary step we need to understand particle dynamics on these spacetimes, and how it is affected by the presence of the Universal and Killing horizons.

The problem is that the natural ray trajectories to consider in these spacetimes correspond to particles which have non-trivial dispersion relations as discussed in section \ref{dispersion}, which adds complications as these trajectories are no longer metric geodesics, and are not determined by the spacetime geometry alone. How are the rays affected by the presence of Universal and Killing horizons? Does the Universal horizon affect the ray trajectories of modified dispersion relations, in a way analogous to the Killing horizon affecting the relativistic rays? If so, what is the effect of the Killing horizon? Can we say anything about which surface is relevant for Hawking radiation? The purpose of this chapter is to clarify these questions.

This chapter is arranged as follows: In Sec.~\ref{subsec:trajectories} we review the behavior of relativistic rays at the Killing horizon and, as a warm-up, we perform a study of slices of constant ``\aether\ time''. We then proceed to the main body of our investigation in Sec.~\ref{physicaltrajectories} where we look at the ray trajectories associated to modified dispersion relations in these spacetimes. 
In addition to studying the behavior of rays near the Universal and Killing horizons, we also provide a notion of surface gravity for the Universal horizon, and compare it to those already existing in the literature~\cite{Berglund:2012bu, Berglund:2012fk,Cropp:2013zxi}. 
We conclude in Sec.~\ref{raytracingdiscussion} with a summary and discussion of the implications of our work, and indicate some possible future directions. Appendix~\ref{gendispersion} discusses the universality of temperature of the Universal horizon.

\section{Physical trajectories in an Einstein--\Aether\ black hole}
\label{physicaltrajectories}

We analyze the motion (determined by the group velocity) of particles endowed with modified, Lorentz-violating, dispersion relations in the black-hole geometries discussed in Sec.~\ref{ae-bhs}. 
Modified dispersion relations arise due to the interaction of these particles with the {\ae}ther. Thus the trajectories are \emph{not} simply the geodesics determined from the spacetime metric.  

It will prove beneficial to first review the standard description of particle trajectories in a black-hole spacetime, with particular attention to their behavior close to the Killing horizon. 
We do that in   Secs.~\ref{sec:standardpeeling} and ~\ref{sec:RaeT}. After discussing the modified dispersion relations in Sec.~\ref{subsec:modDisRel}, we will introduce an appropriate notion of the conserved particle energy in Sec.~\ref{subsec:conservation law}, which will then be needed in Sec.~\ref{subsec:trajectories} where we finally construct the ray trajectories.

\subsection{Ray tracing and peeling in purely metric black holes}
\label{sec:standardpeeling}

Let us then first briefly recap some aspects of ray tracing, taking as a simplest example the Schwarzschild spacetime. (See also~\cite{Poisson}.) In Eddington--Finkelstein coordinates introduced in \ref{eq:ef}
\begin{equation}
\d s^2= -\left(1-\frac{2M}{r}\right)\d v^2+2\,\d v\,\d r+ r^2\d \Omega_2^2,
\end{equation}
the outgoing rays will be given by 
\begin{equation}
\left.\frac{\d v}{\d r}\right|_{\mathrm{out}}=\frac{2}{1-\frac{r}{2M}}
\end{equation}
with the ingoing rays travelling at $45^{\circ}$. 

In these coordinates a peeling-off of rays from the Killing horizon can be seen (Fig.~\ref{schwarz}). 
It is this peeling that is related to the large increase in frequency/energy as one traces back along an outgoing mode in Hawking radiation~\cite{Jacobson:1999zk}.

\begin{figure}[!htb]
\centering
\includegraphics[scale=0.6]{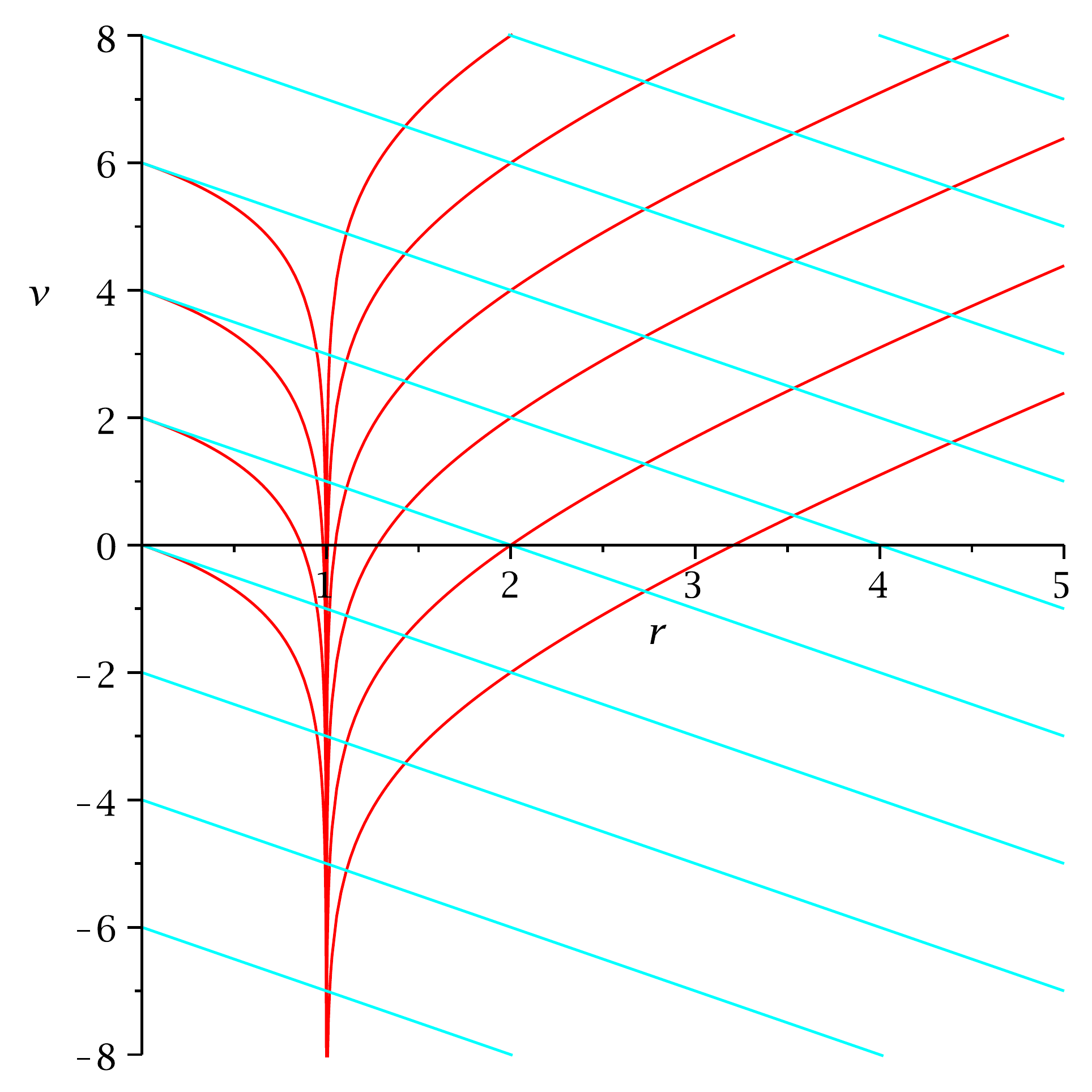}
\caption{Ingoing (blue) and outgoing null rays (red) for standard Schwarzschild black hole, horizon at r=1.}
\label{schwarz}
\end{figure}

In particular one can associate a surface gravity notion to this peeling by a suitable expansion of the ray behaviour close to the Killing horizon. This takes the form

\begin{equation}
\left.\frac{\d r}{\d v}\right|_{\mathrm{out}}=\left.\frac{\d r}{\d v}\right|_{\KH}+\left.\frac{\d}{\d r} \frac{\d r}{\d v}\right|_{\KH}(r-r_{\KH}) +\mathcal{O}(r-r_{\KH})^2.
\label{eq:standarddrdv}
\end{equation}
At the Killing horizon, the first term on the right hand side of this equation vanishes, and the second term defines the peeling surface gravity
\begin{equation}
\kappa_{\rm peeling}\equiv \left. \frac{1}{2}\frac{\d}{\d r} \frac{\d r}{\d v}\right|_{\KH}
\end{equation}

There are other ways to define surface gravity, as noted in the previous chapter, all of which coincide for stationary black holes in general relativity, but this particular version is that most closely linked to the trajectories of particles, and therefore of most utility in a ray tracing study. 
Furthermore, this version of surface gravity  is closely linked to Hawking radiation when the degeneracy between definitions is broken (see~\cite{Barcelo:2010xk, Cropp:2013zxi}).


Strengthened by this brief review of the standard case, we can now consider the propagation of rays in the presence of the {\ae}ther.  
However, before doing so, we find it useful to perform a study of the slices of constant khronon field $\tau$ in our spacetime, as these are closely related to the universal horizon and also can be seen as the ray trajectories of physical particles in the limiting case of an infinite propagation speed (with respect to the aether).

\subsection{Rays of constant \aether\ time}
\label{sec:RaeT}

\begin{figure}[!h]
\centering
\includegraphics[scale=0.6]{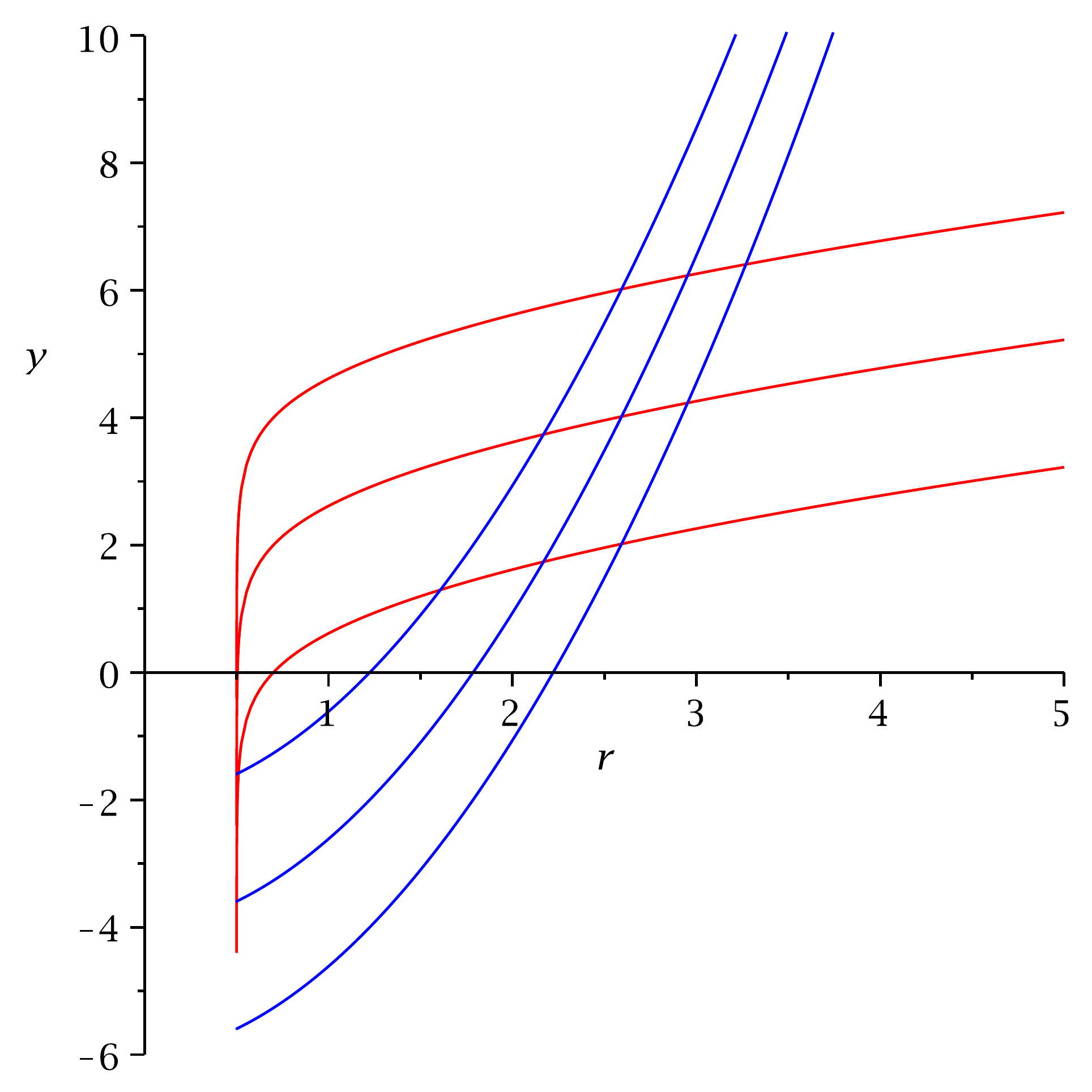}
\caption{Lines of constant $\tau$ (red) and $\sigma$ (blue) for the $c_{123}=0$ solution. The Killing horizon is at $r=1$ and the Universal horizon at $r=1/2$.}
\label{fig:tau-sigma}
\end{figure}

As noted, we have a clear notion of causality in these theories: Nothing can travel backwards in \aether\ time. 
The \aether\ time can be related to the metric one (what we might call the Killing time, given that we are dealing with static metrics) given that the \aether\ one-form is
\begin{equation}
u=u_v\d v+u_r\d r= u_v\left(\d v +\frac{u_r}{u_v} \,\d r\right)= u_v\d \left( v +\int\frac{u_r}{u_v}\,\d r \right).
\end{equation}
Now, using the fact that $u^a$ has a unit norm, it can verified that there exists a function $\tau $ such that $u=\d \tau/||\d \tau||$. The explicit form of $\tau$ is given by
\begin{equation}
\tau =v+\int \frac{u_r}{u_v} \;\d r.
\label{constantkhronon}
\end{equation}
Likewise, we can define a spatial $\sigma$ coordinate, corresponding to constant $s$ slices, $s=\d \sigma/||\d \sigma||$, and in a similar manner, find
\begin{equation}
\sigma =-v+\int \frac{s_r}{s_v} \;\d r.
\end{equation}
We plot the constant $\tau$,~$\sigma$ surfaces in Fig.~\ref{fig:tau-sigma} for the $c_{123}=0$ solution.
A peeling-like behavior at the Universal horizon is evident, and indeed a notion of $\kappa_\mathrm{peeling}$ can be associated to the constant $\tau$ slices.  
As realistic rays must travel forward in \aether\ time they must have paths as steep as, or steeper than, the constant $\tau$ slices. Thus the constant $\tau$ slices, corresponding to infinite velocity with respect to the \aether\, will provide a lower bound to the peeling properties (and so the peeling surface gravity) of physical rays. 

We can easily calculate the value of this surface gravity which, as noted, will be the relevant one for rays propagating with infinite group velocity $\rm v_g$. 
Generically any particle propagating in our spacetimes will have a four-velocity that can be given in the orthonormal frame provided by $u^a$ and $s^a$ as,
\begin{equation}
V^a=u^a+{\rm v}_g\, s^a . 
\label{eq:fourV}
\end{equation}
The trajectory for an instantaneously propagating ray would then be given by
\begin{equation}
\frac{\d v}{\d r} = \frac{V^v}{V^r}= \lim_{{\rm v}_g \to \infty} \,\frac{u^v+{\rm v}_g s^v}{u^r+{\rm v}_g s^r}= \frac{s^v}{s^r}.
\label{eq:dvdr}
\end{equation}
Then, Taylor expanding the trajectory close to the horizon one gets
\begin{align}
\left.\frac{\d r}{\d v}\right|_{\mathrm{out}}=\left.\frac{\d r}{\d v}\right|_{\UH}+\left.\frac{\d}{\d r} \frac{\d r}{\d v}\right|_{\UH}(r-r_{\UH}) +\mathcal{O}(r-r_{\UH})^2.
\end{align}
A straightforward calculation based on Eq.~\eqref{eq:s} shows that $({\d r}/{\d v})|_{\UH}=0$ for both the aforementioned solutions. (This happens because $s^r|_{UH}=0$, which can be checked by plugging in either of the explicit formulae given in Sec.~\ref{ae-bhs}). 
This is also what one should expect from the behavior in Fig.~\ref{fig:tau-sigma}. Consequently,
\begin{align}
\kappa_{\UH}\equiv \left.\frac{1}{2}\frac{\d}{\d r} \frac{\d r}{\d v}\right|_{\UH}
\label{eq:kappapeelUH}
\end{align}
where $\kappa_{\UH}$ is by definition the ``surface gravity" corresponding to the peeling-off property of the infinite velocity modes close to the horizon. 

%

\subsection{Modified dispersion relations}
\label{subsec:modDisRel}

As discussed in section \ref{dispersion} the theories we are considering naturally have dispersion relations. For radial motion one can then generically expect modified dispersion relations of the form
\begin{equation}
\omega^2=c^2\, \ks^2+\ell^2\, \ks^4+\ell^4\, \ks^6 + \dots
\label{eq:mod-disp}
\end{equation}
where $\omega\equiv-(k \cdot u)$ is the energy in the \aether\ frame (note that $\omega > 0$ for propagating particles) and (for radial motion) $\ks\equiv (k \cdot s)$ is the spatial component of $k_a$ orthogonal to the \aether\ field. 
The length scale $\ell$ here is the UV Lorentz-violating scale for matter which we do not need to necessarily identify with the Planck scale. Letting $\ell\to0$ one would recover standard Lorentz-invariant dispersion relations.%

Note that we are considering only UV modifications of the matter dispersion relations, while of course \emph{a priori} one might also expect radiative corrections to produce IR modifications (for e.g., inducing particle-dependent coefficients of the order $\ks^2$ terms), see for e.g. Ref.~\cite{Collins:2004bp}.
We do not explicitly consider these potential complications as they are highly constrained phenomenologically,  and furthermore there are known mechanisms that suppress such terms (see, for e.g., Ref.~\cite{Liberati:2013xla} and references therein). 
Also, in general a modified speed of light will not effect our results except for the fact that different particles will perceive different Killing horizons (placed at different radii). 

Hence, we shall only consider $\omega^2=c^2 \,\ks^2+\ell^2 \,\ks^4$, which we shall further brutally simplify by looking at the expansion for low $k$
\begin{equation}
\label{eq:cubicDisRel}
 \omega=c \,\ks+\frac{1}{2c}\ell^2\, \ks^3.
\end{equation}
We do this mainly for simplicity, (in particular, by solving the cubic, $\ks(\omega)$ can be written down in a closed form of reasonable length).  Henceforth we will absorb the factor of $2c$ into the suppression factor.  Later on we shall consider the possibility of more general dispersion relations and show that our main results are independent of the specific form of the dispersion relation. 

\subsection{A notion of conserved energy}
\label{subsec:conservation law}

 The modified dispersion relation gives us \emph{one} equation between $\omega$ and  $\ks$. In order to find their explicit values on spacetime we need a \emph{second} equation relating them. This equation is a conservation equation, which says that $(k \cdot \chi)$ is constant on the whole spacetime. We now give two arguments whic are valid for general modified dispersion relations. \par
For the first derivation, start by, defining differential operators acting on a generic quantity $X$ by setting
\begin{equation}
\nabla_1 X = u^a \partial_a X, \qquad\hbox{and} \qquad \nabla_2 X = \sqrt{\nabla_a[(g^{ab}+u^au^b)\nabla_b X]}. 
\end{equation}
These are respectively temporal derivatives in the direction of the \aether, and spatial derivatives on constant \aether-time hypersurfaces.  
Then to a given dispersion relation, $\omega=f(\ks)$, we can naturally associate the differential operator
\begin{equation}
D(x;\nabla) =   -(\nabla_1)^\dagger (\nabla_1)+ [f(-i\nabla_2)]^\dagger [f(-i\nabla_2)]\, .
\end{equation}
We are working with a static spacetime, and therefore the Killing equation implies that in terms of the Killing time coordinate
\begin{equation}
[-i\partial_t, D(x,\nabla) ] = 0.
\end{equation}
Both of these are Hermitian operators, so the vanishing of their commutator implies simultaneous diagonalizability.
This implies that $-i\partial_t$ is explicitly diagonalizable, so one may write
\begin{equation}
\Phi(t,r) = e^{i\Omega t}\; \Phi(r),
\end{equation}
with $\Omega$ a \emph{position independent} constant.

Now pick a particular tetrad based on completing the zwei-bein $(u^a,\,s^a)$. Using spherical symmetry one can decompose the 4-momentum as
\begin{equation}
k_a = e^A{}_a k_A = f(\ks) \;u_a + \ks \;s_a\, ,
\end{equation}
from which we can read
\begin{equation}
- \Omega = k_t =  f(\ks) \;u_t + \ks \; s_t, 
\end{equation}
which implicitly defines $\ks(r)$ via
\begin{equation}
-\Omega =  f(\ks(r)) \;u_t(r) + \ks(r) \;s_t(r).
\end{equation}
In particular, we now have the statement
\begin{equation}
\nabla_a \left(k_b\, \chi^b\right) = \nabla_a \Omega = 0.
\end{equation}
The fact that we have this position-independent constant means we can solve for $\ks(r)$ and $\omega(r)$. Then we can integrate the group velocity to obtain the ray trajectory of the particle explicitly. 
Physically, $\Omega$ is the Killing energy at infinity, where also $\Omega=\omega$. That is, when we talk about high (low) $\Omega$ rays, we mean rays that arrive at infinity with high (low) Killing energy. 

An alternative derivation of this result uses that, in the geometric-optics approximation, the dynamics of the wavepacket is described as a point particle and is governed by some Hamiltonian,
\begin{align}
\label{eq:hamiltonian}
\mathcal{H}=\frac{1}{2}\mathcal{G}^{\alpha \beta}(x,k)\, {k}_{\alpha} {k}_{\beta}
\end{align}
where $k_\mu$ are the generalized momenta conjugate to the position coordinates $x^\mu$ of the particle (the centre of the wave-packet), and $\mathcal{G}^{\alpha \beta}$ is a metric on the phase space, which is generally a function of both position and momentum.
 Since the underlying field theory is diffeomorphism invariant, the $\mathcal{H}$ that descends from it is independent of the parameter along the trajectory of the particle. This implies that $\mathcal{H}$ is a constant on shell. Its value is the squared mass, which we take to be zero. Then Eq.~\eqref{eq:hamiltonian} is nothing but the dispersion relation \cite{Misner:1974qy, Girelli:2006fw}.

Now let the coefficients of the spacetime metric $g_{ab}$ in the Eddington-Finkelstein $\{v,r\}$ coordinates be independent of $v$, i.e., the coordinate vector field $\chi^a := \left(\frac{\partial}{\partial v}\right)^a$ is a Killing vector field. If the underlying dynamical field is Lie-dragged by $\chi$, the corresponding $\mathcal{H}$ has to respect this symmetry too, hence there is no explicit $v$-dependence in $\mathcal{H}$. 

Now, Hamilton's equation 
\begin{equation}
 \frac{\textstyle \rm{d}k_a}{\textstyle \rm{d}\lambda} = -\frac{\textstyle \partial \mathcal{H}}{\textstyle \partial x^a}
\end{equation}
then implies that $k_v$, the momentum conjugate to $v$, is a constant, i.e., $k_v := k_a \chi^a = - \Omega$, where $\Omega$ is a constant.

Suppose now that the dispersion relation is given in the {\ae}ther frame as $\omega = f(k_s)$, where $\omega=-k_a u^a$ and $k_s = k_a s^a$. The Hamiltonian is then $\mathcal{H}=\frac{1}{2}\left(\omega^2 - f(k_s)^2\right)$. Hamilton's equation for the evolution of the position is 
\begin{align}
\frac{\mathrm{d} x^a}{\mathrm{d} \lambda} &= \frac{\partial \mathrm{H}}{\partial k_a} \nonumber \\
 &= -\omega u^a - f(k_s) f'(k_s) s^a,
\end{align}
where $'$ denotes the derivative w.r.t the argument. Now noting that  $|f'(k_s)|$ is just the group velocity $\rm{v}_g$ we get, after dividing $\frac{\mathrm{d}r}{\mathrm{d}\lambda}$ by $\frac{d v}{\mathrm{d}\lambda}$, the equation describing the trajectory of the particle,
\begin{align}
\frac{\mathrm{d} r}{\mathrm{d} v} = \frac{ u^r \pm \mathrm{v}_g s^r}{ u^v \pm \mathrm{v}_g s^v},
\end{align}
 where $+(-)$ sign is for the outgoing(ingoing) particle.

\subsection{Physical ray trajectories}
\label{subsec:trajectories}
We now are set to solve for the trajectory, with our cubic dispersion relation in Eq.~\eqref{eq:cubicDisRel}
\begin{equation}
\omega=\ks+\ell^2\ks^3,
\end{equation}
and the conservation equation,
\begin{equation}
\label{eq:conservation eqn}
-\Omega =  \omega \,(\chi \cdot u) \pm \ks\, (\chi \cdot s).
\end{equation}
The $\pm$ refers to whether the mode is outgoing or ingoing, respectively. 
In what follows we will focus on the outgoing modes.
As $\Omega$ is the energy in the \aether\ frame at spatial infinity,  we require that $\Omega$ be non-negative. \par
What happens to $\ks$ and $\omega$ near the Universal horizon? 
On the Universal horizon we have $\chi \cdot u = 0$ and $\chi \cdot s = \|\chi \|$. We might naively but incorrectly argue from Eq.~\eqref{eq:conservation eqn} that $ - \Omega = \ks \|\chi \|$, but this is inconsistent because the RHS is positive while the LHS is negative. 
This just emphasizes the fact that something singular is happening on the Universal horizon. Let us then seek to better understand the divergence structure of $\ks$ on the Universal horizon.

Note  that $u$ is everywhere timelike while $\chi$ is timelike outside, null on, and spacelike inside the Killing horizon. 
Also from the fact that $(\chi\cdot u)=-1$ at infinity and becomes zero only at the Universal horizon, we deduce that this product is  negative everywhere outside the Universal horizon, we hence write it as $-|\chi \cdot u|$. 
For the sign of $\chi \cdot s$ one has to specify a choice of the $s$ basis (inward or outward pointing). We choose $s$ in Eq.~\eqref{eq:s} so that  $\chi \cdot s$  is  positive everywhere  outside the Universal horizon. 
We hence write it as $| \chi \cdot s |$. So we may rewrite Eq.~\eqref{eq:conservation eqn} as
 \begin{equation}
\label{eq:conservation eqn2}
 \Omega = (\ks+ \ell^2 \ks^3)\, |\chi \cdot u | - \ks\, |\chi \cdot s|.
 \end{equation}
Let us parametrize the singular behavior of $\ks$ at the Universal horizon as
\begin{equation}
 \ks \sim \frac{a(r)}{| \chi \cdot u |^\gamma},
 \end{equation}
where $a(r)$ is regular at the Universal horizon and $\gamma$ is a constant to be determined. 
Substituting this ansatz in Eq.~\eqref{eq:conservation eqn2}, finiteness of the LHS implies 
\begin{equation}
\gamma=\frac{1}{2}\,; \qquad \hbox{and} \qquad a(r_{\UH})=\dfrac{\sqrt{ |\chi \cdot s|}_{\UH}}{\ell}.
\end{equation}
Therefore, close to the Universal horizon $\ks \sim 1/\sqrt{ |\chi \cdot u|}$.

\begin{figure}[!htb]
\centering
\includegraphics[scale=1.2]{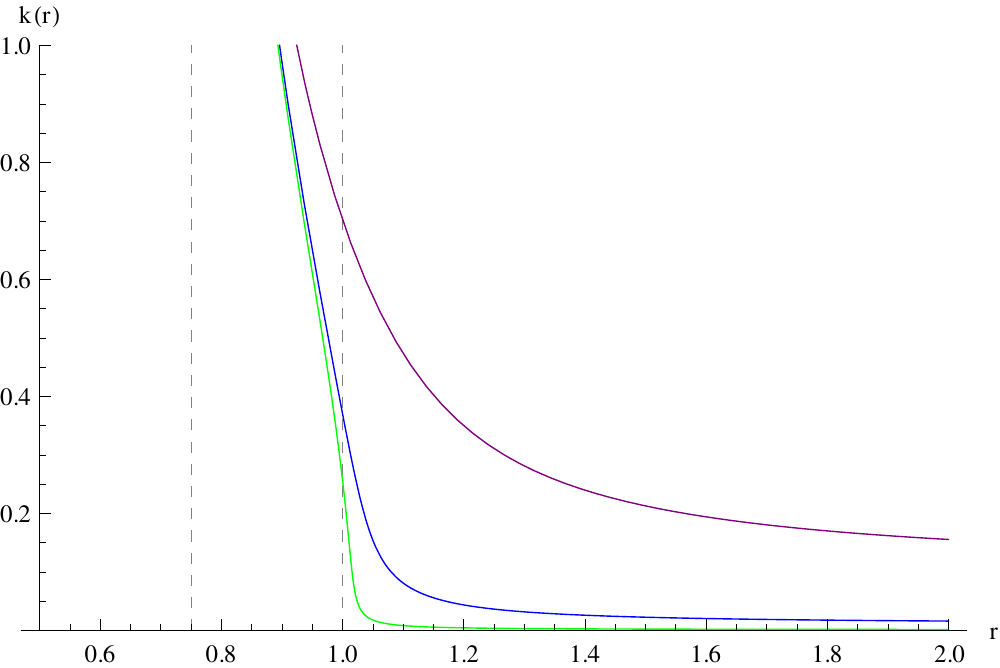}
\includegraphics[scale=1.2]{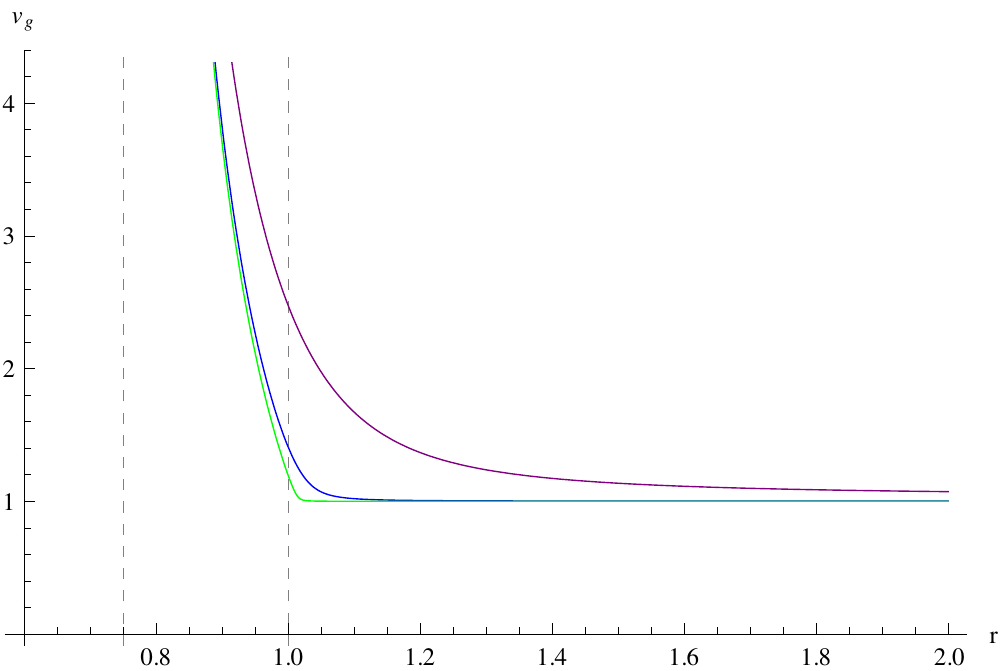}
\caption{The wavenumber $k$ (left) and group velocity (right) as functions of $r$  for the $c_{14}=0$ solution, for energies $\Omega = 10^{-1}$ (purple), $\Omega = 10^{-2}$ (blue) and $\Omega = 10^{-3}$ (green). The Lorentz-violating scale is fixed at $\ell=1$. The Killing horizon is at $r=1$, and the Universal horizon is at $r=0.75$. }
\label{fig:k}
\end{figure}

Let us now move away from the Universal horizon and consider the trajectory throughout the whole spacetime. To do this we shall basically follow the procedure of section \ref{sec:RaeT}, with the difference that the group velocity $\partial \omega/\partial \ks$ is no longer infinite, but is now a function of $r$ which diverges at the Universal horizon. 
From our dispersion relation in Eq.~\ref{eq:cubicDisRel} we find the group velocity to write the four velocity of the particle as in Eq.~\eqref{eq:fourV}, $V^a = u^a + {\rm{v}}_g \; s^a$.  
In the Eddington-Finkelstein coordinate system denoted by $\{v,r\}$, the trajectory of the particle can again be obtained by solving the differential equation
\begin{equation}
\label{eq:trajectory}
\frac{\d v}{\d r}=\frac{V^v}{V^r}=\frac{u^v (r)+ {\rm v}_g(\ks)\; s^v(r)}{u^r (r)+ {\rm v}_g(\ks) \; s^r(r)}.
\end{equation}
When one aims at plotting the above trajectories in spacetime it is clear that the only difficulty arises from the $\ks$-dependence of ${\rm v}_g=1+3 \ell^2\, \ks^2$, as the 3-momentum $\ks$ is \emph{not} a conserved quantity along the path. 
This is where the conservation equation, Eq.~\eqref{eq:conservation eqn2}, comes in as it allows us to solve for $\ks$ as a function of $r$ and $\omega$,
\begin{eqnarray}
\ks(r)&=&\frac{\left[(12)^{\frac{1}{2}}\left( 9\Omega\ell +\sqrt{\frac{\textstyle 12(\chi\cdot s-\chi\cdot u)^3}{\textstyle(\chi\cdot u)}+(9\Omega\ell)^2}\right)(\chi\cdot u)^2\right]^{\frac{1}{3}}}{6(\chi\cdot u) \ell} \\
&+&\frac{(12)^{\frac{2}{3}}(\chi\cdot u)(\chi\cdot u-\chi\cdot s)}{6(\chi\cdot u) \ell \left[ \left( 9\Omega\ell +\sqrt{\frac{\textstyle 12(\chi\cdot s-\chi\cdot u)^3}{\textstyle (\chi\cdot u)}+ (9\Omega\ell)^2}\right)(\chi\cdot u)^2\right]^{1/3}} \nonumber
\end{eqnarray}
where the functions $\chi\cdot u$ and $\chi \cdot s$ can be read off for either of the two particular solutions given in section \ref{ae-bhs}. 
(The particular form of $\ks(r)$ given above of course depends crucially on the assumed cubic dispersion relation. For more general dispersion relations, while $\ks(r)$ certainly exists, it may be difficult to exhibit an explicit formula.)

\begin{figure}[!htb]
\centering
\includegraphics[scale=1.1]{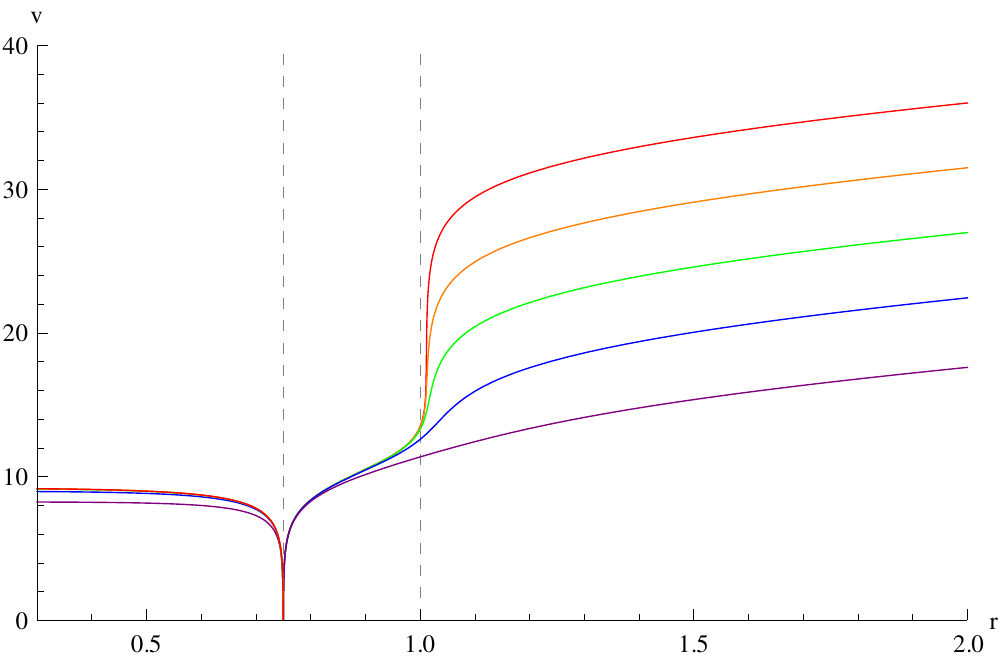}
\includegraphics[scale=1.1]{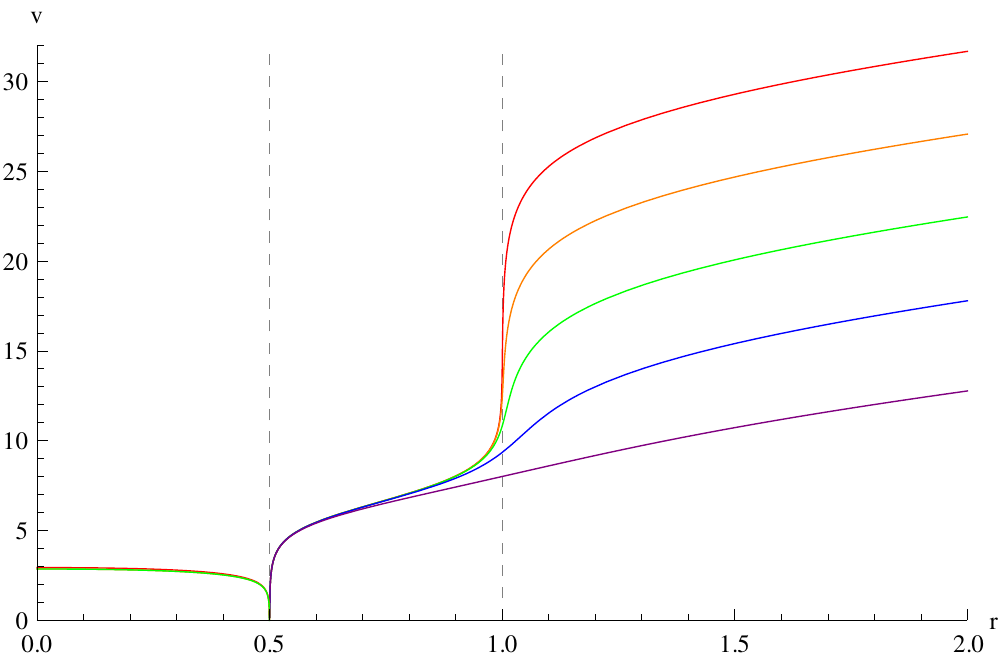}
\caption{Trajectories of the outgoing particle in $v$--$r$ Eddington--Finkelstein coordinates. \newline Energies of $\Omega = 0.1$ (purple), $\Omega=10^{-2}$ (blue), $\Omega=10^{-3}$ (green), $\Omega=10^{-4}$ (orange) and $\Omega=10^{-5}$ (red). For these parameters of the black hole the  $c_{123}=0$ solution (left) has Universal horizon at $r_{\UH}=0.75$, while for the $c_{14}=0$ solution (right)  the Universal horizon is at $r_{\UH}=0.5$. For both situations $r_{\KH}=1$. Behaviour at the Universal horizon is universal while behaviour at the Killing horizon at $r_{\KH}=1$ depends on energy.}
\label{fig:lingering}
\end{figure}

We can now numerically integrate Eq.~\eqref{eq:trajectory} and plot  trajectories for different values of the conserved energy $\Omega$.   The result is shown in Fig.~\ref{fig:lingering}. We see that the particles which arrive at infinity with a high energy (i.e., the trajectories of high $\Omega$) hardly feel the presence of the Killing horizon. The particles  which arrive at infinity with a low energy (i.e., the trajectories for low $\Omega$) instead feel the presence of the Killing horizon acutely: they hover close to it for a considerable interval before escaping to infinity.

\section{Near-horizon physics}
In this section we study the behavior of rays close to the Universal and Killing horizons. In Sec.~\ref{subsec:nearUH} the peel-off behavior of all rays, low or high $\Omega$, at the Universal horizon will let us define the surface gravity of the Universal horizon. 
In Sec.~\ref{subsec:nearKH} we attempt to define a surface gravity for the low-$\Omega$ rays close to the Killing horizon. 
In Sec.~\ref{subsec:lingeringKH} we attempt to quantify the extent to which the Killing horizon might reprocess the information in the low-$\Omega$ rays.

\subsection{Near the Universal horizon}
\label{subsec:nearUH}
As seen in Fig.~\ref{fig:lingering}, there is a peeling behavior at the Universal horizon, therefore it is possible to associate with this behavior a notion of surface gravity. Surface gravity for Universal horizons has previously been considered in the literature   (see for example Refs.~\cite{Berglund:2012bu, Cropp:2013zxi}). 
Such derivations are non-trivial, given that the Universal horizon is qualitatively different from the Killing horizon, for which several alternative definitions of surface gravity all agree with each other (see the previous chapter and Ref.~\cite{Cropp:2013zxi} for an exhaustive discussion). 
Let us stress that Refs.~\cite{Berglund:2012bu} and~\cite{Cropp:2013zxi} and chapter \ref{surfgrav} use metric quantities and specific symmetries of spacetime to define the surface gravity of Universal horizons, and are in agreement with each other. 

However, in Ref.\cite{Berglund:2012fk}, the tunneling method~\cite{Parikh:1999mf} was used to claim that the Universal horizon should Hawking radiate at a particular temperature $T_{\UH}$. 
Given this temperature, a surface gravity can be associated to it in the usual way, $\kappa_{\rm thermal}={2\pi \, T}$. The surface gravity  thus defined, however, does not agree with the one derived in Refs.~\cite{Berglund:2012bu, Cropp:2013zxi} and the previous chapter. 
As a consequence of this mismatch it was claimed in Ref.~\cite{Berglund:2012fk} that for Universal horizons the usual relation between surface gravity and temperature might break down.

Given this state of affairs, it is interesting to calculate the surface gravity in our framework, as the relevant quantity governing the observed peeling at the Universal horizon.
As we already understand the divergence at the Universal horizon, the calculation follows easily. We know that ${\rm v}_g \to \infty$ at the Universal horizon. 
This can be easily deduced from the fact that $\ks$ diverges there and that ${\rm v}_g=1+3\ell^2 \ks^2$ (see also Fig.~\ref{fig:k}).
Then from Eq.~\eqref{eq:dvdr} one can easily see that at the Universal horizon one gets
\begin{equation}
\frac{\d v}{\d r} = \frac{s^v}{s^r}.
\end{equation}
Thus the near-horizon peeling is the same as that calculated for the constant-$\tau$ slices at the Universal horizon, and it is independent of the specific form of dispersion relation (as long as the dispersion relation is superluminal with unbounded group velocity, i.e., of the form as in Eq.~\eqref{eq:mod-disp} possibly with additional parity-odd terms). 
This can be confirmed by explicit computation (see appendix~\ref{gendispersion}). The surface gravity \eqref{eq:kappapeelUH} for the two exact solutions at hand is given by
\begin{equation}
\kappa_{c_{14}=0}=\frac{1}{2 r_{\UH}}\,\sqrt{\frac{2}{3(1-c_{13})}},
\label{eq:kappaUH1}
\end{equation}
and
\begin{equation}
\kappa_{c_{123}=0}=\frac{1}{2 r_{\UH}} \,\sqrt{\frac{2-c_{14}}{2(1-c_{13})}},
\label{eq:kappaUH2}
\end{equation}
respectively.

One can now easily check that the temperature calculated in Ref.~\cite{Berglund:2012fk}, and the above surface gravity, are indeed related in the standard way. 
While at the leading order we use the same dispersion relation as in Ref.~\cite{Berglund:2012fk}, we stress that our analysis shows that the peeling surface gravity of the Universal horizon is indeed universal, i.e., independent of the specific form of the superluminal dispersion relation. 
(An explicit demonstration is provided in Appendix~\ref{gendispersion}). 
This strongly suggests that it should be possible to carry out the tunneling calculation of Ref.~\cite{Berglund:2012fk} for general dispersion relations, and that the resulting temperature should be universal.

One might wonder why our analysis agrees with the temperature provided by the tunneling method, but does not agree with the surface gravity previously calculated in Ref.~\cite{Berglund:2012bu, Cropp:2013zxi} and chapter \ref{surfgrav}. 
Again the gist of the problem is that the methods of those references do not capture the role of \aether, and hence do not take into account the non-relativistic nature of the particle dynamics. On the contrary, the calculations presented here rely on \aether\ in an essential way. 
Given that the Universal horizon is an \aether-dependent object, it is our  ``\aether\ sensitive" surface gravity which ends up being related to the temperature of the Universal horizon and has the physical meaning related to the peel-off behavior of ray trajectories.

In connection to this last comment one final remark is due. While it can be shown that the metric notion of surface gravity associated to geodesic peeling is equivalent to self-evidently covariant definitions~\cite{Cropp:2013zxi}, one might wonder if such alternative definitions are available for the peeling surface gravity associated to physical rays as discussed here. A natural candidate for a covariant definition of the surface gravity, which should match the peeling one, is the so called $\kappa_{\rm normal}$, which is equal (modulo a sign) to the normal derivative to the horizon of the redshift factor. For the case of the Universal horizon this simply takes the form
\begin{equation}
u^a\nabla_a (\chi^2)=-2\kappa^{\rm metric}_{\rm normal}
\end{equation}
and can be shown to be equal to the peeling surface gravity for geodesic rays~\cite{Cropp:2013zxi}. However, such a definition obviously does not capture the role of the {\ae}ther in the propagation of the physical rays. We will argue in the next chapter that a natural generalization of the redshift factor $\chi^2$ to our framework is given by $u\cdot\chi$ which is constant (actually zero) on the Universal horizon. Then, following the standard logical steps, one can recognize that the natural generalization of the above formula is $u^a\nabla_a (u\cdot\chi)=2\kappa_{\rm normal}$ evaluated on the Universal horizon. A straightforward calculation shows that using this definition yields the same values as in Eq.~\eqref{eq:kappaUH1}, ~\eqref{eq:kappaUH2}. In fact, using the general form of the metric coefficients for spherically symmetric solutions given in Sec.~\ref{ae-bhs} one can show that $\kappa_{\rm normal}$ as defined by $u^a\nabla_a (u\cdot\chi)|_{UH}=2\kappa_{\rm normal}$ always equals the peeling-off surface gravity $\kappa_{UH}$ as defined by Eq.~\eqref{eq:kappapeelUH}. We therefore have a covariant expression for the surface gravity of the Universal horizon as defined by the peeling-off behavior of rays,
\begin{align}
\label{eq:covkappaUH}
\kappa_{UH}=\frac{1}{2} u^a\nabla_a (u\cdot\chi)\biggr\rvert_{UH}.
\end{align}
Using the Killing equation this can also be written as 
\begin{equation}
\kappa_{UH}=\dfrac{1}{2}\chi \cdot a_u \biggr\rvert_{UH}
\end{equation}
where $a_u$ is the acceleration of the {\ae}ther, $a^b=u^c \nabla_c u^b$.

\subsubsection{More General Dispersion Relations}
\label{gendispersion}

We have picked a superficially rather strange dispersion relation for our explicit numerical calculations. (It is certainly valid for low energies, simply being the expansion of $\omega^2=\ks^2+\ell^2 \ks^4$, where $\ks^4$ is generally expected to be the most relevant term at low energies). 
But we showed earlier that $\ks$ blows up near the Universal horizon, so how many of our conclusions can be carried over for more general dispersion relations?
Let us first look at the behavior near the Universal horizon. We still have the conservation equation
\begin{equation}
\label{appenomega}
\Omega =  \omega(\ks) |(\chi \cdot u)| - \ks |(\chi \cdot s)|.
\end{equation}
Again, on the Universal horizon we have $(\chi \cdot u) = 0$ and $(\chi \cdot s) = \|\chi \|$, we would naively but incorrectly get that
$ -\Omega = \ks \|\chi \|$, 
which is inconsistent because the RHS is positive while the LHS is negative.  This argument holds just as well for a very large and relevant class of superluminal dispersion relations.
Let us now write 
\begin{equation}
\label{knearuh}
\ks \sim \frac{a(r)}{|(\chi\cdot u)|^\gamma},
\end{equation}
near the Universal horizon, with $a(r)$ regular on the Universal horizon. 

Now (temporarily) assume the dispersion relation has the form
\begin{equation}
\label{gendisp}
f(\ks)=\sum_{n=1}^{n=N} b_n \ks^{n}.
\end{equation}
Substituting this ansatz in the conservation equation, the finiteness of $\Omega$ implies that 
%
\begin{equation}
N\gamma-1=\gamma
\end{equation}
so
\begin{equation}
\gamma=\frac{1}{N-1};
\end{equation}
and
\begin{equation}
\label{auh}
 a(r_\UH)^{N-1} = \frac{(\chi \cdot s)_\UH}{b_N}.
\end{equation}
How steeply $k_s$ diverges near the Universal horizon varies, but for superluminal dispersion relations of this form, it always will diverge. This would lead to a divergent group velocity at the Universal horizon. 

In fact, we can generalize this argument even further. As the phase velocity is $v_{\mathrm{phase}}=\omega/k_s$, we can rearrange Eq.~\eqref{appenomega} as 
\begin{equation}
\Omega =  k_s [ v_{\mathrm{phase}} |\chi \cdot u| - |\chi \cdot s| ].
\end{equation}
If we assume the existence of outgoing modes at the universal horizon (meaning that this, rather than some Killing horizon, is the casual barrier), we necessarily have
\begin{equation}
v_{\mathrm{phase}} >  \frac{|\chi \cdot s|}{ |\chi \cdot u|},
\end{equation}
implying the phase velocity diverges at the Universal horizon. 
Now, using L'Hopital's rule, 
\begin{equation}
\lim_{k \to\infty} \frac{\omega(k_s)}{k_s} = \lim_{k \to \infty} \frac{[\d \omega(k_s)/\d k_s]}{[\d k_s/\d k_s]}=\lim_{\to \infty} [\d \omega(k_s)/\d k_s] = \infty.
\end{equation}
So, (assuming sufficient smoothness in $\omega(k_s)$), the group velocity also diverges.

Note other forms of superluminal dispersion exist, for instance extrapolating between two distinct limiting velocities. As Einstein-\Aether\ and \Horava-Lifshitz fundamentally violate Lorentz invariance, we do not expect to move between two Lorentzian regimes in this way, so dispersion relations of the form in Eq.~\eqref{gendisp} are the most relevant (also note for such rays the Universal horizon will not be the casual barrier).
Finally, using Eqs.~\eqref{eq:dvdr} and \eqref{eq:kappapeelUH}, we get a surface gravity which is universal and thus associates the universal temperature with the Universal horizon. 

In \cite{Cropp:2013zxi} it was argued that equations \ref{eq:kappaUH1} and \ref{eq:kappaUH2} should therefore hold for all dispersion relations of this form: however, there is an extra subtlety that went unnoticed. 

Previously we argued that as $v_g \to \infty$ at the universal horizon, we can calculate this as
\begin{equation}
2\kappa =\frac{d}{dr} \left[\frac{s^r}{s^v}\right]|_{UH}
\end{equation}
However, at the universal horizon $s^r$ is also zero (see appendix \ref{aetherelations}), so $v_g s^r$ can be regular, and we cannot first take the limit of infinite velocity then close to universal horizon approximation.

This is easier to see if rewrite everything in terms of $(u\cdot \chi)$ and $(s \cdot \chi)$. A quick check shows
\begin{equation}
 u^r = -(s\cdot \chi) ; \qquad s^r = -(u\cdot \chi)
\end{equation}
\begin{equation}
 u^v=s^v=\frac{1}{(s\cdot \chi)-(u\cdot \chi)}
\end{equation}
which are correct everywhere, so that
\begin{equation}
\frac{u^v+v_gs^v}{u^r+v_g s^r}= \frac{-[1+v_g]}{[(s\cdot\chi)-(u\cdot\chi)][(s\cdot \chi)+(u\cdot\chi)v_g]}.
\end{equation}
Now, pick a dispersion relation. Near the UH, the highest power of $k$ is most relevant, so just work with
\begin{equation}
 \omega =b k^N
\end{equation}
so that
\begin{equation}
 v_g \equiv \frac{d\omega}{d k_s}= Nb k_s^{N-1}
\end{equation}
Further, using equations \ref{knearuh} and \ref{auh}
\begin{equation}
 v_g \approx Nb \left(\frac{(s\cdot \chi)}{ b (|u\cdot \chi|)}\right)^{(N-1)/(N-1)}= N\frac{(s\cdot \chi)}{-(u\cdot \chi)}.
\end{equation}
Plugging this into our formula for the trajectory and simplifying
\begin{eqnarray}
\frac{u^v+v_gs^v}{u^r+v_g s^r} &\approx& \frac{-[1-N\frac{(s\cdot \chi)}{(u\cdot \chi)}]}{[(s\cdot\chi)-(u\cdot\chi)][(s\cdot \chi)-(u\cdot\chi)N\frac{(s\cdot \chi)}{(u\cdot \chi)}]}\nonumber \\
&=& \frac{-[1-N\frac{(s\cdot \chi)}{(u\cdot \chi)}]}{[(s\cdot\chi)-(u\cdot\chi)](s\cdot \chi)(1-N)} \nonumber \\
&\approx&  \frac{N}{1-N}\frac{1}{(s\cdot \chi)(u\cdot \chi)}.
\end{eqnarray}

So that
\begin{equation}
 2\kappa= \left(\frac{N-1}{N}\right)\frac{d}{dr}[(s\cdot \chi)(u\cdot \chi)]_{UH}
\end{equation}
Compare this to what we had before:
\begin{equation}
\frac{u^v+v_gs^v}{u^r+v_g s^r} \approx \frac{s^v}{s^r} = -(u\cdot \chi)[(s\cdot \chi)-(u\cdot\chi)] \approx -(u\cdot \chi)(s\cdot \chi)
\end{equation}
where the approximation is near the universal horizon, and therefore
\begin{equation}
 2\kappa_{old}= -\frac{d}{dr}[(s\cdot \chi)(u\cdot \chi)]_{UH}.
\end{equation}
Differing by a factor of $(\frac{N-1}{N})$ from the correct surface gravity.

This means that \ref{eq:kappaUH1} and \ref{eq:kappaUH2} are only correct in the case $N \to \infty$. It can be argued however, that this is the most natural situation.
Some thought will show that if we have species $A$ and $B$ which both have polynomial dispersion relations of order $N_A$ and $N_B$ we could revive the violation of the GSL using Hawking radiation at the universal horizon. However, as both species have to interact gravitationally, it is difficult to construct a protection mechanism to stop terms in the dispersion relation of $A$ being generated in the dispersion relation of $B$ by loop corrections, such that the highest term is the same for both.

\subsection{Near the Killing horizon}
\label{subsec:nearKH}
In this section, we will work with the particular solution corresponding to $c_{123}=0$. As for the other solution, calculations become unpleasantly long and do not give any additional insight. 

At the Killing horizon, ($r=r_0+r_u$), the standard metric-determined surface gravity, (which can be found by any of the standard methods), is:
\begin{equation}
\kappa_{\KH, \mathrm{metric}}=\frac{r_0+2r_u}{4(r_0+r_u)^2}.
\end{equation}
Now at the Killing horizon, for the cubic dispersion relation
\begin{equation}
\ks|_{\KH}=\frac{[2(r_0+r_u)]^{1/3}\Omega^{1/3}}{(r_0+2r_u)^{1/3}\ell^{2/3}},
\end{equation}
\begin{equation}
\omega|_{\KH}=\frac{[2(r_0+r_u)]^{1/3}\Omega^{1/3}}{(r_0+2r_u)^{1/3}\ell^{2/3}}+\left[\frac{[2(r_0+r_u)]}{(r_0+2r_u)}\right]\Omega.
\label{eq:omegaOmega}
\end{equation}
While the group velocity is
\begin{eqnarray}
{\rm v}_g|_{\KH}&=&1+\frac{3\ell^{2/3}\Omega^{2/3}(2(r_0+r_u)^{2/3})}{(r_0+2r_u)^2/3}\nonumber \\
&=&1+3\ell^{2/3}\Omega^{2/3}\left[1+\left(\frac{r_0}{r_0+2r_u}\right)^{2/3}\right],
\end{eqnarray}
which is still close to $1$ for small values of $\ell \Omega$, so for a dispersion relation only modified at high energies we are still almost relativistic until inside the Killing horizon. 

Can we define an approximate {\ae}ther-sensitive notion of surface gravity? We can certainly expand the ray trajectory in a Taylor series around the Killing horizon, trying to follow the standard peeling surface gravity calculation as given in Sec.~\ref{sec:standardpeeling}.
However, as these rays have a non-relativistic dispersion relation the first term in this expansion \eqref{eq:standarddrdv} is no longer zero, and there is a net non-zero Killing-horizon-crossing velocity:
\begin{equation}
v_{\KH} = \left.\frac{\d r}{\d v}\right|_{\KH}.
\end{equation}
For the specific cubic dispersion relation considered above we have
\begin{equation}
\left.\frac{\d r}{\d v}\right|_{\KH}=\frac{3}{2}\,\frac{(r_0+2r_u)^2(2\ell\Omega)^{2/3}}{2(r_0+r_u)^{4/3}\left[2(r_0+2r_u)\right]^{2/3}+3(2\ell\Omega)^{2/3}(r_0+r_u)^{2/3}}.
\end{equation}
This is an outward-pointing, radial group velocity of the particle at the Killing horizon. Note that it has the correct relativistic limit, i.e., it vanishes as $\ell\to 0$. (The particular form above depends very much on the assumed cubic dispersion relation, but the fact that $v_{\KH} = ({\d r}/{\d v})|_{\KH}\neq 0$ is generic.)

In analogy with the usual treatment we can still use the second term of the Taylor expansion for defining a peeling surface gravity of the Killing horizon. The trick is to note that near the Killing horizon
\begin{align}
\left.\frac{\d r}{\d v}\right|_{\mathrm{out}}=\left.\frac{\d r}{\d v}\right|_{\KH}
+\left.\frac{\d}{\d r} \frac{\d r}{\d v}\right|_{\KH}(r-r_{\KH}) +\mathcal{O}(r-r_{\KH})^2.
\end{align}
Then if we compare two nearby trajectories $r_1(v)$ and $r_2(v)$, we see
\begin{align}
\left.\frac{\d (r_1-r_2)}{\d v}\right|_{\mathrm{out}}=\left.\frac{\d}{\d r} \frac{\d r}{\d v}\right|_{\KH}(r_1-r_2) 
+\mathcal{O}[(r_1-r_{\KH})^2, (r_2-r_{\KH})^2].
\end{align}
So this \emph{difference} certainly exhibits exponential peeling near the Killing horizon, with
\begin{equation}
\kappa_{\KH} = \frac{1}{2}\left.\frac{\d}{\d r} \frac{\d r}{\d v}\right|_{\KH}.
\end{equation}
For the specific cubic dispersion relation considered above we have
\begin{equation}
\kappa_{\KH}= \frac{r_0+2r_u}{4(r_0+r_u)^2}-{\frac{\left[2(r_0+2r_u)\right]^{1/3} \left(5r_0+3r_u\right){(\ell \Omega)^{2/3}}}{4\left(r_0+r_u \right)^{7/3}}} +\mathcal{O}(\Omega^{4/3}).
\label{eq:killingkappa}
\end{equation}
This is the closest notion we can construct to the usual metric surface gravity $\kappa_{\KH, \mathrm{metric}}$ at the Killing horizon once we are in the presence of modified dispersion relations.  
We see that it shows an $\Omega$ dependent correction to $\kappa_{\KH, \mathrm{metric}}$. For  $\ell \to 0$, this value agrees with the standard $\kappa_{\KH, \mathrm{metric}}$.  (Again, the particular form above depends very much on the assumed cubic dispersion relation, but the general features are generic.)


We can also consider a quantitative comparison of the magnitude of the surface gravity at the Universal and Killing horizons. In particular, the ratio of Universal horizon surface gravity (see Eq.~\eqref{eq:kappaUH2} for the $c_{123}=0$ solution), and Killing horizon surface gravity (Eq.~\eqref{eq:killingkappa}), is 
\begin{equation}
\frac{\kappa_{\UH}}{\kappa_{\KH}}=\frac{2(r_0+r_u)^2}{r_0^2} +\mathcal{O}(\Omega^{2/3}).  
\end{equation}
Hence, the surface gravity of the Universal horizon is higher than that of the Killing horizon. 

\subsection{Lingering near the Killing horizon}
\label{subsec:lingeringKH}


From the ray trajectories plotted in Fig.~\ref{fig:lingering} it is apparent that the low-$\Omega$ rays are strongly affected by the presence of the Killing horizon. 
These rays linger close to the Killing horizon before escaping out to infinity. It is conceivable that there is some sort of reprocessing going on in the vicinity of the Killing horizon. The ratio of the  time scale of lingering with respect to the intrinsic time scale associated to the ray seems like a good quantity to quantify the degree of reprocessing at the Killing horizon.  
In particular, one might try to compare how many ``ray cycles" at the Killing horizon, $\tau_{\mathrm{intrinsic}}=1/\omega |_{\KH}$, are contained in the lingering time, as this might give a good estimate of the extent to which the rays are significantly reprocessed. 

\begin{figure}[!htb]
\centering
\includegraphics[scale=1.0]{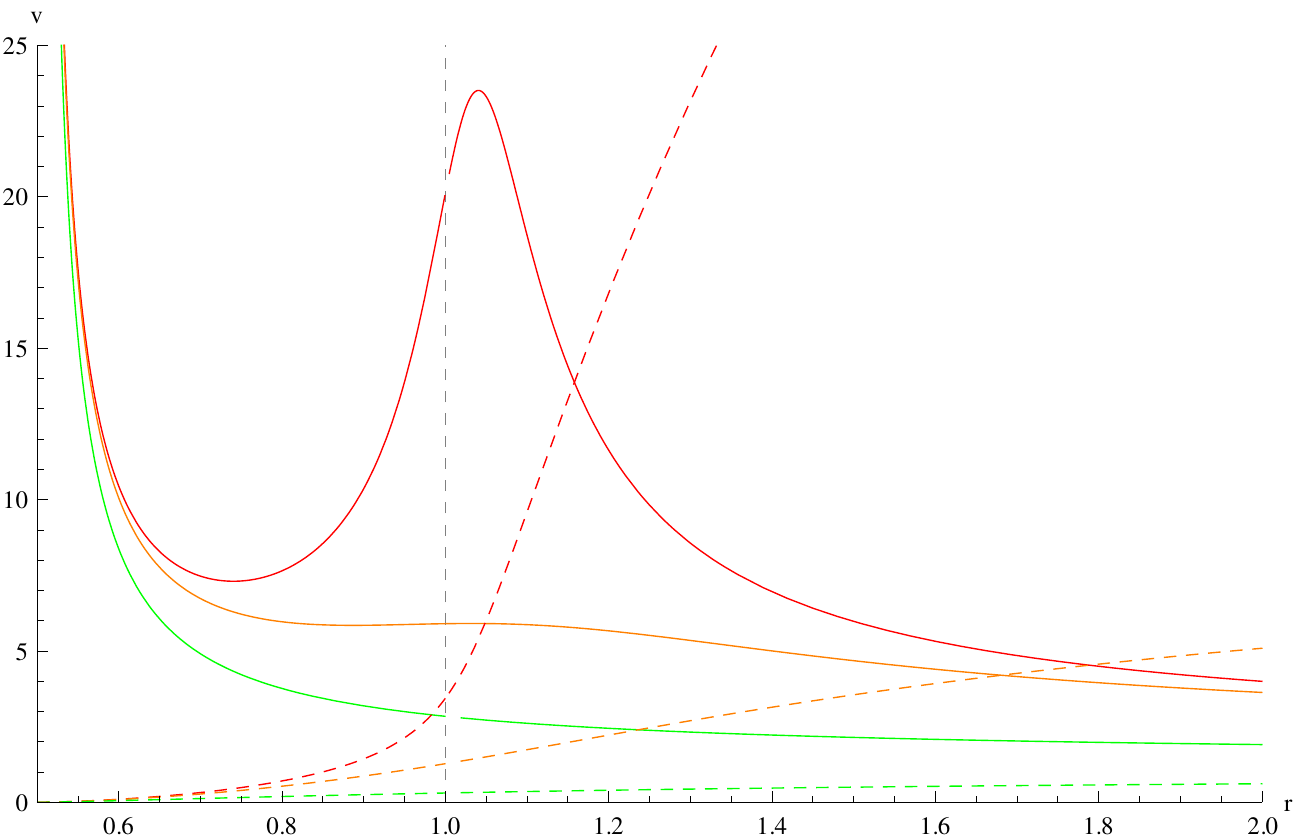}
\caption{$\ell/v$ (solid) and $\tau_{\mathrm{intrinsic}}$ (dashed), for $c_{123}=0$ solution, at $\Omega\ell=0.01$ (red), $\Omega\ell=0.1$ (orange), $\Omega\ell=1$ (green). Note the the peak is outside the Killing horizon}
\label{fig:times}
\end{figure}

An educated guess could then be to consider the ratio
\begin{equation}
{\mathcal{R}} = 
\frac{\tau_{\mathrm{linger}}}{\tau_{\mathrm{intrinsic}}}=  \,\ell \,\omega\,\left.\frac{\d v}{\d r}\right|_{\KH} ,
\end{equation}
where we have defined the lingering time as the ``width'' of the horizon, (for which $\ell$ is the simplest choice, though in the presence of modified dispersion relation a broadening of the horizon was found in, e.g., Ref.~\cite{Finazzi:2010yq}), times the crossing velocity. 
There is a problem with this, however. The lingering is \emph{outside} the Killing horizon, and the rays, even those with low energies, cross the horizon with a high coordinate velocity quite independently from how long they have lingered close to it. 
This can be seen in Fig.~\ref{fig:times}, which shows the time scales $1/\omega$ and $\ell/v$, (the time taken to go distance $\ell$ at instantaneous velocity $\d r/ \d v$). 
On the other hand, the previously defined $\kappa_{\KH}$ carries information about the concavity of the ray. If a particle has lingered it enters concave up, while those with high $\Omega$, will enter concave down.


Alternatively, we could consider the dispersion-dependent part of the Killing surface gravity in Eq.~\eqref{eq:killingkappa}, and write
\begin{eqnarray}
\varkappa &\equiv& |\kappa_{\KH}-\kappa_{\KH, \mathrm{metric}}|\nonumber \\
&=&{\frac{[2(r_0+2r_u)]^{1/3} \left(5r_0+3r_u\right){\ell^{2/3}\Omega^{2/3}}}{2\left(r_0+r_u \right)^{7/3}}}+ \mathcal{O}(\Omega^{4/3}),
\end{eqnarray}
and then identify $\tau_\mathrm{linger}=1/\varkappa$. Note that this $\tau_\mathrm{linger}$ goes to zero for large $\Omega$, and becomes infinite as $\Omega$ goes to zero. 
Using Eq.~\eqref{eq:omegaOmega} for $\omega$, in the small $\Omega$ limit we now get
\begin{eqnarray}
{\mathcal{R}} = \frac{\tau_\mathrm{linger}}{\tau_\mathrm{intrinsic}}&=& \frac{\omega}{\varkappa}\approx
\frac{2(r_0+r_u)^{8/3}}{(r_0+2r_u)^{2/3}(5r_0+3r_u)} \frac{1}{\ell^{4/3}\Omega^{1/3}}.
\end{eqnarray}
So, rays with small $\Omega$  remain close enough to the Killing horizon for long enough to be significantly reprocessed. 

%



\section{Discussion}
\label{raytracingdiscussion}

For the known Einstein-\AEther\ black holes, the relevant causal barrier is not the Killing horizon but the Universal horizon. 
This is the surface that admits an exact peeling behavior for ray trajectories, and it is also this is the surface for which the associated notion of surface gravity leads to the temperature obtained from tunneling methods. 
In this sense these results lend support to the findings of Refs.~\cite{Berglund:2012bu, Berglund:2012fk, Mohd:2013zca}, which point towards the Universal horizon being the surface relevant for thermal emission.


We have seen indications of the reprocessing of the low-energy particles at the Killing horizon. This seems to suggest that any thermal spectrum from the Universal horizon will be modified by the presence of the Killing horizon. 

To summarize, our work supports a picture of thermal radiation from the Universal horizon, with an associated temperature determined by the peeling surface gravity \eqref{eq:kappaUH1} in the standard way ($T=\kappa/2\pi$). 
The Killing horizon appears instead as a ``reprocessing/scattering surface" that  distorts the low-energy part of the original thermal spectrum in an energy and species (i.e., dispersion relation) dependent way. 

A shortcoming is that we cannot (with the current analysis) predict exactly what spectrum an observer at infinity will see from such a black hole. 
Further, there are several complications that, while not making the calculation impossible, will add an extra level of difficulty to determining such a spectrum. 
Even assuming that  Hawking radiation is produced at the Universal horizon one would still need to  include the role of the Killing horizon in reprocessing the outgoing low-energy modes. 

Nonetheless we feel that the clearer picture we have now, of the ergoregion behind the Universal horizon, the peeling-off behavior at the Universal horizon, and reprocessing of low-energy rays at the Killing horizon, shines new light on the thermodynamic character of black holes in Lorentz-violating theories. 
It is perhaps  too early to apply Eddington's rule-of-thumb to these theories.

\chapter{Hints of Hawking Radiation}
\label{Hawking}
\epigraph{Well, the thing about a black hole - its main distinguishing feature - is it's black. And the thing about space, the colour of space, your basic space colour, is black. So how are you supposed to see them?}{Holly, \it{Red Dwarf}, Season III}

\noindent This chapter will discuss some possible lines of inquiry into the existence and temperature of Hawking radiation from the universal horizon. As per the title, this will merely be hints of what may go into such a calculation, an exploration of some of the difficulties and possible approaches to the problem.

Consider black hole spacetime, as in the previous chapter, with a universal horizon inside a Killing horizon. What can we say about whether \emph{either} or \emph{both} of these horizons radiate? There are several basic ingredients necessary for Hawking radiation. However, studies related to the robustness of Hawking radiation show that these conditions only need to be valid at low energies. 
Firstly, we need a causal barrier enclosing an ergoregion, from which the energy needed to fuel Hawking radiation can be mined. 
In addition, we need an exponential peeling of rays from the causal barrier, which links the ingoing and the outgoing rays, as discussed in~\cite{Barcelo:2010xk}. Finally, we need an Unruh state, which is a vacuum state for the ingoing observer. \par
We have seen that the Universal horizon is a true causal barrier for particles of any energy, with a peel-off associated to all superluminal particles. It is then tempting to associate a temperature with the Universal horizon determined by the surface gravity of the previous chapter in Eq.~\eqref{eq:kappaUH1} and Eq.~\eqref{eq:kappaUH2},
 \begin{equation}
T_{c_{14}=0}=\frac{1}{4\pi r_{\UH}}\,\sqrt{\frac{2}{3(1-c_{13})}},
\end{equation}
and
\begin{equation}
T_{c_{123}=0}=\frac{1}{4\pi r_{\UH}} \,\sqrt{\frac{2-c_{14}}{2(1-c_{13})}}.
\end{equation}
This is in agreement with Ref.~\cite{Berglund:2012bu}, where the temperature associated with the Universal horizon was found using the Parikh-Wilczek tunneling formalism~\cite{Parikh:1999mf}. But it is not entirely clear as to where the energy for the Hawking radiation is mined from.

In the case of Killing horizon in a static geometry, one has an intuitive picture of pair-creation close to the horizon: by energy conservation one particle must have positive energy while the second has negative energy. The positive energy particle escapes to infinity, while the negative energy particle which is classically not allowed to exist outside the horizon,  tunnels through the horizon and appears inside where it can now exist as a real particle because the Killing vector is spacelike there. 
From the point of view of the outside observer however, the negative energy has gone in, the black hole has lost mass which has been carried away as energy in the Hawking radiation. What is the analogous story in case of the Universal horizon? Is there an analogue of the ergoregion behind the Universal horizon? \par
The relevant notion of energy in case of conventional Hawking radiation from a Killing horizon is the Killing energy. For a Universal horizon on the other hand, the relevant notion of energy is the energy measured in the {\ae}ther frame, $\omega=-k_a\, u^a$. 
Indeed, it is in the {\ae}ther frame that we have the modified dispersion relation, Eq.~\eqref{eq:mod-disp}. Now consider the pair-creation process right outside the Universal horizon. The particle created with positive $\omega$ escapes to infinity. As for the negative $\omega$ particle of the pair, from the equation for the conservation of energy  Eq.~\eqref{eq:conservation eqn} we have that for an ingoing mode with a negative energy outside the Universal horizon,
\begin{align}
\omega=\frac{-\Omega+\ks (\chi \cdot s)}{(\chi \cdot u)}.
\end{align}
We note that $\omega$ changes sign at the Universal horizon, with $\chi \cdot u$ now playing the role of the redshift factor $\chi^2$ in the standard calculation. This is so because $\chi \cdot u$ is negative outside, zero on, and positive inside the Universal horizon. Thus the negative energy particle of the pair after tunnelling through the Universal horizon appears inside with a positive $\omega$ and  propagates thereafter as a real particle. 
This suggests that the region behind the Universal horizon is analogous to ergoregion in the standard case and the picture that the pair-production close to horizon is responsible for the  Hawking radiation carries over to the Universal horizon. The only change being that the appropriate notion of energy is no longer the Killing energy but the energy measured in the {\ae}ther frame.  \par
There have been studies, mostly in the context of analogue spacetimes, which show that Hawking radiation from the Killing horizon is robust with respect to UV modifications of the dispersion relation, which will be discussed further below. 
In fact, the horizon does not even need to be an event horizon, but only some effective horizon~\cite{Barcelo:2006uw,Barcelo:2010xk}. This conclusion seems to be supported by Fig.~\ref{fig:lingering} where the low $\Omega$ rays seem to be peeling off the Killing horizon. 
However, there is a crucial difference. The cited studies do not have a Universal horizon whose presence will modify the boundary conditions for the modes. 
Once one realizes that there is in fact a Universal horizon one finds that, all the rays, irrespective of their energy, actually peel off the Universal horizon. 
The Killing horizon then seems to play the role of an efficient scattering surface for the low energy rays. The lesson is that the findings of analogue gravity can not be blindly applied to the case at hand. 
The reason, as we will fully discuss in the next chapter, is simply that we do not (yet) have an analogue spacetime modeling the Universal horizon. Hence the intuition from the analogue gravity program must be exercised with caution. 

Finally, a crucial aspect of the derivation of Hawking radiation is the onset of an Unruh-type quantum vacuum state after the gravitational collapse. While there is evidence that Universal horizons form in spherically symmetric collapse~\cite{Saravani:2013kva}, we do not yet know the nature of the quantum vacuum state that is established at the end of this process. 
The experience based on black hole physics in general relativity~\cite{Barcelo:2007yk} would suggest that the quantum vacuum state for free falling observers (i.e., the Unruh state) should be established only at the Universal horizon because the universal and exact ``redshift"  associated with the peel-off behavior would  erase any leftover renormalized stress energy tensor exponentially fast at the horizon once it is formed. 
Furthermore, the evidence found in Ref.~\cite{Berglund:2012fk}, and in \cite{Cropp:2013sea} and the previous chapter, strongly suggests that thermal radiation will be produced at the Universal horizon and will escape the Killing horizon.  
This in turn suggests that the state experienced by the free falling observers at the Killing horizon would be non-generic, and would typically be different from the Unruh vacuum. 

In this chapter, we will first present on of the standard calculations of Hawking radiation, then discuss a potential physical problem with the high energies involved in standard calculations. The discussion will then turn to analogue spacetimes with dispersion and how such systems have played in understanding the robustness of Hawking radiation. 

Next, we shall briefly present two existing calculations for Hawking radiation from the universal horizon, and discuss some potential issues with both approaches. Finally we shall focus on a way forward to do a calculation addressing questions or drawbacks in the existing approaches. 

\section{Hawking Radiation -- The standard picture}
\label{standardhawking}

As mentioned in section \ref{bhs} of the introduction, a major turning point in our understanding of black holes was Hawking's discovery \cite{Hawking:1974sw} that black holes radiate. We now work through a derivation of this radiation to understand it better. Today, there are many ways we can arrive at this result, and this derivation is loosely based on chapter six of \cite{Jacobson:2003vx}. The setup is as shown in \ref{fig:hawking}

\begin{figure}[!htb]

\begin{minipage}[c]{0.7\textwidth}

\begin{tikzpicture}[scale=1.4]
       
\draw[decorate,decoration=zigzag] (0,-2)--(0,5);

\draw (2,-2)--(2,5);
\node [above] at (2,5) {$H$};

\draw (2,0.5) to [out=90,in=220] (3,3);
\node [right] at (2,1.5) {$T$};

\draw (3,3) to [out=40,in=220] (6,5);
\node [below] at (6,5) {$P$};

\draw (3,3) to (5,0);
\node [below] at (5,0) {$R$};


\draw (0,4) to (5,4); 
\node [above] at (1,4) {$\Sigma_f$};

\draw (0,1) to (5,1);
\node [below] at (1,1) {$\Sigma_i$};


\draw (4,-2) to [out=120,in=320] (2,1);
\node [right] at (4,-2) {observer};

\draw (2,1) to [out=140,in=350] (0,2);

\end{tikzpicture}
\end{minipage}\hfill
\begin{minipage}[c]{0.3\textwidth}

\hspace{-2cm}

\caption{Black hole long after collapse. Tracing back in time, outgoing $P$ is split into reflected, $R$, and transmitted, $T$, parts.}

  \end{minipage}

\label{fig:hawking}
\end{figure}

We want the expectation value $\langle\Phi|N(P)|\Phi\rangle$ of the number operator, $N(P)= a^\dag (P) a(P)$, where $P$ is the outgoing wavepacket far from the black hole. This is split into a near-horizon outgoing part $T$ and a reflected wavepacket, $R$. Using stationarity, boundary conditions of no incoming radiation, and the Unruh vacuum state, one can express this as
\begin{equation}
\langle\Phi|N(P)|\Phi\rangle=\langle\Phi|a^\dagger(T) a(T)|\Phi\rangle.
\end{equation} 
Splitting $T$ into positive and negative frequency parts ($T^+$ and $T^-$) with respect to an infalling observer with proper time $\tau$ (who sees everything in the ground state at horizon crossing), we can further write
\begin{equation}
\langle\Phi|N(P)|\Phi\rangle=-\langle T^-, T^-\rangle_{\Sigma_i},
\end{equation} 
where $\Sigma_i$ is a spacelike slice sufficiently after the black hole has formed. 
So we need to understand the form of the wavepacket and be able to split it into positive and negative frequency parts. 

Look at wavepacket $T$ emerging from, and close to, the Killing horizon. As the wavepacket is outgoing it must be a function of the retarded time $u=t-r_*$.
\begin{equation}
 T \sim \exp(i\omega u),
\label{Tdependence}
\end{equation}
with $u$ divergent at the horizon. 

Consider how this wavepacket is seen by the infalling observer. We relate $u$ and the free-fall time by noting for the line element
\begin{equation}
 ds^2 = -e(r)dudv+r^2d\Omega^2,
\end{equation}
(here we certainly want to use something horizon-penetrating but other than this, the coordinate choice is not crucial). The path of the infalling radial timelike observer with proper time, $\tau$, is given by
\begin{equation}
 -e(r)\frac{du}{d \tau}\frac{dv}{d\tau}=-1.
\end{equation}
Implying
\begin{equation}
 \frac{d u}{d \tau} \sim \frac{1}{e(r)} \sim \frac{1}{r-r_s}.
\end{equation}
This measures the pile-up of rays (described by $u=\mathrm{const}$). Further,
\begin{eqnarray}
\exp\left(-\frac{r_*}{r_s}\right)&=&\exp(\frac{r+r_s\ln(r/r_s-1)}{r}) \nonumber \\
&=&\exp\left(\frac{r_s}{r}\ln\left(\frac{r}{r_s}-1\right)\right) \nonumber \\
&\sim& \frac{1}{r-r_s},
\end{eqnarray}
where the last relation is only true \emph{near the horizon}. From the definition of $r_*$, 
\begin{equation}
 \exp(-r_*/r_s)=\exp\left(\frac{u-v}{2r_s}\right).
\end{equation}
Near the horizon, $u$ diverges and $v$ is finite, and using $\kappa= 1/2r_s$.
\begin{equation}
 \exp(-r_*/r_s) \sim \exp(\kappa u)
\end{equation}
Therefore
\begin{equation}
 \frac{du}{d \tau} \sim \exp(\kappa u).
\end{equation}

Integrating,
\begin{equation}
 \tau = \int \exp(-\kappa u) d u =-\tau_0 \exp(-\kappa u),
\end{equation}
so that 
\begin{equation}
 u=-\frac{1}{\kappa}\ln(-\tau/\tau_0).
\end{equation}

Pick the free-fall observer who falls across horizon at $\tau=0$. For this observer wavepacket has a dependence on the proper time $T(\tau)$,
and vanishes for $\tau > 0$ (inside horizon). 
Then we split $T(\tau)$ into positive and negative frequency parts. This is simple because of the form of the wavepacket is
\begin{equation}
T \sim \exp\left(i\frac{\Omega}{\kappa}\ln (-\tau)\right)
\end{equation}
and we just need to work with the analytic continuation of logarithm, where across $\tau=0$ the positive frequency extension is obtained by replacing $\tau$ with $\tau + i\pi$ and the negative frequency extension by replacing $\tau$ with $\tau - i\pi$

Define $\bar{T}$ by flipping the wavepacket across the horizon, so it only has support inside, i.e. $\bar{T}(\tau)= T(-\tau)$ for $\tau>0$. Then the combinations
\begin{eqnarray}
T^+ &=& c_+ (T+\exp\left(\frac{-\pi \Omega}{\kappa}\right)\bar{T}); \nonumber \\
T^- &=& c_- (T+\exp\left(\frac{+\pi \Omega}{\kappa}\right)\bar{T})
\end{eqnarray}
are the positive and negative frequency components. We can match with T outside horizon to fix 
\begin{eqnarray}
c_- &=& \frac{1}{1-\exp\left(\frac{2\pi \Omega}{\kappa}\right)}; \nonumber \\
\frac{c_+}{c_-} &=& \exp\left(\frac{2\pi \Omega}{\kappa}\right).
\end{eqnarray}
Finally, this leads us to
\begin{equation}
\langle T^-, T^-\rangle = \frac{\langle T, T\rangle}{1-\exp(2\pi\Omega/\kappa)}, 
\end{equation}
which is a thermal spectrum with a greybody $<T, T>$.

\subsection{Which infalling observer?}
In the standard calculation of \cite{Jacobson:2003vx} presented above, one uses the infalling observer at horizon crossing to define a vacuum state. If we are working within the framework of general relativity, this is completely unambiguous: pick a ingoing timelike (radial) geodesic. Here, however, we appear to have (at least) two natural notions for an infalling observer: An infalling (metric) geodesic observer, and infalling {\ae}thereal observer. We could make a more complicated choice by giving our observer a non-trivial coupling to the \aether\ via a dispersion relation, and then showing the results are valid for a large range of dispersion relations. This would be overkill, as we see the key results don't depend on the type of infalling observer within reasonable restrictions.

Which should we work with, and how much will this affect the conclusions drawn? Notice the important point from the calculation above: we want to extract the divergent part and therefore what we need (and what we have for the geodesic observer) is that $\d x^a/\d \tau$ is finite and regular at horizon crossing.

For this section we will use $\tau$ for aether time and $\phi$ for proper time of an \aether\ observer. We define the aether observer by an unnormalized vector parallel to the aether, 
\begin{equation}
\bar{u}_a= \nabla_a \tau
\end{equation}
so that the proper time, $\phi$, along an infalling {\ae}thereal observer is given by
\begin{equation}
 g^{ab}\bar{u}_a\nabla_b \phi=  g^{ab}\partial_a \tau \partial_b \phi =1
\end{equation}
working in EF coordinates where
\begin{equation}
 g^{vv}=0; \quad g^{vr}=1; \quad g^{rr} =e(r)
\end{equation}
 and further,
\begin{equation}
 \tau= v+\int \frac{(s\cdot \chi)-(u\cdot \chi)}{(u\cdot \chi)} dr.
\end{equation}
\begin{eqnarray}
 g^{ab}\partial_a \tau \partial_b \phi &=& \partial_v \tau \partial_r \phi +\partial_r \tau \partial_v \phi +e(r)\partial_r \tau \partial_v \phi \nonumber \\
 &=& \left[ 1+e(r)\left(\frac{(s\cdot\chi)-(u\cdot\chi)}{(u\cdot\chi)}\right)\right]\partial_r\phi + \frac{(s\cdot\chi)-(u\cdot\chi)}{(u\cdot\chi)}\partial_v \phi \nonumber \\&=&1 
\end{eqnarray}
We have a solution of the form
\begin{equation}
 \phi(r, v)=\int \frac{1}{F(r)} d r +C_1 \left( v -\int \frac{G(r)}{F(r)} d r \right) +C_2,
\end{equation}
where the $C$s are arbitrary constants. Near the universal horizon 
\begin{equation}
F(r) \to 0 , \qquad \frac{G(r)}{F(r)} \to 1-(s\cdot\chi)^2, 
\end{equation}
and therefore
\begin{equation}
 \frac{dv}{d\phi}=\frac{1}{C_1}; \qquad \frac{dr}{d\phi}= \frac{1}{(C_1((s\cdot\chi)^2-1))}.
\end{equation}
Assuming $C_1 \neq 0$, $dv/d\phi$ and $dr/d\phi$ are regular and non-zero on horizon, and correspondingly one can pick 
$C_2$ such that at $v_0$ where the ray crosses the horizon $\phi=0$, as we had previously with the geodesic observer ($\tau=0$ with $dv/d\tau$ and $dr/d\tau$ regular). 

As long as we choose an infalling observer that is related in a similarly regular way, it will not affect the final spectrum calculated.

\section{The Transplanckian Problem}

Calculations of Hawking radiation involve using outgoing rays close to the event horizon, which are highly blueshifted (consider the form of eqn \ref{Tdependence}). To see how much consider and outgoing ray of frequency $\omega_1$ measured by an observer crossing the horizon at $v_1$. At $v_2=v_1 +\Delta v$ the ray is blueshifted by an exponential factor 
\begin{equation}
\frac{\omega_1}{\omega_2}=\exp(\kappa \Delta v)
\end{equation}
For a astrophysically reasonably sized black hole and $\Delta v$ of order of second this requires blueshifting into the regime beyond the Planck mass. 

This naturally raises the question of validity of Hawking's calculation, as this would mean trusting the symmetries of general relativity, the UV structure of the QFT for the relevant field, and the structure of spacetime up to such energies. A number of investigations have now shown that, at leading order Hawking radiation is robust against some modifications to the standard picture. 
For an interesting discussion of this issues see \cite{Jacobson:1999zk, Barbado:2011ai}.

\section{Hawking Radiation Calculations with Dispersion}

Recall that Hawking radiation calculations relies only on the kinematics of the spacetime, not the dynamics. Therefore, if Hawking radiation is independent of the UV completion of gravity, we should also see such an effect in analogue spacetimes. Alternatively, given that we know what the UV physics of condensed matter systems is, we should see where the calculation breaks down or is altered. 

Consider a simple fluid set-up, so that at the horizon $v=c_s$. Tracing back along an outgoing ray, it is blueshifted until it reaches the point where the dispersion relation is significant enough to change the group velocity from $c_s$ down to below the local speed of fluid flow. If the group velocity is below the fluid flow speed, the ray is dragged in. This means when tracing back, the Hawking quanta came originally from far outside the black hole, moves in and has ``mode conversion" takes place near the horizon, changing to an outgoing mode. This mode conversion depends, at the lowest order, on the surface gravity of the black hole, and hence the spectrum is approximately thermal at the temperature of Hawking radiation.  

There is an extra level of complexity with superluminal modes --- if one is considering something similar to a Schwarzschild black hole, by the argument above, outgoing modes come from \emph{inside} the horizon, and should emerge from the singularity, Therefore one must impose some sort of boundary condition inside the horizon (and possibly on the singularity itself), which is difficult to justify without a full-fledged theory of quantum gravity. However, one may study such systems in the analogue framework where, the easiest set up is connecting two regions of constant velocity, one superluminal and one subluminal. On both of these asymptotic regions one may now happily place sensible boundary conditions without invoking new physics.  

Numerous studies of such systems have taken place (see \cite{Jacobson:1991gr, Unruh:1994je, Finazzi:2012iu, Coutant:2011in, Macher:2009nz, Macher:2009tw, Finazzi:2011jd, Coutant:2014cwa, Coutant:2014wga, Coutant:2009cu}), and have generally concluded that, at leading order, the Hawking result is unchanged. Additionally, it has also been found that some new exotic effects can exist due to the presence of dispersion, such as black holes lasers \cite{Corley:1998rk, Finazzi:2010nc, Michel:2013wpa, Steinhauer:2014dra}. It has been understood that the horizon should be considered of finite width, rather than a 2D surface \cite{Finazzi:2010yq}. Interesting results also concern white hole horizon stability in the presence of modified dispersion relations \cite{Barcelo:2005fc}.

\section{Hawking Radiation with a Universal Horizon}

At the time of writing, two calculations exist computing the radiation from the universal horizon, which are in wild disagreement with each other. The first \cite{Berglund:2012fk} used a tunnelling method a l{\'a} Parikh and Wilczek \cite{Parikh:1999mf}. In this approach one considers particle creation near the horizon, and the contribution from both positive energy outgoing particles escaping from just inside the horizon and negative energy particles falling inside the horizon. One does this by considering
\begin{equation}
\phi(x) =\phi_0 \exp \left(iS\left[ \phi(x)\right]\right);
\end{equation}
\begin{equation}
S\left\lfloor \phi(x)\right\rfloor= \mp \Omega t + \int^r k_r(r') \d r'
\end{equation}
The outgoing ray will be singular at the horizon, giving an imaginary contribution. Thus tunnelling probability is given by
\begin{equation}
2\mathrm{Im} S = \mathrm{Im} \oint k_{r(o)}^+ (r) \d r
\end{equation}
where $+$ indicates positive energy and $(o)$ outgoing rays. 

By this technique \cite{Berglund:2012fk} found a thermal spectrum using a quartic dispersion relation. Importantly nothing in this construction relies on the null character or other special aspects of event horizons, and so is easily adaptable. However, it clearly dependent on the exact pole structure for $S$ near the horizon. For this reason it was impossible to extend this technique to arbitrary dispersion relations, clearly an undesirable feature.  Another drawback is nothing in this technique can tell us the effect of the Killing horizon: whether it also radiates or if it effects the shape of the spectrum via a frequency dependent greybody. 

The second attempt \cite{Michel:2015rsa} takes inspiration from condensed matter calculations with dispersion. The authors take an idealized collapse of a null shell of matter to form a black hole, allowing them to place boundary conditions on the flat space section inside the shell. Looking at the area near both the shell and the universal horizon, one calculates the overlap along the shell of the stationary modes and the outwards propagating modes. This overlap decreases at high energies (which are the relevant ones due to the high blue-shift near the horizon) and becomes negligible, and thus the relevant Bogliubov coefficient is suppressed.

This is used to argue that the universal horizon does not effect the standard calculation with dispersion for a Killing horizon, and the temperature measured at infinity is thus, at first order, the one of the Hawking radiation from the Killing horizon, and is determined by the surface gravity associated to the Killing horizon.

The advantage to this approach is that, although a quartic dispersion relation was used, it is clear the result should hold for general superluminal dispersions. 

However, questions remain about this approach. In particular, two features give pause: looking along the ingoing shell, the \aether\ is discontinuous, which raise the question of whether the vacuum is continuous through this shell. It is also less clear that outside the shell the vacuum corresponds to physical observers. It seems the observers measuring this vacuum would have to be moving infinitely fast close to the universal horizon. While infinitely fast observers are possible in these theories, this does not seem to correspond to vacuum seen by an infalling observer in any simple (regular) way.  

\section{Adiabaticity}
Particle creation, such as Hawking radiation, can generally be expected to occur when an adiabatic or WKB condition breaks down (see discussion in \cite{Barcelo:2007yk, Schutzhold:2013mba, Schutzhold:2008tx}). Such concerns, as noted in \cite{Schutzhold:2013mba}, are dependent on the frame chosen, so one must make sure one considers the breakdown of adiabaticity with respect to sensible observers. 

Which observers are sensible to pick in the case of \AE\ gravity? We will consider the breakdown of adiabaticity of the \aether\ frequency with respect to the \aether\ time, that is we want $\dot \omega/\omega^2$ for $\omega = k_au^a$.
From the ray tracing results for we have $\omega(k_s)$ for one dispersion relation, so we can easily find $d\omega/dr$. Then we can combine
\begin{equation}
\dot \omega =\frac{d \omega}{d r}\left(\frac{d\tau}{dr}\right)^{-1}
\end{equation}
We can use the formulae (\ref{constantkhronon}) for the khronon slices and (\ref{eq:trajectory}) for the outgoing null rays to write this as
\begin{equation}
\dot \omega = \frac{d \omega}{d r}\left(\frac{u^v+v_gs^v}{u^r + v_gs^r} +\frac{u_r}{u_v}\right)^{-1}.
\end{equation}
Simplify slightly using results from appendix \ref{aetherelations} so that
\begin{equation}
 \dot \omega=-\frac{d\omega}{dr}\left((\chi\cdot u)(u^r+v_gs^r)\right).
\label{adiabaticparameter}
\end{equation}
Now, we have explicit formaulae for everything appearing in eqn \ref{adiabaticparameter}, as we have $\omega(k(r))$ from the previous chapter. 

Note that $\dot \omega/ \omega^2$ depends explicitly on both $\kappa_\UH$ and $\kappa_\KH^{\mathrm{metric}}$, as a consequence of $k_s(r)$ being a function of both parameters (see appendix \ref{aetherelations}). However, once we pick a given sector (the $c_i$ coefficients) of the theory  we are in a one-parameter family of solutions, and one can express both these surface gravities in terms of the mass.

\begin{figure}[!htb]
\centering
\includegraphics[scale=0.6]{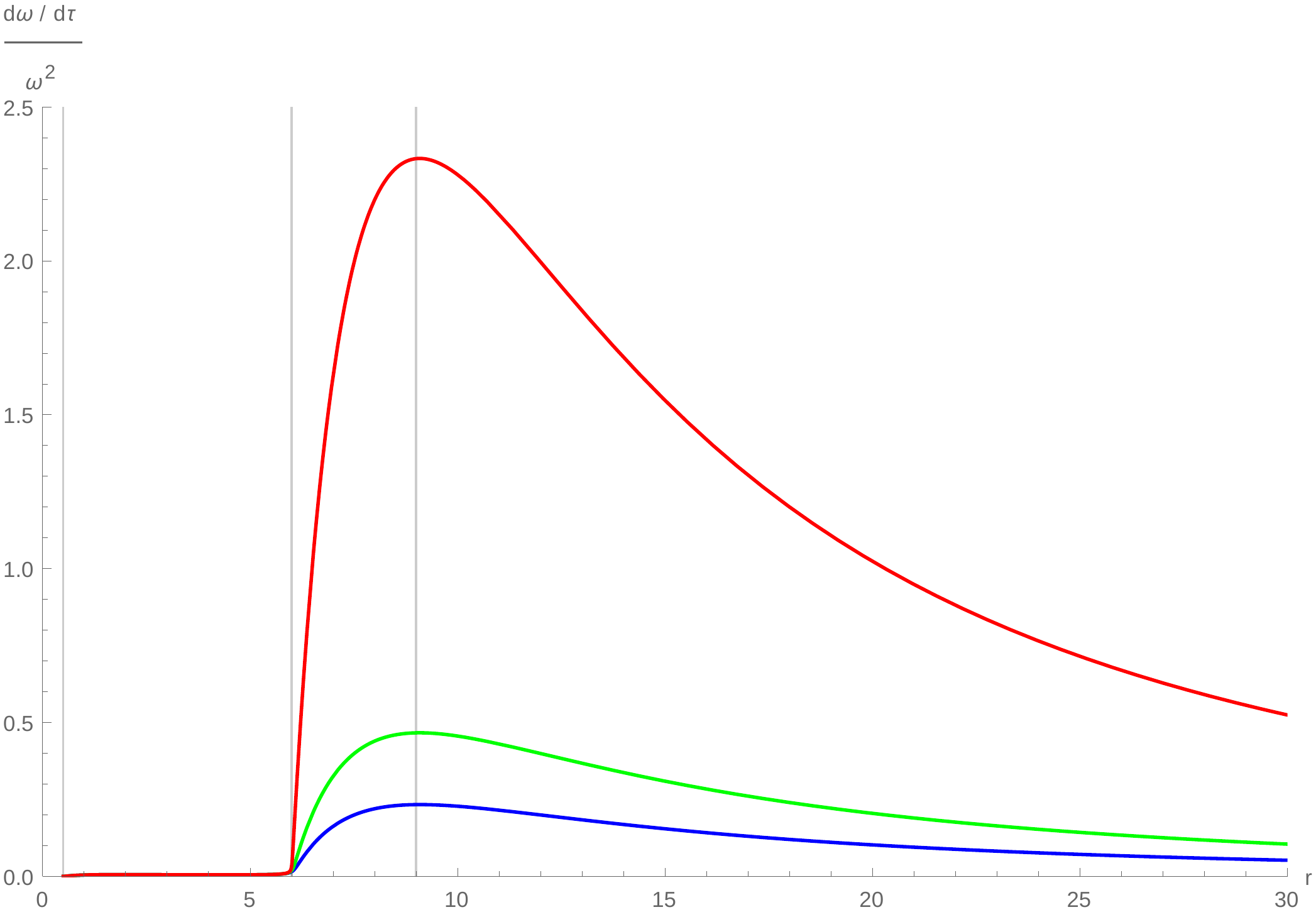}
\caption{Breakdown of adiabaticity occurs when $\frac{\dot \omega}{\omega^2} > 1$. Here shown for $\Omega=0.01$ (Red), $\Omega=0.05$ (Green), $\Omega=0.1$ (Blue). universal (r=6) and Killing horizons (r=9) marked in black. Notice the peak at the Killing horizon.}
\label{adiabatic}
\end{figure}

One may notice the adiabaticity only seems to be broken near or outside the Killing horizon (see figure \ref{adiabatic}). One might try to take this as evidence that particles are created at the Killing horizon and not the universal horizon. However, the key point for the observer at infinity is not where such particles are created, but what controls the temperature. In this sense the dependence of the adiabaticity on both surface gravities could be taken as a hint that the temperature of radiation will not be purely fixed by the properties of the Killing horizon.  

\section{Euclidean continuation}

One alternative way to find the temperature of Hawking radiation is using the Euclidean continuation and finding the deficit angle near the horizon. This relies on the fact that periodicity in imaginary time corresponds to thermality in Lorentzian signature \cite{frolov2011introduction}. 

The standard technique is as such: expand the Scharzschild black hole near the Killing horizon, so that
\begin{equation}
f(r)= 2\kappa (r-r_\KH); \qquad \kappa \equiv \frac{1}{2}\frac{\d}{\d r} f(r),
\end{equation} 
and change coordinates to
\begin{equation}
\rho = \sqrt{\frac{2}{\kappa} \left(r-r_\KH\right)},
\end{equation}
so the near horizon metric takes the form
\begin{equation}
\d s^2 = -c^2\kappa^2 \rho^2 \d t^2 + \d \rho^2 + r(\rho)^2 \d\Omega^2.
\end{equation}
Now preform an analytic continuation $\tau = it$, so
\begin{equation}
\d s^2 = \rho^2 \d (\kappa \tau)^2 + \d \rho^2 + r(\rho)^2 \d\Omega^2.
\end{equation}
This has the form of polar coordinates on the flat $\rho- \tau$ plane if the angular variable $1/\kappa \tau$ has the correct periodicity, that is periodic in $\tau$ with periodicity $\frac{\kappa}{2\pi}$. Therefore the spacetime has temperature 
\begin{equation}
T=\frac{\kappa}{2\pi}.
\end{equation}

Could we adapt such a construction to the universal horizon? At first sight, it seems obvious we cannot: this is a geometrical construction relying on the vanishing of $f(r)$. In our spacetimes this happens precisely at the Killing horizon, and nowhere else. Instead of this approach, let us take a disformal transformation, as in \eqref{disformal}, for an arbitrary velocity, s.
\begin{eqnarray}
\d s^2 &=& \left(-f(r) + (1-s^2)u_t^2\right)\d t^2 + (1-s^2)u_tu_r \d t\d r \nonumber \\
&+& \left(\frac{1}{f(r)} +(1-s^2)u_r^2\right)\d r^2 +r^2\d\Omega^2
\end{eqnarray}
Using the fact that both original metric and \aether\ are spherically symmetric and time-independent, we can transform by 
\begin{equation}
\d t = \d T - \frac{(1-s^2)u_t u_r}{-f(r) + (1-s^2)u_t^2}\d r
\end{equation}
and the use the unit condition,
\begin{equation}
-\frac{1}{f(r)}u_t^2 + f(r)u_r^2=-1
\end{equation}
to get this back to diagonal form
\begin{equation}
\d s^2 =-s^2F(r)\d t^2 +\frac{1}{F(r)}\d r^2 + r^2\d\Omega^2;  \qquad F(r)= f(r) - (1-s^2)u_t^2.
\end{equation}
We now have a speed-dependent horizon, $r_\KH(s)$, which we can solve for for either of the exact solutions in section \ref{ae-bhs}. We can now follow through the standard steps, expanding near the horizon
\begin{equation}
F(r, s)= 2s\kappa (s) (r-r_\KH(s)) \qquad \kappa \equiv \frac{1}{2}\frac{\d}{\d r} F(r, s)|_{\KH(s)},
\end{equation} 
and preforming the same (now s-dependent) transformations, to arrive at
\begin{equation}
\d s^2 = s^2\rho(s)^2 \d (\kappa(s) \tau)^2 + \d \rho(s)^2 + r(\rho(s))^2 \d\Omega^2,
\end{equation}
which in turn implies 
\begin{equation}
T(s) = \frac{\kappa(s)}{2\pi s}.
\end{equation}
As we saw in section \ref{ae-bhs}, the universal horizon corresponds to the Killing horizon of a disformally transformed metric in the limiting case where the velocity, $s$, goes to infinity. Does $T(s)$ give a sensible result in this limit? 

Take the exact solution for $c_{123} = 0$ (similar results can be obtained for the other exact solution). 
\begin{equation}
r_\KH(s) =\frac{r_0(1+s)-2r_u}{2s}
\end{equation}
and
\begin{equation}
\kappa(s) =\frac{2s^3(r_0+ 2r_u)}{\left(r_0(1+s)+ 2r_u\right)^2}
\end{equation}
and so
\begin{equation}
T(s)=\frac{2s^2(r_0+ 2r_u)}{2 \pi \left(r_0(1+s)+ 2r_u\right)^2}
\end{equation}

\begin{equation}
T_\UH = \lim_{s \to \infty} T(s)  =\frac{2(r_0+ 2r_u)}{2\pi r_0}.
\end{equation}
 
This is exactly the expected Hawking temperature if the relevant surface gravity is provided by the $\kappa_{\mathrm{peeling}}$ calculated in the previous section. 

\section{Ansatz{\"e} for Hawking radiation calculations}

We could try naive extensions of the approach in section \ref{standardhawking}. Consider, for instance, 
\begin{equation}
P \sim \exp(i(-\Omega v + \int k_r dr))
\label{Pform}
\end{equation}
where $\Omega$ is our conserved energy conjugate to v and $k_r$ is defined by
\begin{equation}
k_s \equiv k_as^a =k_v s^v +k_r s^r \quad \Rightarrow \quad k_r=\frac{k_s+\Omega s^v}{s^r},
\end{equation} 
We define 
\begin{equation}
 u_{\mathrm{eff}}=v-\frac{1}{\Omega}\int k_r dr
\end{equation}

One physical issue with this approach is that when we are calculating $\frac{d u_{\mathrm{eff}}}{d\tau}$ this is no longer a measure of the density 
(and hence pile-up) of rays, as $u_{\mathrm{eff}}$ is \emph{not} constant along the rays. Of course one can write a form that does this, namely, 
\begin{equation}
\label{raytracingform}
 P \sim \exp\left[-i\Omega\left( v - \int \frac{u^v+v_g s^v}{u^r+v_g s^r} d r\right)\right],
\end{equation}
and is clearly related to the equation describing the ray from our ray tracing paper. This form also ensures that $P$ is only a function of an appropriate retarded time, so outgoing. 

From such a form, we can continue with the construction of section \ref{standardhawking} to find
\begin{equation}
 T \sim \exp\left(i\frac{\Omega}{\kappa_p}\ln(-\tau)\right)
\end{equation}
and hence
\begin{equation}
 \langle T^-, T^- \rangle=\frac{\langle T, T\rangle}{1-\exp(2\pi\Omega/\kappa_p)}.
\end{equation}
It therefore seems like part of the problem is the breaking of the relationship between $k_r$ and the coordinate describing ray pile up. However, there is hardly any justification for the form \ref{raytracingform} other than the fact it is related to the physical ray structure we provided. As such it is unclear if this would be consistent with a standard WKB approach. 

If, instead one starts with eqn \ref{Pform}, one runs into a pole structure which becomes difficult to work with mathematically, and interpret physically.

\section{Ways Forward}
Overall, there appears to be several open issues. Firstly, the right boundary conditions and picking the right vacuum state, and secondly finding a calculation that can handle a different pole structure at the horizon. 

Given the complexities of the only collapse scenario, it seems best to work with a static geometry or, in the same style of \ref{standardhawking}, a geometry long after collapse. 

Following techniques of \cite{Unruh:2004zk}, we can take a dispersive field equation and, expanding around the universal horizon now, instead of the Killing horizon, approximately solve this equation near the horizon, for the several possible modes near the universal horizon. We could then link to the solution far away, and match what leads to the outgoing particle at infinity. 

One difficulty with following the techniques of \cite{Unruh:2004zk} is that their set-up was within the analogue framework and involved linking two asymptotically flat regions. Therefore the boundary conditions of such a set-up cannot be followed blindly. 

It is clear that a full understanding of Hawking radiation in spacetime is still lacking, and that gaining such a full understanding would give us much insight into the process of Hawking radiation. 
 
\chapter{Relativistic Bose--Einstein condensates for mimicking Killing and Universal horizons}
\epigraph{In 1972, I was asked to give a colloquium to the physics department at Oxford. In trying to describe what a black hole was, I came up with an analogy. Imagine you are a blind fish, and are also a physicist, living in a river. At one place in the river, there is a particularly virulent waterfall, such that at some point in the waterfall, the velocity of the water over the waterfall exceeds the velocity of sound in the water.}{Bill Unruh}

\section{Introduction}

\noindent Analogue spacetimes, as we have discussed previously, can often have more freedom in their configurations than those spacetimes that are vacuum or simple matter solutions in general relativity. But what about theories with new degrees of freedom? As we have seen with our investigations of universal horizons, some features cannot be understood solely in terms of the metric. 

In this chapter we will consider how to couple a fluid to external fields so as to incorporate effects due to extra scalar or vector degrees of freedom. As a convenient system we will consider relativistic Bose--Einstein condensates (rBECs). We pick up this system as it is clear how to couple fields, and as we have in mind \AE\ gravity, we want a relativistic system which we can naturally couple a four-vector to. 

These rBECs have been less investigated than their non-relativistic counterparts, for obvious reasons of lack of existing experimental capabilities. To date, the only explicit analogue spacetime developed with rBECs is the $k=-1$ FLRW cosmological solution \cite{Fagnocchi:2010sn} (but see also a ``draining bathtub" geometry \cite{Bilic:1999sq} for relativistic acoustics). As static black holes are some of the most useful and simple analogue spacetimes, we investigate black holes and how to relate black hole features to the properties of the rBEC. In the first part of this chapter, we will look explicitly at the difficulties in modelling the Schwarzschild solution, and examine a simplified flow leading to a canonical acoustic black hole. 

In the second part of this chapter, we will look at coupling an external electromagnetic field to the rBEC to simulate an \aether\ field. We will see the simplest option, while allowing the incorporation of vorticity in analogue spacetimes, does not supply us with an observable \aether\ field. We will then consider other couplings and other fields which allow the interpretation of an \aether\ field. Finally, we shall try modelling a universal horizon in these analogue spacetimes

\section{Modelling Universal horizons in standard acoustic geometries}
\label{digression}

Let us first reassure ourselves that we truly cannot incorporate features such as the universal horizon into the standard picture of analogue models. It is interesting to note that in the menagerie of acoustic spacetimes, a specific causal structure similar to that of these Einstein--\AEther\ black holes was found (see Fig.~26 of Ref.~\cite{Barcelo:2004wz}). 
This spacetime was named the ``unphysical black hole" as it needed an unphysical, diverging, fluid flow in order to mimic the presence of a singularity. Nonetheless, the resemblance with the causal structure shown in Fig.~\ref{fig:conformal} appears striking.
However, no concept of Universal horizon has ever surfaced in acoustic spacetimes (or any of the other analogue models for that matter).  It is therefore interesting to clarify why this is the case. 

Of course, it is trivial to put the metric of the two previously described black-hole solutions of Einstein--{\AE}ther theory into the standard Painleve--Gullstrand form,
\begin{equation}
\d s^2=-\left(1-\frac{ {\rm v}^2 }{c_s^2}\right)\d t^2 +2\,\frac{{\rm v}}{c_s}\,\d t\, \d r + \left|\d \vec{x}\right|^2,
\end{equation}
where ${\rm v}$ is the flow velocity and $c_s$ the speed of sound for the corresponding acoustic geometry. In this geometry there is a natural notion of the preferred frame, the frame in which the fluid is at rest. So the corresponding {\ae}ther field is most naturally taken to be
\begin{equation}
u^a=\left\lbrace u^t, \frac{\rm v}{c_s}\right\rbrace.
\end{equation}
The zeroth component is determined by the requirement that the {\ae}ther field is unit timelike, which gives
\begin{equation}
u^t=\frac{{\rm v}^{2}\pm\sqrt{ {\rm v}^4-c_s^4+{\rm v}^2\,c_s^2} }{c_s^2-{{\rm v}^2}}.
\end{equation}
Regularity at the Killing horizon (the surface where ${\rm v}=c_s$) fixes the sign to be minus. Now let us calculate --- when is the Killing vector orthogonal to the {\ae}ther?  We have 
\begin{equation}
\chi \cdot u = \pm \frac{ \sqrt{ {\rm v}^{4}+c_s^4-{\rm v}^2 c_s^2} } {c_s^2} 
=\pm \frac{ \sqrt{ ({\rm v}^{2}-c_s^2)^2 + ({\rm v} c_s)^2} } {c_s^2} .
\end{equation}
But this, being proportional to a sum of squares, is never zero for real-valued $\rm{v}$. Therefore, there is no Universal horizon. 

One might similarly try with a relativistic fluid, as in \cite{Visser:2010xv}, and which will be more fully described below, which has a metric given by
\begin{equation}
 G_{ab}=\Omega \left[g_{ab}+\left(1-\frac{c_s^2}{c^2}\right)\frac{v_av_b}{c^2}  \right]
\end{equation}
where $G_{ab}$ is the effective metric, $g_{ab}$ the background metric (which we pick to be Minkowski, as appropriate for any laboratory set-up) and $g^{ab}v_av_b=-c^2$. If we set the {\ae}ther to be our naturally preferred frame, $v$, then the norm constraint on the {\aether} is imposed by the choice $\Omega=1$.
If we want to model an \AE\ black hole, we have timelike Killing vector, and the condition for a Universal horizon is
\begin{equation}
 \chi^av^bG_{ab}=v^bG_{tb}=\Omega\left[-v^t+\left(1-\frac{c_s^2}{c^2}\right)\frac{v^2 v_t}{c^2} \right]=0 
\end{equation}
Which is tantamount to
\begin{equation}
 v^t=-\left(1-\frac{c_s^2}{c^2}\right)v^t
\end{equation}
which again is impossible to satisfy. 

Why is there no Universal horizon in these cases? In the acoustic geometry, the \aether\ field is determined by the same flow, ${\rm v}$, that determines the metric. 
There is simply not enough freedom to have additional structures such as Universal horizon, where the geometry is well behaved, but the \aether\ time has run to infinity. One needs a separately defined dynamical {\ae}ther to have a Universal horizon. 
This of course does not mean that we cannot capture this structure within the context of analogue models. One simply needs more degrees of freedom in order to do so.  It is clear that we truly need to add some new degrees of freedom to model the {\ae}ther. 

\section{Relativistic BECs: a new system for analogue black holes}\label{uncoupledrbec}

In this section we will give a brief description of a relativistic Bose--Einstein condensate. For further details, see \cite{Fagnocchi:2010sn}. We start with the Lagrangian density for a complex scalar field $\phi(\mathbf{x},t)$ which can be given by
\begin{eqnarray}\label{lrbec}
\mathcal{L}&=&-\eta^{\mu\nu}\partial_{\mu}\phi^{\dagger}\partial_{\nu}\phi-\left(\frac{m^{2}c^{2}}{\hbar^{2}}+V(t,\mathbf{x})\right)\phi^{\dagger}\phi-U(\phi^{\dagger}\phi;\lambda_{i})
\end{eqnarray}
where $m$ is the mass of bosons, $V(t,\mathbf{x})$ is an external potential, $c$ is the speed of light, $U$ is a self-interaction term and $\lambda_{i}(t,\mathbf{x})$ are the coupling constants. 

The Lagrangian (\ref{lrbec}) is invariant under the global $U(1)$ symmetry and has a conserved current 
\begin{equation}
j^{\mu}=i(\phi^{\dagger}\partial^{\mu}\phi-\phi\partial^{\mu}\phi^{\dagger}), 
\label{jcurrent}
\end{equation}
related to a conserved ensemble charge $N-\bar{N}$, where $N(\bar{N})$ is the number of bosons (anti-bosons).

In the case where there are no self-interactions ($U=0$) and no external potential ($V=0$), the average number of bosons $n_{k}$ in the state of energy $E_{k}$ can be written as 
\begin{equation}\label{nb}
N-\bar{N}=\Sigma_{k}[n_{k}-\bar{n}_{k}], 
\end{equation}
where 
\begin{eqnarray}
n_{k}(\mu,\beta)&=&1/\left\lbrace{\exp}[\beta(|E_{k}|-\mu)]-1\right\rbrace, \nonumber \\
\bar{n}_{k}(\mu,\beta)&=&1/\left\lbrace{\exp[\beta(|E_{k}|-\mu)]-1}\right\rbrace
\end{eqnarray}
and $\mu$ is the chemical potential, $T\equiv1/(k_{B}\beta)$ is the temperature and the energy of the state $k$ is given by $E^{2}_{k}=\hbar^{2}k^{2}c^{2}+m^{2}c^{4}$.

The relation between the conserved charge density $n=(N-\bar{N})/\Omega\,$ (where $\Omega$ is the volume of the system) and the critical temperature is
\begin{equation}\label{nd}
n=C\int^{\infty}_{0}dkk^{2}\frac{\sinh(\beta_{c}mc^{2})}{\cosh(\beta_{c}|E_{k}|)-\cosh(\beta_{c}mc^{2})}, 
\end{equation}
where $C=1/(4\pi^{3/2}\Gamma(3/2))$.

The non-relativistic and ultra-relativistic limits can be obtained directly from (\ref{nd}). The non-relativistic limit,  $k_{B}T_{c}\ll{mc^{2}}$,  is given by 
\begin{equation}
k_{B}T_{c}=\frac{2\pi\hbar^{2}}{n}\left(\frac{n}{\zeta(3/2)}\right)^{2/3}, 
\end{equation}
where $\zeta$ is the Riemann zeta function. The ultra-relativistic limit, $k_{B}T_{c}\ll{mc^{2}}$, implies
\begin{equation}
(k_{B}T_{c})^{2}=\frac{\hbar^{3}c\Gamma(3/2)(2\pi)^{3}}{4m\pi^{3/2}\Gamma(d)\zeta(2)}n.
\end{equation}

The condensation of the relativistic Bose gas occurs when $T\ll{T_{c}}$. In this phase, it is possible to uncouple the BEC ground state from its perturbations. To separate this state from its perturbation, analogously to section \ref{BEC} we can make use of the mean-field approximation, performing the substitution $\phi=\varphi(1+\psi)$, where $\varphi$ is the classical background field satisfying the equation
\begin{equation}\label{nlkg}
\Box\varphi-\left(\frac{m^{2}c^{2}}{\hbar^{2}}+V\right)\varphi-U'\varphi=0, 
\end{equation}
and $\psi$ is a fluctuation. The nonlinear Klein-Gordon equation (\ref{nlkg}) gives the dynamics of the relativistic condensates.

It is also convenient decompose the degrees of freedom of the complex scalar classical field in terms of the Madelung representation, $\varphi=\sqrt{\rho}e^{i\theta}$. Using this representation, the continuity equation and the condensate equation (\ref{nlkg}) are
\begin{eqnarray}\label{cemrnc}
\partial_{\mu}(\rho{u^{\mu}})&=&0, \\
-u_{\mu}u^{\mu}&=&c^{2}+\frac{\hbar^{2}}{m^{2}}\left[V(x^{\mu})+U'(\rho;\lambda_{i}(x^{\mu}))-\frac{\Box\sqrt{\rho}}{\rho}\right],
\end{eqnarray}
where 
\begin{equation}\label{hypersurface}
 u^{\mu}=\frac{\hbar}{m}\partial^{\mu}\theta
\end{equation}
 is the fluid four-velocity of the condensate.

The quantum perturbation $\psi$ satisfies 
\begin{eqnarray}\label{lpemr}
\left[i\hbar{{u}^{\mu}}\partial_{\mu}-T_{\rho}-mc_{0}^{2}\right]\psi=mc_{0}\psi^{\dagger}, 
\end{eqnarray} 
where $c_{0}^{2}\equiv\frac{\hbar^{2}}{2m^{2}}\rho{U''}$ is related to the interaction strength and 
\begin{equation}
T_{\rho}\equiv-\frac{\hbar^2}{2m}\left(\Box+\eta^{\mu\nu}\partial_{\mu}ln\rho\partial_{\nu}\right)
\end{equation}
is a generalized kinetic operator, the relativistic equivalent of the quantum potential (cf.  equation \ref{non-relqp}). It is instructive to obtain a single equation for the field $\psi$. This can be done taking the Hermitian conjugate of (\ref{lpemr}) and using the result to eliminate $\psi^{\dagger}$. After some manipulation, the equation describing the propagation of the linearized perturbations is
\begin{eqnarray}\label{lpe}
\left\lbrace\left[i\hbar{{u}^{\mu}}\partial_{\mu}+T_{\rho}\right]\frac{1}{c_{0}^{2}}\left[-i\hbar{{u}^{\nu}}\partial_{\nu}+T_{\rho}\right]-\frac{\hbar^{2}}{\rho}\eta^{\mu\nu}\partial_{\mu}\rho\partial_{\nu}\right\rbrace\psi=0,
\end{eqnarray}
which does not depend on the external potential $V$. This is the relativistic generalization of the Bogoliubov-de Gennes equation. The rBEC perturbations has many interesting physical situations. An extensive analysis of the dispersion relation of such perturbations was carried out in \cite{Fagnocchi:2010sn}. Here, we are concerned with the low momentum massless modes,
\begin{equation}
|k|\ll\frac{mu^{0}}{\hbar}\left[1+\left(\frac{c_{0}}{u^{0}}\right)^{2}\right]. 
\end{equation}
In this limit, we can disregard the quantum potential $T_{\rho}$, and the description of the acoustic disturbances propagation is governed by an acoustic metric. This can be seen by applying the above conditions, the equation (\ref{lpe}) reduces to
\begin{eqnarray}\label{lpnqp}
\left[{{u}^{\mu}}\partial_{\mu}\left(\frac{1}{c_{0}^{2}}{{u}^{\nu}}\partial_{\nu}\right)-\frac{1}{\rho}\eta^{\mu\nu}\partial_{\mu}\left(\rho\partial_{\nu}\right)\right]\psi=0.
\end{eqnarray}

Using the continuity equation (\ref{cemrnc}), we can rewrite the equation (\ref{lpnqp}) as
\begin{eqnarray}
\partial_{\mu}\left[-\rho\eta^{\mu\nu}+\frac{\rho}{c_{0}^{2}}u^{\mu}u^{\nu}\right]\partial_{\nu}\psi=0.
\end{eqnarray}

Now, taking the usual approach, we express this as a Klein-Gordon in a curved metric, which is given by
\begin{eqnarray}\label{ramnr}
g_{\mu\nu}=\frac{\rho}{\sqrt{1-u_{\alpha}u^{\alpha}/c_{0}^{2}}}\left[\eta_{\mu\nu}\left(1-\frac{u_{\alpha}u^{\alpha}}{c_{0}^{2}}\right)+\frac{u_{\mu}u_{\nu}}{c_{0}^{2}}\right]. 
\end{eqnarray}
Sometimes it is more convenient express the acoustic metric (\ref{ramnr}) as
\begin{equation}\label{ram}
g_{\mu\nu}=\rho\frac{{c}}{c_{s}}\left[\eta_{\mu\nu}+\left(1-\frac{c_{s}^{2}}{c^{2}}\right)\frac{v_{\mu}v_{\nu}}{c^{2}}\right],
\end{equation}
where $v^{\mu}=cu^{\mu}/||u||$ is the normalized (with respect to Minkowski) four-velocity and $c_{s}$ is the speed of sound, which is defined by 
\begin{equation}
c_{s}^{2}=\frac{c^{2}c_{0}^{2}/||u||}{1+c_{0}^{2}/||u||^{2}}.
\end{equation}

\section{Analogue Black holes for rBECs}

Other than the FLRW solution studied in \cite{Fagnocchi:2010sn}, few solutions have been studied to date. Here we investigate black hole solutions in these analogue models (but see also the analysis of horizons in \cite{Bilic:1999sq}).

\subsection{Schwarzschild black hole}\label{schwarz}

Can we match the Schwarzschild solution with the form \eqref{ram}? If we take
\begin{equation}
 v_t^2=\frac{c^2-c_s^2f(r)}{\left(1-\frac{c_s^2}{c^2}\right)}=c^2+\frac{2M}{r\left(1-\frac{c_s^2}{c^2}\right)}.
\end{equation}
and 
\begin{equation}
 v_r^2=\frac{c_s^2\left[1-f(r)\right]}{(1-\frac{c_s^2}{c^2})}.
\end{equation}
(note that this obeys the unit timelike constraint on $v$ by construction), we see the corresponding acoustic metric is 
\begin{eqnarray}
 \d s^2&=&-\frac{c_s^2}{c^2}\left(1-\frac{2M}{rc_s^2}\right)dt^2+\frac{2}{c^2}\sqrt{\frac{2M}{r}\left(c^2-c_s^2+\frac{2M}{r}\right)}\d t\d r \nonumber \\
 &+&\left(1+\frac{2M}{rc^2}\right)\d r^2 +r^2\d\Omega^2
\label{schwarzmetric}
\end{eqnarray}
(note the similarity to the metric in section 4 of \cite{Baccetti:2012ge}). This can be seen to be a form of the Schwarzschild metric by starting with \ref{schwarzmetric} and transforming to a new coordinate, $T$ by
\begin{equation}
\d t=\d T-\frac{\sqrt{\frac{2M}{r}\left(c^2-c_s^2+\frac{2M}{r}\right)}}{\frac{c_s^2}{c^2}\left(1-\frac{2M}{rc_s^2}\right)}\d r
\end{equation}
to bring the metric into standard diagonal form.

However, at this point we have not taken into neither account the continuity equation \eqref{cemrnc}, nor the hypersurface orthogonal condition \eqref{hypersurface}. These are not satisfied with this velocity flow.  

What we can do is this; pick $c_s$ to be constant, and set $v_a$ as above. This means that 
\begin{equation}
 u_a= g(r)v_a
\end{equation}
But the hypersurface orthogonal condition \eqref{hypersurface} implies 
\begin{equation}
g(r)=\frac{K}{v_t(r)},
\end{equation}
so $u$ is fixed up to a multiplicative constant. Then
\begin{equation}
\rho(r)=\frac{\bar{K}}{u_rr^2},
\label{eq:}
\end{equation}
which will appear in the conformal factor.

Thus, as in the non-relativistic case, we have a metric that is not identical to, but conformal to the Schwarzschild solution. For many purposes this will be enough, as many features (causal features, for example), are conformally invariant.

\subsection{Canonical acoustic metric}
\label{CAM}

From above, we see that, as in the non-relativistic analogue metrics, working with the exact Schwarzschild picture involves some complexity. Depending on the use to which one wishes to put the analogue spacetime to, it is advisable to either work with a metric conformal to Schwarzschild, or to take a simplified flow. We take the latter approach here, following closely on \cite{Visser:1997ux}.

Let us assume we have an incompressible fluid, $\rho=$ constant. In four dimensions, for spherical symmetry, the continuity equations implies 
\begin{equation}
 \rho\partial_r (r^2 u_r)=0 \quad \Rightarrow \quad u_r=\frac{KR_0^2}{r^2}
\end{equation}
And from the hypersurface orthogonality condition $u_t=\bar{K}$. Picking $\bar{K}=c, K=-c$ (where we have the negative sign to indicate ingoing flow), we have
\begin{equation}
 u_a =c \left\lbrace 1, -\frac{R_0^2}{r^2}, 0, 0 \right\rbrace
\end{equation}
and the metric becomes
\begin{eqnarray}
 \d s^2&=&\left[-1+\left(1-\frac{c_s^2}{c^2}\right)\left(\frac{1}{1-\frac{R_0^4}{r^4}}\right)\right]\d t^2-\left[\left(1-\frac{c_s^2}{c^2}\right)\left(\frac{R_0^2}{r^2}\frac{1}{1-\frac{R_0^4}{r^4}}\right)\right]\d t \d r \nonumber \\
&+& \left[1+\left(\frac{1-\frac{c_s^2}{c^2}}{\frac{R_0^4}{r^4}-1}\right)\right]\d r^2 +r^2\d\Omega^2
\end{eqnarray}
This can be converted to Schwarzschild coordinates by
\begin{equation}
 \d T=\d t-\frac{\frac{R_0^2}{r^2}\frac{1-c_s^2/c^2}{1-R_0^4/r^4}}{-1+\frac{ 1-c_s^2/c_2}{1-R_0^4/r^4}}\d r, 
\end{equation}
bringing the metric into the diagonal form
\begin{equation}
\d s^2=\left[-1+\left(1-\frac{c_s^2}{c^2}\right)\left(\frac{1}{1-\frac{R_0^4}{r^4}}\right)\right]\d t^2+\frac{{c_s^2}/{c^2}}{\left[-1+\left(1-\frac{c_s^2}{c^2}\right)\frac{1}{1-\frac{R_0^4}{r^4}}\right]}\d r^2 +r^2\d\Omega^2
\end{equation}

Now we check if this velocity flow has a Killing horizon. As there is no time dependence, and no rotation this is defined where $g_{tt}=0$, which is true when
\begin{equation}
 r=\sqrt{\frac{c}{c_s}}R_0 \equiv r_\KH
\end{equation}
so we have a Killing horizon at a positive $r$, as needed.

Similarly to the metric in section 8 of \cite{Visser:1997ux}, there is a singular sphere when $r^4=1$, which is precisely where $ u_a =\left\lbrace c, -c, 0, 0 \right\rbrace$, so becomes null with respect to the lab Minkowski metric. This will not concern us so long as we are only working with physics outside this singular surface.

\subsection{Killing Horizon in Relativistic Acoustic Metrics}

Let us consider again the relativistic acoustic metric of \ref{ram}, and make the assumption of stationary, so that there is a Killing vector associated to time translation isometries given by
\begin{equation}
 \chi^2=g_{tt}=-1+\left(1-\frac{c_s^2}{c^2}\right)\frac{v_t^2}{c^2}
\end{equation}
as $\chi^a=(1, \overrightarrow{0})$. Note that the condition that this becomes null, corresponds to the horizon in the static case. For the stationary case this is the condition for the ergosurface. From this, and using the unit norm constraint, 
\begin{equation}
 v_r^2|_{KH}=\frac{c_s^2}{\left(1-\frac{c_s^2}{c^2}\right)}, \qquad  v_t^2|_{KH}=\frac{c^2}{\left(1-\frac{c_s^2}{c^2}\right)}.
\end{equation}
This seems somewhat at odd with the usually non-relativistic flow condition for the horizon. however, in terms of charge current we have
\begin{equation}
 g_{tt} \propto -\left(1-\frac{u_au^a}{c_0^2}\right)+\frac{u_t^2}{c_0^2}=-1+\frac{u_r^2}{c_0^2}
\end{equation}
so
\begin{equation}
 u_r^2|_{KH}=c_0^2
\end{equation}
which appears more similar to the usual expression. Let us note that a similar approach is presented in \cite{Bilic:1999sq}, where the the velocity, $v_i$, used is related ours at horizon a rescaling of $\frac{c_s}{\sqrt{1-c_s^2/c^2}}$, so these conditions are in agreement with one another.

\subsection{Surface Gravity}

What of the surface gravity in such spacetimes? Here we will be dealing only with static Killing horizons (though this is easily extendable to stationary spacetimes), so we will not concern ourselves with the complications of \cite{Cropp:2013zxi} and chapter 2, though such concerns will be equally relevant when considering non-Killing horizons in the context of relativistic fluids. 

Taking the one of the standard definitions discussed in \cite{Cropp:2013sea} (in terms of inaffinity of null geodesics), the surface gravity, $\kappa$ of the Killing horizon is given by
\begin{equation}
 \chi^a\nabla_a \chi^b =\kappa\chi^b.
\end{equation}
Which we can directly calculate with the metric and the timelike Killing vector to give
\begin{equation}
 \kappa= \frac{c_s}{2}\frac{d}{dr}\left[\frac{(c^2-c_s^2)v_t^2}{c^2}\right]
\label{kapparbec}
\end{equation}
Note that taking $c_s, v_r \ll c$ we can recover the standard definition for non-relativistic fluid flow. Also note again the relation between our v and the one used in the expression for surface gravity in \cite{Bilic:1999sq}, as above. Taking this rescaling into account our result also agrees with that of \cite{Bilic:1999sq}. 
Finally, in terms of the charge current, we can re-write (\ref{kapparbec}) as
\begin{equation}
 \kappa=\frac{c_s}{2}\frac{d}{dr}\left(\frac{c_s^2}{c_0^2}u_t^2\right)
\end{equation}


\section{Coupling an External Field to the rBEC}
\label{coupledrbec}

To mimic universal horizons in analogue gravity framework it is necessary to input an additional structure to the rBEC system. As discussed in the introduction of this chapter, neither non-relativistic nor relativistic acoustic metrics give us enough freedom to simulate a region where arbitrarily fast supersonic modes can be trapped. One possible way to circumvent this problem is to take into account the role of an external background field coupled to the condensate. This will allow us to uncouple the fluid four-velocity from another preferred system, and associate to the latter the role of an acoustic {\aether} field, while the analogue metric would be determined by the charge current $u$. To this end, we will consider a electrically charged rBEC.
The minimal prescription implies that 
\begin{eqnarray}\label{emmc1}
\phi'&=&e^{-\frac{iq}{c\hbar}\alpha(x)}\phi, \\\label{emmc2}
\partial_{\mu}&\rightarrow&D_{\mu}=\partial_{\mu}+\frac{iq}{c\hbar}A_{\mu},
\end{eqnarray}
where $A^{\mu}$ is the gauge field. 
The $U(1)$ gauge invariant Lagrangian describing the interaction of the complex scalar field $\phi$ and the electromagnetic field can be written as
\begin{eqnarray}
\mathcal{L}=-\left(D_{\mu}\phi\right)^{\dagger}\left(D_{\mu}\phi\right)-m^{2}\phi^{\dagger}\phi-\lambda(\phi^{\dagger}\phi)^{2}-\frac{1}{4}F_{\mu\nu}F^{\mu\nu},
\end{eqnarray}
where $F_{\mu\nu}\equiv\partial_{\mu}A_{\nu}-\partial_{\nu}A_{\mu}$ is the electromagnetic field strength. The Noether theorem leads to a locally conserved current $j^{\mu}$ which is given by
\begin{equation}
j^{\mu}=i[\phi({D}^{\mu}\phi)^{\dagger}-\phi^{\dagger}({D}^{\mu}\phi)]=i(\phi\partial^{\mu}\phi^{\dagger}-\phi^{\dagger}\partial^{\mu}\phi)+\frac{q}{c\hbar}|\phi|^{2}A^{\mu}. 
\end{equation}
So that the conserved charge is
\begin{equation}
\mathcal{Q}=i\int{d^{3}x}\left[\phi({D}_{0}\phi)^{\dagger}-\phi^{\dagger}({D}_{0}\phi)\right], 
\end{equation}
which we can associate to a bosonic chemical potential $\mu$. 

The momentum canonically conjugate to $\phi$ is
\begin{equation}
\pi=\frac{\partial\mathcal{L}}{\partial\dot{\phi}}=\dot{\phi}^{\dagger}+iqA_{0}\phi.
\end{equation}
Since we are treating the fields $\phi$ and $\phi^{\dagger}$ independently, the Hamiltonian density will be given by
\begin{equation}
\mathcal{H}=\pi\dot{\phi}+\pi^{\dagger}\dot{\phi}^{\dagger}-\mathcal{L},
\end{equation}
and the partition function is
\begin{eqnarray}
\mathcal{Z}&=&\mathcal{N}\int(DA)({D}\pi)^{\dagger}({D}\pi)({D}\phi)^{\dagger}({D}\phi)\times {\det\left(\frac{\partial{F}}{\partial{\omega}}\right)\delta{F}}\nonumber \\
&&\exp\left\lbrace\int^{\beta}_{0}d\tau\int{d^{3}x}\left[\pi\dot{\phi}+\pi^{\dagger}\dot{\phi}^{\dagger}-\left(\mathcal{H}-\mu{Q}\right)\right]\right\rbrace.
\end{eqnarray}
Integrating the momenta away, we arrive at
\begin{eqnarray}
\mathcal{Z}=\mathcal{N}\int(DA)(D\phi)^{\dagger}(D\phi)\exp\left[\int^{\beta}_{0}d\tau\int{d^{3}x}\mathcal{L}_{eff}\right]\times{\det\left(\frac{\partial{F}}{\partial{\omega}}\right)\delta{F}} 
\end{eqnarray}
where
\begin{eqnarray}
\mathcal{L}_{\mathrm{eff}}&=&\left[\partial_{\mu}-\frac{iq}{c\hbar}\left(A_{\mu}+\frac{\mu}{q}\delta_{\mu0}\right)\right]\phi^{\dagger}\left[\partial^{\mu}+\frac{iq}{c\hbar}\left(A^{\mu}+\frac{\mu}{q}\delta^{\mu0}\right)\right]\phi-m^{2}\phi^{\dagger}\phi\nonumber \\
&-&\lambda(\phi^{\dagger}\phi)^{2}-\frac{1}{4}F_{\mu\nu}F^{\mu\nu}-e\mathcal{J_{\mu}}A^{\mu}  \nonumber\\
&=&\left(D_{\mu}\phi\right)^{\dagger}\left(D^{\mu}\phi\right)+i\mu\left(\phi\partial_{0}\phi^{\dagger}-\phi^{\dagger}\partial_{0}\phi\right)+2q\mu{A^{0}}\nonumber \\
&-&V(\phi)-\frac{1}{4}F_{\mu\nu}F^{\mu\nu}-e\mathcal{J_{\mu}}A^{\mu} 
\end{eqnarray}
where
\begin{equation}
V(\phi)=-\left(\mu^{2}-m^{2}\right)\phi^{\dagger}\phi+\lambda(\phi^{\dagger}\phi)^{2} 
\end{equation}
is the effective potential. It is clear from the above potential that in order to spontaneous break the $U(1)$ local symmetry, the condition $\mu^{2}>m^{2}$ must be satisfied.

Now, the equation of motion for $\phi$ is
\begin{eqnarray}
&&\left[\Box-\frac{m^{2}c^{2}}{\hbar^{2}}+2i\mu\partial_{0}-\mu^{2}+2\frac{iq}{c\hbar}A^{\mu}\partial_{\mu}\right. \nonumber \\
&+&\left.\frac{iq}{c\hbar}\partial_{\mu}A^{\mu}-\frac{q^{2}}{c^{2}\hbar^{2}}A^{\mu}A_{\mu}-2e\mu{A_{0}}-U'(\rho;\lambda)\right]\phi=0,
\end{eqnarray}
Making a shift $A_{\mu}\rightarrow{A}_{\mu}-\frac{\mu}{q}\delta_{\mu0}$, we can factor out the chemical potential dependence and express the field equation as
\begin{equation}
\left[\Box-\frac{m^{2}c^{2}}{\hbar^{2}}+2\frac{iq}{c\hbar}A^{\mu}\partial_{\mu}+\frac{iq}{c\hbar}\partial_{\mu}A^{\mu}-\frac{q^{2}}{c^{2}\hbar^{2}}A^{\mu}A_{\mu}-U'(\rho;\lambda)\right]\phi=0,
\end{equation}
which is an equation that describes a charged relativistic Bose--Einstein condensate. Following exactly the same steps of section \ref{uncoupledrbec}, we get the gauge invariant linearized perturbation equation
\begin{eqnarray}\label{gcrbec}
\left\lbrace\left[i\hbar{f^{\mu}}\partial_{\mu}+T_{\rho}\right]\frac{1}{c_{0}^{2}}\left[-i\hbar{f^{\nu}}\partial_{\nu}+T_{\rho}\right]-\frac{\hbar^{2}}{\rho}\eta^{\mu\nu}\partial_{\mu}\rho\partial_{\nu}\right\rbrace\psi=0,
\end{eqnarray}
where 
\begin{equation}
f^{\mu}\equiv{u}^{\mu}+\frac{q}{mc}A^{\mu}. 
\label{eq:fcurrent}
\end{equation}
is the gauge invariant four-velocity in the Madelung representation satisfying the conservation equation 
\begin{equation}\label{cemr}
\partial_{\mu}\left[\rho\left({u}^{\mu}+\frac{q}{mc}{A}^{\mu}\right)\right]=0.  
\end{equation}

The equation (\ref{gcrbec}) is a generalization of the linearized perturbation propagation equation in a charged relativistic Bose--Einstein condensate. 

It is clear from the charged conserved current that the role of the gauge field $A^{\mu}$ is basically perform a shift in the fluid four-velocity $u^{\mu}$.

\subsection{Modified Dispersion Relation}

As we commented at the end of the previous section, the equation describing the charged linearized perturbations has the same structure that the uncoupled case. This is useful since we can make use of the previous results of section \ref{uncoupledrbec}. In the ray optics limit/eikonal approximation, 
the Fourier decomposition of equation (\ref{gcrbec}) gives us
\begin{eqnarray}\label{gdr}
&&\left[\left({u^{0}}+\frac{q}{mc}A^{0}\right)\frac{w}{c}-\left({u^{i}}+\frac{q}{mc}A^{i}\right)k_{i}-\frac{\hbar}{2m}w^{2}+\frac{\hbar}{2m}\mathbf{k}^{2}\right]\nonumber\\
&\times&\left[-\left({u^{0}}+\frac{q}{mc}A^{0}\right)\frac{w}{c}+\left({u^{i}}+\frac{q}{mc}A^{i}\right)k_{i}-\frac{\hbar}{2m}w^{2}+\frac{\hbar}{2m}\mathbf{k}^{2}\right]\nonumber\\
&&-\left(\frac{c_{0}}{c}\right)^{2}w^{2}+c_{0}^{2}\mathbf{k}^{2}=0.
\end{eqnarray}
We can further simplify (\ref{gdr}) by going to the frame where $f^{i}=u^i+\frac{q}{mc}A^i=0$, which give us as solution
\begin{equation}\label{disp}
{w}_{\pm}^{2}=c^{2}\left\lbrace\mathbf{k}^{2}+2\left(\frac{m}{\hbar}\right)^{2}\alpha^{00}\pm2\left(\frac{m}{\hbar}\right)\sqrt{\left(\frac{m\alpha^{00}}{\hbar}\right)^{2}+\alpha^{00}\mathbf{k}^{2}-c_{0}^{2}\mathbf{k}^{2}}\right\rbrace.
\end{equation}
where
\begin{eqnarray}
\alpha^{00}=\left(u^{0}+\frac{q}{mc}A^{0}\right)\left(u^{0}+\frac{q}{mc}A^{0}\right)+c_{0}^{2}\mathbf{k}^{2}. 
\end{eqnarray}

Again, as we are interested in a description of phonons in an effective geometry, the relevant excitations to our problem are the gapless excitations in the low momentum regime 
\begin{equation}\label{klimit}
|\mathbf{k}|\ll\frac{m}{\hbar}\left(u^{0}+\frac{q}{mc}A^{0}\right)\left[1+\left(\frac{c_{0}}{u^{0}+\frac{q}{mc}A^{0}}\right)^{2}\right], 
\end{equation}
which gives us
\begin{eqnarray}\label{fmdr}
w^{2}&\approx&c^{2}\left\lbrace\frac{\left(\frac{c_{0}}{u^{0}}\right)^{2}}{1+\left(\frac{c_{0}}{u^{0}}\right)^{2}+2\frac{q}{mc}\frac{A^{0}}{u^{0}}+\left(\frac{q}{mc}\right)^{2}\left(\frac{A^{0}}{u^{0}}\right)^{2}}\mathbf{k}^{2}\right.\nonumber\\
&+&\left.\frac{\left[1+2\frac{q}{mc}\frac{A^{0}}{u^{0}}+\left(\frac{q}{mc}\right)^{2}\left(\frac{A^{0}}{u^{0}}\right)^{2}\right]^{2}}{\left[1+\left(\frac{c_{0}}{u^{0}}\right)^{2}+2\frac{q}{mc}\frac{A^{0}}{u^{0}}+\left(\frac{q}{mc}\right)^{2}\left(\frac{A^{0}}{u^{0}}\right)^{2}\right]^{3}}\frac{\mathbf{k}^{4}}{4(mu^{0}/\hbar)^{2}}\right\rbrace.
\end{eqnarray}
At low momentum range, is clear from (\ref{fmdr}) that our system can have supersonic modes. 

\subsection{Coupled Relativistic Acoustic Metric}\label{coupledmetric}

As in the uncoupled case, it is possible neglect the quantum potential $T_{\rho}$ in the eikonal approximation. With that, our new relativistic effective metric describing the propagation of the massless excitation is
\begin{eqnarray}\label{nrbec}
\partial_{\mu}\left[\frac{\rho}{c_{0}^{2}}{u^{\mu}}{u^{\nu}}-\rho\eta^{\mu\nu}+2\frac{\rho}{c_{0}^{2}}\frac{q}{mc}{u^{(\mu}}A^{\nu)}+\frac{\rho}{c_{0}^{2}}\frac{e^{2}}{m^{2}c^{2}}A^{\mu}A^{\nu}\right]\partial_{\nu}\psi=0. 
\end{eqnarray}

One might now hope to interpret the first two terms as a d'Alembertian, to which we can associate the standard acoustic metric (\ref{ram}), and the other terms as the interaction (which adds a force term, rather than free geodesic motion) of an acoustic perturbation with an analogue {\aether} field defined as
\begin{equation}
a^{\mu}\equiv{c}_{s}\frac{A^{\mu}}{\sqrt{-g_{\alpha\beta}A^{\alpha}A^{\beta}}} 
\end{equation}
However there is a problem with this approach, that $A$ is a gauge field, and as such, physically unobservable. The only gauge invariant quantity is $f^\mu$, which is the conserved current, and we can always write the metric
\begin{equation}
g_{\mu\nu}=\frac{\rho}{\sqrt{1-f_{\alpha}f^{\alpha}/c_{0}^{2}}}\left[\eta_{\mu\nu}\left(1-\frac{f_{\alpha}f^{\alpha}}{c_{0}^{2}}\right)+\frac{f_{\mu}f_{\nu}}{c_{0}^{2}}\right]. 
\label{gcoupledbec}
\end{equation}
So we appear to be back to the original scenario, where our only possible \aether\ , $f^\mu$, is also what defines the metric. 
Does this mean the coupling makes essentially no change?

\subsection{Vorticity}
To derive a metric, normally one must assume the vorticity is zero (consider what we have done in section \ref{BEC}, or the standard fluid derivation in \cite{Barcelo:2005fc}), which places restrictions on the form of the flows chosen. In particular, working within physical acoustics, it is a long-standing problem to introduce vorticity into this the analogue spacetime framework \cite{PerezBergliaffa:2001nd}. We will show that systems coupled this way can have non-zero vorticity, in a natural and simple manner. 

Consider the full metric, incorporating both the flow and the coupling, of equation \ref{gcoupledbec}, formed with $f^\mu = u^\mu +\frac{q}{mc} A^\mu$, and consider the vorticity tensor
\begin{eqnarray}
\omega_{\mu\nu}&\equiv& h^\rho_\mu h^\sigma_\nu\nabla_{[\rho}f_{\sigma]} \nonumber\\
&=&\left(\delta^\rho_\mu-f^\rho f_\mu\right)\left(\delta^\sigma_\nu-f^\sigma f_\nu\right) \nabla_{[\rho}f_{\sigma]} \nonumber\\
 &=&\frac{2q}{mc} \left(\delta^\rho_\mu-f^\rho f_\mu\right)\left(\delta^\sigma_\nu-f^\sigma f_\nu\right) F_{\mu\nu}.
\end{eqnarray}
Sot the vorticity will automatically be non-zero as long as  $F_{\mu\nu} \neq 0$.

If we want to split the potential into $\left\{\phi, \vec A \right\}$, we can see this is direct proportional to the magnetic field. 

\section{Coupling a Proca Field}
 
Previously we had the problem of gauge freedom, and our \aether\ field candidate was gauge-dependent, and therefore unphysical. Fundamentally this comes from the original assumption of gauge coupling --- this is a nice simple and physically realizable option, but is evidently too simple for our purposes. Furthermore the physical motivation for our investigation is more theoretical in nature as nor rBEC has yet been realized in a laboratory. This allows us to consider more general fields and couplings which might not be so simple to realize experimentally as gauge coupling. 
Given our previous investigation, a very economic approach could be to consider again a vector field but remove the obstruction related to gauge invariance. 

Therefore instead of the standard electromagnetic field, we can try to couple a Proca field to our rBEC. This is a massive vector field,
\begin{equation}
\mathcal{L} =\frac{1}{4}F^2_{\mu \nu} +\frac{m^2 c^2}{8\pi \hbar^2} A^2. 
\end{equation}
obviously the mass term is \emph{not} gauge invariant. However, the field equations imply that the Proca field must obey
\begin{equation}
m_A^2 \partial_\mu A^\mu =0
\end{equation}
which is the Lorentz condition, albeit this is no longer a gauge condition. For more details see \cite{martinshaw}.

We couple this field via
\begin{equation}
\mathcal{L} =-\frac{1}{2}\eta^{\mu\nu}\left(\partial_\mu \phi^\dag\right)\left(\partial_\nu \phi\right)-\frac{1}{4}F^2_{\mu \nu} +\frac{m^2 c^2}{8\pi \hbar^2} A^2+ j_\mu A^\mu +V(\phi),
\end{equation}
where $j^\mu$ is given by \ref{jcurrent}.

\subsection{Metric and Dispersion Relation}

Similarly to section \ref{coupledrbec} we can write the equation for the perturbation as 
\begin{eqnarray}
&&\left[u^\mu \partial_\mu\frac{1}{c_0}u^\nu \partial_nu+\frac{q}{mc}u^\mu \partial_\mu\frac{1}{c_0}A^\nu \partial_nu +\frac{q}{mc}A^\mu \partial_\mu\frac{1}{c_0}u^\nu \partial_nu\right. \nonumber \\
&+&\left.\frac{q^2}{m^2c^2}A^\mu \partial_\mu\frac{1}{c_0}A^\nu \partial_nu - \frac{1}{\rho}\eta^{\mu\nu}\partial_\mu \rho\partial_\nu\right]\psi=0
\end{eqnarray}
Now, using the conservation equation,
\begin{equation}
\partial_\mu \left(\rho u^\mu\right)=0,
\end{equation}
the Lorentz condition (now a field equation), and the condition that $\rho$ is a constant (we must now assume this separately, as the conserved current is $\rho u$ not $\rho f$), we can rewrite the perturbation equation as
\begin{eqnarray}
\partial_{\mu}\left[\frac{\rho}{c_{0}^{2}}{u^{\mu}}{u^{\nu}}-\rho\eta^{\mu\nu}+2\frac{\rho}{c_{0}^{2}}\frac{q}{mc}{u^{(\mu}}A^{\nu)}+\frac{\rho}{c_{0}^{2}}\frac{e^{2}}{m^{2}c^{2}}A^{\mu}A^{\nu}\right]\partial_{\nu}\psi=0. 
\end{eqnarray}

Given that $\rho u$ is still our conserved current, we want write the first two terms as the metric --- consider setting up the flow and hence the metric and turning on the coupling. As the flow does not change we regard it as natural to interpret the metric as the one associated to $u$ with no dependence on $A$ . We can interpret the other terms as a force term incorporating the coupling to the Proca field. 

As for the dispersion relation, we can borrow from the gauge coupling as the perturbation is insensitive to the mass term, so 
\begin{equation}
{w}_{\pm}^{2}=c^{2}\left\lbrace\mathbf{k}^{2}+2\left(\frac{m}{\hbar}\right)^{2}\alpha^{00}\pm2\left(\frac{m}{\hbar}\right)\sqrt{\left(\frac{m\alpha^{00}}{\hbar}\right)^{2}+\alpha^{00}\mathbf{k}^{2}-c_{0}^{2}\mathbf{k}^{2}}\right\rbrace.
\end{equation}
The key points here being that this the $t, r$ lab frame, and that we have higher order terms due to the quntum potential. 

\section{Analogue Universal horizons}

Now we turn our attention to the question of whether we can mimic the \aether\ and universal horizons. 

\subsection{Coupled canonical acoustic metric}
We could attempt to mimic the exact solutions of section \ref{ae-bhs}. For instance, for the $c_{123}=0$ solution we can follow the steps of section \ref{schwarz}, using
\begin{equation}
f(r)=1-\frac{r_0}{rc_s^2}-\frac{r_u(r_0+r_u)}{r^2c_s^4}.
\end{equation}
We remain with the same problem as before, we can match up to a conformal factor, with a non-constant $\rho$. We therefore take the simpler approach, using the canonical acoustic metric. This hopefully encompasses much of the physics needed for many uses of a analogue model without the complexity of the exact solution.

Let us then take the metric from section \ref{CAM}, and try to pick an external field such that we mimic a black hole with a universal horizon.

We want to mimic a static \aether\ , so we choose $A$ to be a function of $r$ only. 
Now the electrostatic Proca equations in 4D for spherical symmetry, are
\begin{equation}
\partial_r (r^2 A^r)=0 \quad \Rightarrow \quad A^r=\frac{C_2R_0^2}{r^2}.
\end{equation}
and 
\begin{equation}
\nabla^2 A_t - m_A^2 A_t =j_t.
\end{equation}
where $j$ is the conserved current, and so directly related to $u$. We wish to solve this for the simplest scenario we can, so we solve this using the current associated to the metric of \ref{CAM}. Here, $\rho$ and $u_t$ are both constants, and therefore so is $j$, so that the relevant equation is 
\begin{equation}
\frac{2r\frac{\d A_t}{\d r}+ r^2 \frac{\d^2 A_t}{\d r^2}}{r^2} - m_A^2 A_t =K, 
\end{equation}
so that
\begin{equation}
A_t =\frac{C_1 \exp\left(-m_Ar\right)}{r} - \frac{K}{m_A^2}
\end{equation}

\subsection{Universal Horizon} 

Universal horizon is where $\chi \cdot a=0$

Killing vector is $\chi^a=(1, 0)$, $a^a=\frac{A^a}{\sqrt{-g_{cd}A^cA^d}}$

\begin{eqnarray}\label{UHcondition}
 \chi^aa^bg_{ab}&=&\frac{1}{\sqrt{-g_{cd}A^cA^d}}\left[A^r\left(1-\frac{c_s^2}{c^2}\right)\frac{v_tv_r}{c^2}+A^t\left(-1+\left(1-\frac{c_s^2}{c^2}\right)\frac{v_t^2}{c^2}\right)\right]\nonumber \\
&=&\frac{1}{\sqrt{-g_{cd}A^cA^d}}\left[-A^t+\left(1-\frac{c_s^2}{c^2}\right)\left(\frac{A^rv_rv_t+A^tv_t^2}{c^2}\right)\right]
\end{eqnarray}
So the condition for the universal horizon is 
\begin{equation}
 A^t=\left(1-\frac{c_s^2}{c^2}\right)\frac{(A\cdot v)v_t}{c^2}=-\left(1-\frac{c_s^2}{c^2}\right)\frac{(A\cdot v)v^t}{c^2}.
\end{equation}
Note the inner product is taken with the Minkowski metric. 
Plugging the form of $A$ into this, we see we have a universal horizon if
\begin{equation}
{\frac {C_1\,{e^{-mr}}}{r}}-{\frac {K}{{m}^{2}}}-{\frac {\left( \frac{c_s^2}{c^2}-1 \right) \left(C_1\,e^{-mr}m^2r^{3}+C_2\,R_0^2m_A^2-Kr^4 \right) }{ \left(R_0^4-r^4\right) m_A^2c^2}}=0
\label{uhproca}
\end{equation}
which cannot be solved exactly for $r$. 

Let us pick a specific set of coefficients ($C_1 =R_0, K=R_0^2, C_2=\frac{1}{2}, m_A=\frac{1}{R_0}$), so that our full solution is
\begin{equation}
 u_a =c \left\lbrace 1, -\frac{R_0^2}{r^2}, 0, 0 \right\rbrace
\end{equation}
and
\begin{equation}
A_a = \left\lbrace \frac{R_0 \exp\left(-r/R_0\right)}{r} -1, -\frac{R_0^2}{2r^2}, 0, 0 \right\rbrace.
\end{equation}
This is a rather artificial example as there is no reason the mass scale of the Proca field should be at all related to a length scale of the metric. However, this is certainly enough to show the existence of a universal horizon. Indeed, we see the condition \ref{uhproca} is true at $r \approx 0.8\sqrt{\frac{c_s}{c}}R_0$ while the Killing horizon is $r = \sqrt{\frac{c_s}{c}}R_0$.

\begin{figure}[!htb]
\centering
\includegraphics[scale=0.6]{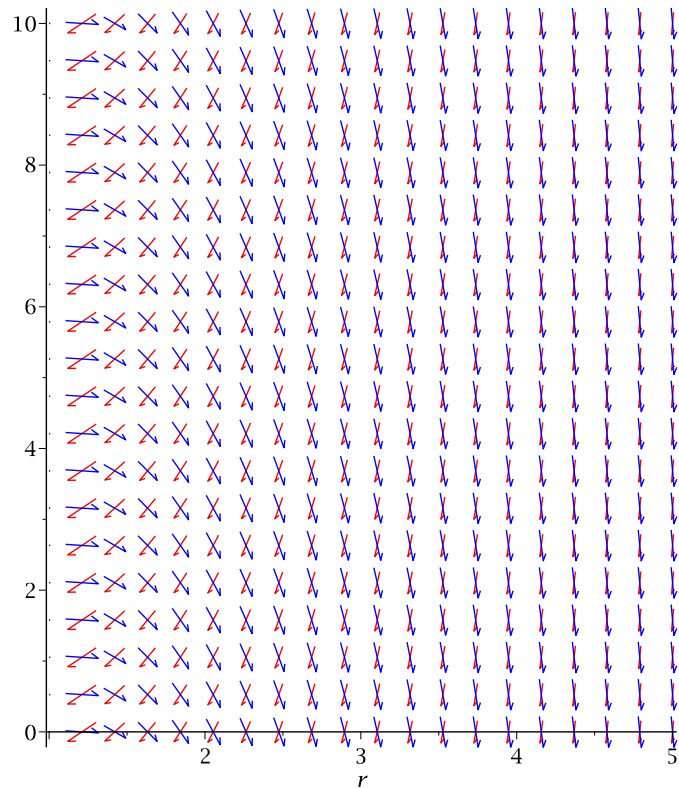}
\caption{$v_a$ (red) and $a_a$ (blue). One can see the universal horizon as the location where $a$ is horizontal}
\label{blueadownredvdown}
\end{figure}

One can see the universal horizon by plotting $v, a$ as in figure \ref{blueadownredvdown}. 

\subsubsection{Ray Tracing}

To understand whether this surface truly blocks perturbations, we work adapt the ray tracing picture of \cite{Cropp:2013sea}. Our dispersion relation, defined at infinity, is given by, for $k_r \ll 1$
\begin{equation}
 \Omega=b_1k_r+b_3k_r
\end{equation}
with $b_{1, 3}$ constants, $\Omega=k_a\chi^a$, $k_r=k_a r^a$. 
As we will have to work inside the Killing horizon (where timelike and spacelike directions flip), we choose to work with the tetrad frame given by $a^a$ and a unit vector orthogonal to $a^a$, $s^a$. Taking
\begin{equation}
 \omega \equiv k_aa^a ; \qquad k_s \equiv k_as^a
\end{equation}
with the transformations between the two given by
\begin{equation}
 \Omega= \omega (a\cdot \chi) \pm k_s (s\cdot \chi) ; \qquad k_r=\frac{k_s-\Omega(s\cdot \chi)}{s^r}
\end{equation}
From which we can derive a relation on $k_s(\Omega, r)$ and $\omega=f(k_s, \Omega, r)$. Explicitly, 
\begin{eqnarray}
k_s&=& \frac{1}{6}\frac{(\chi\cdot a)}{b_3}\sqrt[3]{\left(12\sqrt{3}\sqrt{\frac{27\Omega^2b_3+4b_1^3}{b_3}}+108\Omega\right)b_3^2}\nonumber\\
&-&\frac{2b_1(\chi\cdot a)}{\sqrt[3]{\left(12\sqrt{3}\sqrt{\frac{27\Omega^2b_3+4b_1^3}{b_3}}+108\Omega\right)b_3^2}}+\Omega(\chi\cdot s)
\end{eqnarray}
which, differently from \cite{Cropp:2013sea} and chapter 3, is regular on the universal horizon. However, 
\begin{equation}
v_g\equiv \frac{d\omega}{dk_s}=-\frac{-3\Omega^2b_3(\chi\cdot s)^2+6\Omega b_3k_s(\chi\cdot s)-(\chi\cdot s)(\chi\cdot a)^3-b_1(\chi\cdot a )^2-3b_3k_s^2}{(\chi\cdot a)^4}
\end{equation}
Which diverges at the universal horizon.
In this frame we can now again write
\begin{equation}
 \frac{dt}{dr} = \frac{a^t+v_g s^t}{a^r+v_g s^r}
\end{equation}
For infinite group velocity, this is
\begin{equation}
  \frac{dt}{dr} = \frac{s^t}{s^r}= \frac{a_r}{a_t}
\end{equation}
where the second equality is a consequence of the fact that $s$ is orthogonal to $a$. This means that infinite velocity rays (in the tetrad frame) travel along constant khronon surfaces. This is the basis of the blocking mechanism of the universal horizon, as, by construction, no constant khronon surface can cross it. 







\section{Discussion}

Relativistic Bose Einstein condensate provide an exciting analogue system, that is relativistic (with limiting speed $c_s$) at low energies, and at high energies (with limiting speed $c$). Thus they provide and interesting theoretical testbed for studying effects present in Lorentz-violating theories of gravity. 

We have explored incorporating new structure into analogue geometries. The hope is to break the degeneracy between the what sets the preferred frame and what sets the metric, allowing us to reproduce an \aether\ field and the universal horizon. 

To that end, we have explored coupling a vector field to the a rBEC. The simplest, minimally coupled gauge field does not provide enough new structure to allow this, by nature of being a gauge field. This does however allow a simple way to incorporate vorticity into analogue systems.

To mimic the \aether\ we introduce a Proca field, removing the problematic gauge freedom, and found a solution with a universal horizon. However, currently, getting a physical grasp of how the blocking mechanism works is proving elusive, and further investigations are needed. It seems that the Proca field coupling to charged excitations can force them to propogate along field lines. As a matter of fact,  one might imagine that if excitations of any energy cannot propogate against the preferred direction imposed by the Proca field then at most they can move along a surface orthogonal to it at high energies, i.e. the khronon. At present this is speculation, albeit a plausible one. 

\chapter{Conclusions and future directions}

Black holes have been extensively studied within the framework of general relativity for the past 99 years. Throughout this time, much has been understood about general relativity by studying such simple interesting systems. However, general relativity is not the final word of gravity, especially in light of problems with unifying it with quantum mechanics. 

Given how useful they have proven for our understanding, figuring out the key aspects of black holes in other theories of gravity is expected to gain us both an understanding of how theories of gravity can work, and what our understanding of a black hole is. 

This thesis explored two lines of inquiry into understanding black holes beyond general relativity, via analogue models and Lorentz violating modified gravity. 

Analogue gravity gives us a testbed in which to explore aspects of kinematics (and sometimes dynamics) of curved spacetime. For black holes this has given us insight into lines of inquiry such as, for instance, the robustness of Hawking radiation. As could be expected, given the differing dynamics of the set up, black holes in such systems do not obey the same symmetries as black holes in general relativity. 

In particular, for a axisymmetric black holes, enhanced symmetry on-horizon means a possible wealth of definitions, representing different physical properties, of surface gravity are reduced to just one. Analogue gravity provides a natural setting in which to explore such definitions, giving us a picture of the differences in terms of in-horizon velocity flows and shears. But such concerns are broader than just the analogue gravity programme: the simplicity and symmetry of the axisymmetric solutions of general relativity are crucially dependent on the field equations, and can be expected to fail for general theories of gravitation. Given the role surface gravity plays within the thermodynamic picture of black holes, elucidating which surface gravity we want to study and the differences between them becomes important.

Within the analogue framework, generically modified dispersion relations are present, and at high enough energies sonic black holes no longer act as barriers; the horizon is merely a low-energy effect. One might imagine there would be similar consequences in any theory with Lorentz violation. However, in some theories, a new barrier is present, and understanding these universal horizons has made a significant part of this thesis. 

It could be hoped that universal horizons would provide an escape from issues plaguing black hole thermodynamics in Lorentz violating theories. However, to understand whether this is the case we must understand the physics of this new type of horizon. Can we associate an entropy to it? What is its surface gravity?
Does it radiate? What is the role of the Killing horizon?

We have explored these issues in several ways. Ray tracing is one of the least fraught ways to study such spacetimes; many methods developed to study black holes rely on the \emph{pure metric} nature of general relativity. The very definition of a universal horizon is dependent of non-metric aspects of these theories, and such aspects cannot be ignored. Such rays are not geodesic (except possibly in the rainbow metric/Finslerian approach). These rays peel off the universal horizon, and as such can have a surface gravity associated to this peeling, while a lingering is found for the low-energy rays near the Killing horizon. Together with previous calculations done on tunnelling at the universal horizon, this suggests a temperature, and radiation can be associated to the universal horizon while some reprocessing of mode conversion may still occur at the Killing horizon. However, collapse scenario calculation brings this picture into doubt, and much is still yet to be fully understood. 

To shed light on this, we have explored some avenues for a new calculation of the Hawking effect at the universal horizon, and have identified two main points of difficulty: setting the correct boundary condition for the vacuum state, and the more complicated pole structure at the universal horizon. Working through a calculation which, unlike the tunnellling approach, will tell us the effect of the Killing horizon would be most valuable. 

Another way we may seek to understand the universal horizon is by returning to the analogue framework, and seeking out a way to model this new feature. The difficulty is that, in the standard approach there is not enough freedom to separately set the preferred frame and the geometry, a freedom which is crucial for the existence of the universal horizon. 

We have sought to tackle this problem by using a Bose--Einstein condensate (which provides some obvious methods to couple fluids to another field) and, given the four-dimensional nature of the \aether\ have opted to use a relativistic Bose Einstein condensate. 

These systems have been less studied than their non-relativistic counterparts, given the possibilities the latter offer for real experimental set-ups. Therefore we take some time to explore the relativistic acoustic metric and its properties which have been less elucidated in the literature. In particular, the simplest black holes system is devised, and the closest match to the Schwarzschild black hole derived. We then explore the coupled system and its physics, to provide and \aether\ for our system. In particular, the coupling of a Proca field to a realistic BEC seems a promising avenue in this sense


Much of this thesis has been developing a toolset for understanding general black holes. When we move to new theories, and new features within these theories which may change our notion of black hole, how do we begin to study such objects? What changes? what notions can be kept and what discarded? Given the breadth of such questions, there obviously remain many aspects unanswered and unexplored. 

However, exploring different aspects of this problem, and forging these pieces and hints together, we begin to gain a greater understanding of these strange black holes beyond general relativity.

\appendix

\titleformat{\chapter}[frame]{\normalfont}{\filright APPENDIX \thechapter}{8pt}{\Huge\rm\filcenter}

\chapter{Some relations for spherically symmetric, static \aether\ black holes}
\label{aetherelations}

For convenience several relations between \aether\ and metric components are collected here. 

For both the exact solutions of section \ref{ae-bhs}, we can express
\begin{equation}
f(r) = -\chi^2 = (u\cdot \chi)^2 -(s\cdot \chi)^2.
\end{equation}
For concreteness, take Eddington-Finkelstein coordinates, then
\begin{equation}
u_v = (u\cdot \chi)
\end{equation}
\begin{equation}
u_r = \frac{-1}{(u\cdot \chi) + (s\cdot \chi)}
\end{equation}
This makes it explicit that the system $g_{ab}, u_a$ only has two free functions, not three as it may at first appear. 

From the condition that $u$ and $s$ are orthogonal
\begin{equation}
u^ts_t +u^rs_r =0,
\end{equation}
we have 
\begin{equation}
\frac{u^t}{u^r} = -\frac{s_r}{s_t} \quad \mathrm{equivalently} \quad  \frac{u_t}{u_r} = -\frac{s^r}{s^t}
\end{equation}

From this and the unit condition, we can now express
\begin{equation}
s^v =\frac{1}{s\cdot \chi - u\cdot \chi}; \qquad s^r =-u\cdot \chi
\end{equation}

This makes it clear why the khronon runs to infinity at the universal horizon as
\begin{equation}
\tau =v+\int \frac{u_r}{u_v} \;\d r. = v-\int \frac{1}{(u\cdot \chi)((u\cdot \chi) + (s\cdot \chi))} \;\d r.
\end{equation}

\vspace{1cm}

\noindent Taking the $c_{123}=0$ solution we may always express
\begin{equation}
u\cdot \chi =-1 +\frac{r_\UH}{r}; \qquad s\cdot \chi = \frac{r_\KH-r_\UH}{r} 
\end{equation}
or, alternatively, 
\begin{equation}
u\cdot \chi =\frac{\kappa_\UH-\kappa_\KH}{2\kappa_\UH r \sqrt{\kappa_\UH \kappa_\KH}}-1
\end{equation}
and
\begin{equation}
s\cdot \chi = \frac{\frac{1}{2}\frac{\kappa_\UH-\kappa_\KH}{\kappa_\UH\kappa_\KH}-\frac{1}{2}\frac{\kappa_\UH-\kappa_\KH}{\kappa_\UH\sqrt{\kappa_\KH\kappa_\UH}}}{r}
\end{equation}
where $\kappa_\UH$ is the one found via ray tracing, while $\kappa_\KH$ is the standard metric one. 
Consequently, one can see that $k_s(r_0, r_u, r, \Omega)$ can always be re-expressed in terms of $k_s(r_\UH, r_\KH, r, \Omega)$ or $k_s(\kappa_\UH, \kappa_\KH, r, \Omega)$. 

\vspace{1cm}

\noindent Throughout the thesis we have used both $k_r$ and $k_s$, depending on the purpose. The relation between them is defined by
\begin{equation}
k_s \equiv k_as^a =k_v s^v +k_r s^r
\end{equation} 
so
\begin{equation}
 k_r=\frac{k_s+\Omega s^v}{s^r} = \frac{k_s+\frac{\Omega}{s\cdot \chi - u\cdot \chi}}{-u \cdot \chi} 
\end{equation}



\addcontentsline{toc}{chapter}{Bibliography}

\bibliography{myrefs}{}
\bibliographystyle{eprint}

\end{document}